\documentclass[
onecolumn,secnumarabic,amssymb, amsmath, 
superscriptaddress,
nofootinbib,tightenlines,nobibnotes, aps,prd]{revtex4-1}

\usepackage{graphicx}
\usepackage{hyperref}
\usepackage{multirow}
\usepackage{xcolor}
\usepackage{cleveref}

\begin{document}

\preprint{FERMILAB-PUB-24-0074-V}

\title{Global Fit of Electron and Neutrino Elastic Scattering Data to Determine the Strange Quark Contribution to the Vector and Axial Form Factors of the Nucleon}

\author{S.F. Pate\footnote{spate@nmsu.edu}}
\author{V. Papavassiliou}
\author{J.P. Schaub\footnote{current address: South Puget Sound Community College, 2011 Mottman Road SW, Olympia, WA, 98512, USA}}
\author{D.P. Trujillo\footnote{current address: Mercurial AI, 222 Merchandise Mart Plaza, Chicago, IL, 60654, USA}}
\affiliation{Physics Department, New Mexico State University, Las Cruces, New Mexico, 88003, USA}
\author{M.V. Ivanov}
\affiliation{Institute for Nuclear Research and Nuclear Energy, Bulgarian Academy of Sciences, Sofia 1784, Bulgaria}
\author{M.B. Barbaro}
\affiliation{Universit\`a degli Studi di Torino, Turin, Italy}
\affiliation{Istituto Nazionale di Fisica Nucleare, Sezione di Torino, Italy}
\author{C. Giusti}
\affiliation{Istituto Nazionale di Fisica Nucleare, Sezione di Pavia, Italy}

\date{\today}

\begin{abstract}
We present a global fit of neutral-current elastic (NCE) neutrino-scattering data and parity-violating electron-scattering (PVES) data with the goal of determining the strange quark contribution to the vector and axial form factors of the proton.  Previous fits of this form included data from a variety of PVES experiments (PVA4, HAPPEx, G0, SAMPLE) and the NCE neutrino and anti-neutrino data from BNL E734.  These fits did not constrain the strangeness contribution to the axial form factor $G_A^s(Q^2)$ at low $Q^2$ very well because there was no NCE data for $Q^2<0.45$ GeV$^2$. Our new fit includes for the first time MiniBooNE NCE data from both neutrino and anti-neutrino scattering; this experiment used a hydrocarbon target and so a model of the neutrino interaction with the carbon nucleus was required.  Three different nuclear models have been employed: a relativistic Fermi gas model, the SuperScaling Approximation model, and a spectral function model.  We find a tremendous improvement in the constraint of $G_A^s(Q^2)$ at low $Q^2$ compared to previous work, although more data is needed from NCE measurements that focus on exclusive single-proton final states, for example from MicroBooNE.
\end{abstract}

\maketitle

\section{Motivation:  Strange Quark Contribution to Nucleon Structure}

The strange quark contribution to the vector and axial form factors of the nucleon has been a subject of experimental and theoretical research for many decades.  The contribution to the axial form factor, $G_A^s(Q^2)$, became of great interest when it was discovered by the EMC~\cite{AshmanEMC} experiment that the up, down, and strange quarks did not contribute very significantly to the total spin of the nucleon.  Interest grew in the contribution to the electric and magnetic form factors, $G_E^s(Q^2)$ and $G_M^s(Q^2)$, when it was realized that this contribution could be measured using parity-violating electron-scattering from protons and light nuclei~\cite{KAPLAN1988527,MCKEOWN1989140}.

Subsequently, a great number of measurements in deep-inelastic scattering (DIS) of longitudinally-polarized leptons (electrons, positrons, and muons) from longitudinally-polarized targets were undertaken~\cite{RevModPhys.85.655} with the goal to understand the up, down, and strange quark contributions to the nucleon spin, called $\Delta u$, $\Delta d$, and $\Delta s$ respectively.  Many of these measurements (for example \cite{Airapetian:2007mh}) used {\em inclusive} DIS; that is, only the scattered lepton was observed in the final state.  To extract the $u$, $d$, and $s$ quark polarizations from inclusive DIS data, it is necessary to assume SU(3) flavor symmetry and make use of the beta-decay $F$ and $D$ coefficients.  
Other experiments (for example \cite{Airapetian:2004zf,Alekseev:2010ub}) used {\em semi-inclusive} DIS, commonly called SIDIS, where at least the leading hadron was also detected in the final state along with the scattered lepton. The analysis of these data does not require any assumptions about SU(3) symmetry, but do require knowledge of quark fragmentation functions, which must come from the experiments themselves.  Analyses of DIS results usually point to a negative value of $\Delta s$, while the analyses of SIDIS data usually point to a zero value of $\Delta s$.  The tension between these two types of analyses was starkly indicated in the work of de Florian et al.~\cite{deFlorian:2009vb}.

It is possible to access the longitudinal spin contribution of the strange quarks to the spin of the nucleon, $\Delta s$, through a measurement of the strangeness contribution to the axial form factor, $G_A^s(Q^2)$, if measurements are made at sufficiently low $Q^2$: $\Delta s = G_A^s(Q^2=0)$; the ``strangeness'' here is a sum of the contribution of strange and anti-strange quarks.  A program of measurements using neutrino neutral-current elastic scattering (NCES) was also undertaken in parallel to the effort in leptonic DIS.  The E734 experiment at Brookhaven National Laboratory performed a measurement of neutral-current elastic scattering on a hydrocarbon-based target/detector system, and extracted the neutrino-proton and antineutrino-proton cross sections in the momentum-transfer range $0.45 < Q^2 < 1.05$ GeV$^2$ \cite{Ahrens:1986xe}.  By making assumptions about the $Q^2$-behavior of $G_A^s(Q^2)$, they were able to obtain a value for $\Delta s$, but they also found that this value was strongly correlated to the assumptions that were made. The LSND experiment at Los Alamos National Laboratory proposed to measure the ratio of yields of neutral current scattering from protons and neutrons, $\sigma_p^{\rm NC}/\sigma_n^{\rm NC}$, in a liquid scintillator target/detector system \cite{GARVEY1995245}, but the neutron detection efficiency was never understood well enough to allow a useful result. The MiniBooNE experiment at Fermi National Accelerator Laboratory studied neutrino- \cite{PhysRevD.82.092005} and antineutrino-induced \cite{PhysRevD.91.012004} neutral-current elastic scattering in a mineral-oil based target/detector system.  For the neutrino-induced data, two analyses were performed.  In the first analysis only the scintillation light from the final state proton was considered, which meant the kinetic energy threshold could be as low as 50 MeV, but also secondary protons from NCES events on neutrons are included in the yield. In the second analysis, also the Cherenkov light produced by the proton is considered, which raises the kinetic energy threshold to 350 MeV but excludes contributions from NCES events on neutrons.  The antineutrino-induced data only employed the first sort of analysis.

Four programs of measurements meanwhile took place focusing on the strange quark contribution to the electric and magnetic form factors $G_E^s(Q^2)$ and $G_M^s(Q^2)$, using the technique of parity-violating electron scattering (PVES) from protons, deuterons, and $^4$He nuclei.  The SAMPLE experiment~\cite{Beise:2004py} at the MIT/Bates accelerator center focused on backward-scattering of electrons from liquid hydrogen and deuterium targets at very low $Q^2$; in these kinematic conditions the contribution of the strangeness electric form factor may be ignored, and the focus was on determining the strangeness contribution to the magnetic moment, $\mu_s$.  The G0 Experiment~\cite{Armstrong:2005hs, Androic:2009zu} at Jefferson Lab looked at forward-scattering from hydrogen over a wide momentum transfer range $0.1<Q^2<1.0$ GeV$^2$, and also at backward-scattering from hydrogen and deuterium targets at $Q^2$ = 0.221 and 0.628 GeV$^2$.  HAPPEx~\cite{Aniol:2004hp,Aniol:2005zf,Acha:2006my,Aniol:2005zg,Ahmed:2011vp}, also carried out at Jefferson Lab, looked at forward-scattering from protons and $^4$He targets at a few selected values of $Q^2$.  Finally, the PVA4 experiment~\cite{Maas:2004dh,Maas:2004ta,Baunack:2009gy,BalaguerRios:2016ftd} at the Mainz Microtron looked at forward-scattering from hydrogen, and backward-scattering from hydrogen and deuterium, at selected values of $Q^2$.

\begin{figure}[ht]
\centering
\includegraphics[scale=0.5]{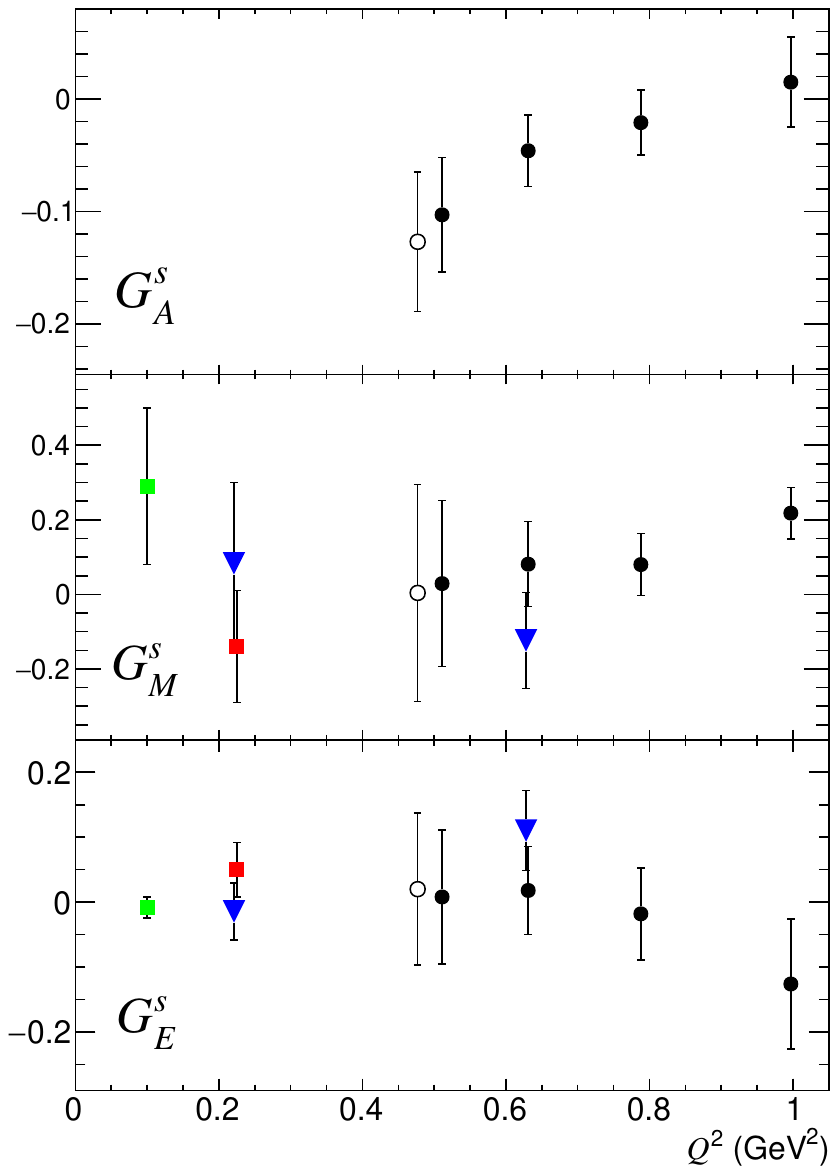}
\caption{Independent determinations of the strangeness form factors of the nucleon using subsets of existing experimental data: Liu et al.\ (green squares) \cite{Liu:2007yi}; Androi\'{c} et al.\ (blue triangles) \cite{Androic:2009zu}; Baunack et al.\ (red squares) \cite{Baunack:2009gy}; Pate et al.\ (open circles use HAPPEx and E734 data, and closed circles use G0-Forward and E734 data) \cite{Pate:2008va}.  This selection of results is representative and not intended to be exhaustive.}
\label{fig:old_points}
\end{figure}

When the first PVES results from HAPPEx~\cite{Aniol:2004hp} at $Q^2 = 0.447$ GeV$^2$ became available, it became possible to combine that measurement with the NCES cross sections from BNL E734~\cite{Ahrens:1986xe} and determine values for all three strangeness form factors $G_E^s$, $G_M^s$, $G_A^s$ at finite $Q^2$, and this was done for the first time in Ref.~\cite{Pate:2003rk}.  Later, when the G0 forward-scattering results became available, the analysis of Ref.~\cite{Pate:2003rk} was extended~\cite{Pate:2008va} to several points in the range $0.45<Q^2<1.0$ GeV$^2$.  These results, and those of other researchers who used only PVES data, are reviewed in Fig.~\ref{fig:old_points}.  In that figure it is clear that the strangeness contribution to the vector form factors, $G_E^s$ and $G_M^s$, is for the most part consistent with zero, and this has been noted in reviews of the PVES measurements, for example in Refs.~\cite{Armstrong:2012bi,MAAS2017209}.  On the other hand, the values for $G_A^s$ seem to indicate a significant $Q^2$-dependence, with the value trending negative with decreasing $Q^2$, suggesting a negative value for $\Delta s$.

In an effort to use all of the available NCES and PVES data to determine $G_E^s$, $G_M^s$, and $G_A^s$ for $Q^2<1\textrm{ GeV}^2$, a 
fitting program was developed using simple models for those form factors, and a preliminary version of such a fit was presented in Ref.~\cite{Pate:2013wra}.  That study made it clear that new exclusive NCES data in the range $Q^2<0.45$ GeV$^2$ could lead to a determination of $\Delta s$.  Unfortunately, the inclusive MiniBooNE NCES data do not fall into that category, because the yields include NC interactions on both protons and neutrons, which weakens the sensitivity to $G_A^s$.  However, the sensitivity is not completely lost and we will see in this paper that inclusion of that data acts to significantly constrain the low-$Q^2$ behavior of $G_A^s$.  

In this paper we will review the formalism for the interpretation of the PVES data on hydrogen, deuterium, and helium-4 from the SAMPLE, G0, HAPPEx, and PVA4 experiments, and that of the NCES data from BNL E734.  We will motivate two simple models that we use for the strangeness contribution to the vector and axial form factors.  The three models used for the neutrino-carbon interaction will be described, as well as the procedure for comparing those calculations to the MiniBooNE data.  The results of the fits of our form factor models to those data will be presented and discussed.

\section{Elastic Electroweak Scattering as a probe of Strangeness Form Factors}

The static properties of the nucleon are described by elastic form factors
defined in terms of matrix elements of current operators.
For example, the matrix element for the electromagnetic
current (one-photon exchange) is expressed as
\begin{eqnarray*}
\raisebox{-.4ex}{${}_N$}\!\left<p'\left| J_\mu^{\gamma}\right|p\right>_N &=&
 \bar{u}(p')\left[\gamma_\mu F_1^{\gamma,N}(Q^2)
  +  i\frac{\sigma_{\mu\nu}q^\nu}{2M}F_2^{\gamma,N}(Q^2)\right]u(p)
\end{eqnarray*}
where the matrix element is taken 
between nucleon states $N$ of momenta $p$ and $p'$,
the momentum transfer is $Q^2=-(p-p')^2$, $u$ is a nucleon spinor,
and $M$ is the mass of the nucleon.  
Similarly, the matrix element of the neutral weak current (one-$Z$ exchange) is
\begin{eqnarray*}
\raisebox{-.4ex}{${}_N$}\!\left<p'\left| J_\mu^{NC}\right|p\right>_N =
\bar{u}(p')\left[\gamma_\mu F_1^{Z,N}(Q^2) \right.
&+& i\frac{\sigma_{\mu\nu}q^\nu}{2M}F_2^{Z,N}(Q^2) \\
&+& \left. \gamma_\mu \gamma_5 G_A^{Z,N}(Q^2)
+\frac{q_\mu}{M}\gamma_5 G_P^{Z,N}(Q^2)\right]u(p).
\end{eqnarray*}
The form factors are respectively the Dirac and Pauli vector
($F_1$ and $F_2$), the axial ($G_A$), and the pseudo-scalar ($G_P$).
Due to the point-like interaction between the gauge bosons ($\gamma$ or $Z$)
and the quarks internal to the nucleon, these form factors can be expressed
as separate contributions from each quark flavor; 
for example, the electromagnetic
and neutral weak Dirac form factors of the proton 
can be expressed in terms of contributions from up, down, and strange quarks:
\begin{eqnarray*}
F_1^{\gamma,p} &=& \frac{2}{3}F_1^u - \frac{1}{3}F_1^d - \frac{1}{3}F_1^s \\
F_1^{Z,p} &=& \left(1-\frac{8}{3}\sin^2\theta_W\right)F_1^u
+\left(-1+\frac{4}{3}\sin^2\theta_W\right)F_1^d
+\left(-1+\frac{4}{3}\sin^2\theta_W\right)F_1^s,
\end{eqnarray*}
where $\sin^2\theta_W=0.23116$ is the square of the sine of the Weinberg mixing angle.
The same quark form factors are involved in both expressions; the 
coupling constants that multiply them (electric or weak charges) correspond
to the interaction involved (electromagnetic or weak neutral).
These measurements are most interesting for low momentum 
transfers, $Q^2< 1.0$~GeV$^2$, as the $Q^2=0$ values of these form factors
represent static integral properties of the nucleon.  It is common to use
in these studies the Sachs electric and magnetic form factors
\begin{equation}
G_E = F_1 - \tau F_2 ~~~~~~~~~~ G_M = F_1 + F_2    \nonumber
\end{equation}
instead of the Dirac and Pauli form factors; here, $\tau=Q^2/4M^2$.
At $Q^2=0$ the electromagnetic Sachs electric form factors take on the
value of the nucleon electric charges ($G_E^{\gamma,p}(0)=1$,
$G_E^{\gamma,n}(0)=0$) and the electromagnetic Sachs magnetic form
factors take on the value of the nucleon magnetic moments
($G_M^{\gamma,p}(0)=\mu_p$, $G_M^{\gamma,n}(0)=\mu_n$). Likewise, the
$Q^2=0$ values of the strange quark contributions to these form factors
define the strange contribution to these static quantities: for
example, the strangeness contribution to the proton magnetic moment is
$\mu_s = G_M^s(Q^2=0)$.  It is also common in these studies to assume
charge symmetry; the transformation from proton to neutron form factors 
is an exchange of $u$ and $d$ quark labels.  In addition, it is generally
assumed that the strange quark distributions in the proton and the
neutron are the same.  Then by combining the electromagnetic 
form factors of the proton
and neutron with the weak form factors of the proton, one may separate
the up, down, and strange quark contributions; for example, the
electric form factors may be written as follows:
\begin{eqnarray*}
G_E^{\gamma,p} &=& \frac{2}{3}G_E^u - \frac{1}{3}G_E^d - \frac{1}{3}G_E^s \\
G_E^{\gamma,n} &=& \frac{2}{3}G_E^d - \frac{1}{3}G_E^u - \frac{1}{3}G_E^s \\
G_E^{Z,p} &=& \left(1-\frac{8}{3}\sin^2\theta_W\right)G_E^u
+\left(-1+\frac{4}{3}\sin^2\theta_W\right)G_E^d
+\left(-1+\frac{4}{3}\sin^2\theta_W\right)G_E^s.
\end{eqnarray*}
To attempt this separation is the motivation behind 
the program of parity-violating
$\vec{e}p$ scattering experiments.

The $Z$-exchange current involves also the axial form factor of the
proton, which in a pure weak-interaction process takes
this form:
$$G_A^{Z,p} = \frac{1}{2}\left(-G_A^u + G_A^d + G_A^s\right).$$
The $u-d$ portion of this form factor is well-known from
neutron $\beta$-decay and other charged-current ($CC$) weak 
interaction processes like $\nu_\mu + n \rightarrow p+ \mu^-$:
$$G_A^{CC} = G_A^u-G_A^d = \frac{g_A}{(1+Q^2/M_A^2)^2}$$ 
where $g_A = 1.2670 \pm 0.0030$ is the axial coupling constant in neutron
decay~\cite{Cabibbo:2003cu} and $M_A = 1.014 \pm 0.014$ is the so-called
``axial mass'' which is a fitting parameter for the data on this form
factor~\cite{Bodek:2007ym}.  The strange quark
portion, $G_A^s$, is a topic of investigation.  In $\nu p$ and $\bar{\nu}p$
elastic scattering, which are pure neutral-current, weak-interaction
processes, there are no significant radiative corrections to be taken
into account~\cite{Marciano:1980pb}, and we may safely neglect heavy
quark contributions to the axial form factor~\cite{Bass:2002mv}.  On
the other hand, since elastic $ep$ scattering is not a pure
weak-interaction process, then the axial form factor does not appear
in a pure form; there are significant radiative corrections which
carry non-trivial theoretical uncertainties.  The result is that,
while the measurement of parity-violating asymmetries in $\vec{e}p$
elastic scattering is well suited to a measurement of $G_E^s$ and
$G_M^s$, these experiments cannot cleanly extract $G_A^s$.

\subsection{Experimental Measurements Sensitive to the Strangeness Form Factors
of the Nucleon} 

There are two principal sources of experimental data from which the
strange quark contribution to the elastic form factors of the proton
may be extracted.  One of these is elastic scattering of neutrinos and
anti-neutrinos from protons; these data are primarily sensitive to the
axial form factor.  The other is the measurement of parity-violating
asymmetries in elastic $\vec{e}p$ scattering; these data are primarily
sensitive to the vector form factors.  This section will describe
these two kinds of experiments.  

\begin{table}[t]
\caption{\label{table_params} Parameters used in this analysis.
  Uncertainties are listed only if they were of significant size and
  were used to generate the uncertainties in the results. The
  uncertainties on the three $R_V$ factors have been increased to 10\% of the value in order to account for their unknown $Q^2$-dependence.}
\begin{tabular}{c|c|c}
\hline
Parameter & Value & Reference \\
\hline
$\alpha$ & $7.2973\times 10^{-3}$ & \cite{PDG2012} \\
$\sin^2\theta_W$ & $0.23116$ & \cite{PDG2012} \\
$G_F/(\hbar c)^3$ & $1.16637\times 10^{-5}/{\rm GeV}^2$ & \cite{PDG2012} \\
$M_A$ & $1.014 \pm 0.014$ GeV & \cite{Bodek:2007ym} \\
$g_A=F+D$ & $1.2670 \pm 0.0030$ & \cite{Cabibbo:2003cu} \\
$3F-D$ & $0.585\pm 0.032$ & \cite{Cabibbo:2003cu} \\
$R^p_V$ & $-0.0520 \pm 0.0052$ & \cite{Liu:2007yi}     \\
$R^n_V$ & $-0.0123 \pm 0.0012 $ & \cite{Liu:2007yi}     \\
$R^{(0)}_V$ & $-0.0123 \pm 0.0012 $ & \cite{Liu:2007yi}     \\
$R^{T=1}_A$ & $-0.26\pm 0.34 $ & \cite{Liu:2007yi}     \\
$R^{T=0}_A$ & $-0.24\pm 0.20 $ & \cite{Liu:2007yi}     \\
$R^{(0)}_A$ & $-0.55\pm 0.55 $ & \cite{Liu:2007yi}     \\
\hline
\end{tabular}
\end{table}

\subsection{Parity-violating Asymmetry in Elastic $\vec{e}p$ Scattering}

The interference between the neutral weak and electromagnetic currents
produces a parity-violating asymmetry in $\vec{e}p$
elastic scattering, which has been the subject of a world-wide
measurement program focused on the determination of the strange
vector (electric and magnetic) form factors.  
For a proton target, the full expression for the parity-violating
electron scattering asymmetry is \cite{Liu:2007yi,Musolf:1994tb}
\begin{eqnarray}
\label{eqn_asymm}
A^p_{PV} &=& -\frac{G_F Q^2}{4\sqrt{2}\pi\alpha}
               \frac{1}{[\epsilon(G^p_E)^2 + \tau(G^p_M)^2]} \nonumber \\
 &\times& \{(\epsilon(G^p_E)^2+\tau(G^p_M)^2)(1-4\sin^2\theta_W)(1+R^p_V) \nonumber \\
 &      & -(\epsilon G^p_EG^n_E + \tau G^p_MG^n_M)(1+R^n_V) \nonumber \\
 &      & -(\epsilon G^p_EG^s_E + \tau G^p_MG^s_M)(1+R^{(0)}_V) \nonumber \\
 &      & -\epsilon'(1-4\sin^2\theta_W)G^p_MG^e_A\},
\end{eqnarray}
where the kinematics factors are
\begin{eqnarray*}
\epsilon & = & \left[1+2(1+\tau)\tan^2(\theta_e/2)\right]^{-1} \\
\epsilon' & = & \sqrt{(1-\epsilon^2)\tau(1+\tau)}.
\end{eqnarray*}
The axial form factor seen in electron scattering, $G^e_A$, as mentioned earlier,
does not appear in its pure form, but is complicated by radiative
corrections:
\begin{equation}
\label{eqn_ga}
G^e_A(Q^2) = G^{CC}_A(Q^2)(1+R^{T=1}_A)+\sqrt{3}G^8_A(Q^2)R^{T=0}_A+G^s_A(Q^2)(1+R^{(0)}_A).
\end{equation}
The $R$ factors appearing in Equations~\ref{eqn_asymm} and
\ref{eqn_ga} are radiative corrections that may be
expressed~\cite{Musolf:1994tb} in terms of standard model
parameters~\cite{PDG2004}.  Because these radiative corrections are
calculated at $Q^2=0$ and have an unknown $Q^2$-dependence, then in
our analysis some additional uncertainty needs to be attributed to
these radiative correction factors; we have assigned a 10\%
uncertainty to take the unknown $Q^2$-dependence into account (see
Table~\ref{table_params}).  Recently, a reevaluation of these
radiative corrections and their uncertainties, in the context of a fit
to world data on parity-violating $\vec{e}p$ scattering, was discussed
in Ref.~\cite{Liu:2007yi}.  Those values differ from the ones we have
used here; however, the use of these slightly different values would
not have significantly changed the results of the work presented here
because of the suppression of the axial terms in the parity-violating
asymmetries at forward angles.

For the vector form factors $G_E^p$, $G_E^n$, $G_M^p$, and $G_M^n$ we
have used the values given by the parametrization of
Arrington and Sick~\cite{Arrington2006PreciseDO} which includes the effects of two-photon exchange. The uncertainties in the vector form
factors do not contribute significantly to the uncertainties in the
results reported here.  For the charged-current (isovector) axial form
factor, $G_A^{CC}$, as already mentioned, we use a dipole form factor
shape where the $Q^2=0$ value is $g_A = 1.2670 \pm
0.0030$~\cite{Cabibbo:2003cu} and the $Q^2$-dependence is given by the
``axial mass'' parameter $M_A = 1.014 \pm
0.014$~\cite{Bodek:2007ym}.  The selection of
a correct parametrization of $G_A^{CC}$ is crucial to the correct
extraction of $G_A^s$ from neutrino neutral-current data because those
data are sensitive to the total neutral-current axial form factor
$G_A^{Z,p} = (-G_A^{CC} + G_A^s)/2$.  Any shift in the value of
$G_A^{CC}$ will produce a shift in the extracted value of $G_A^s$.  We
chose to use the $M_A$ from
Ref.~\cite{Bodek:2007ym} because they used
up-to-date data on the vector form factors and the value of $g_A$ and
performed a thorough re-evaluation of the original deuterium data on
which the value of $M_A$ is traditionally based.  Recently, two modern
neutrino experiments using nuclear targets (oxygen~\cite{Gran:2006jn}
and carbon~\cite{AguilarArevalo:2007ru}) have reported higher effective values of
$M_A$ from an analysis of charge-current, quasi-elastic scattering.
It not clear at this time what impact these new results have for the
value of $M_A$ for the proton.  If a significantly new set of values
for $G_A^{CC}$ for the proton can be established, then the results for
$G_A^s$ presented in this article will need to be re-evaluated.
In this context it is interesting to note that Kuzmin, Lyubushkin, and
Naumov~\cite{Kuzmin:2007kr} have analyzed a broad range of neutrino
charged-current reaction data, on a wide variety of nuclear targets,
and determined a value for $M_A$ in agreement with
Ref.~\cite{Bodek:2007ym}; this supports our use
of the value $M_A = 1.014 \pm 0.014$. 

Appearing in Equation~\ref{eqn_ga} for $G_A^e$ 
is the octet axial form factor
$G_A^8(Q^2)$.  The $Q^2=0$ value of this form factor is the quantity
$(3F-D)/2\sqrt{3}$; we have taken the value of $3F-D$ from Ref.~\cite{Cabibbo:2003cu} (see Table~\ref{table_params}).  We took the 
$Q^2$-dependence of $G_A^8$ to be the same as that of $G_A^{CC}$, i.e.
$$G_A^8(Q^2) = \frac{(3F-D)/2\sqrt{3}}{(1+Q^2/M_A^2)^2}$$
but this is an assumption.  This form factor is multiplied by the
radiative correction factor $R_A^{T=0}$ to which we have already assigned a
10\% uncertainty because we did not know its $Q^2$-dependence;
as a result, we assigned no additional uncertainty to $G_A^8$.

The parity-violating asymmetry may be written as a linear combination
of the strange electric form factor ($G_E^s$), the strange magnetic
form factor ($G_M^s$), and the strange axial form factor ($G_A^s$), as
follows:
$$A^p_{PV} = A_0^p + A_E^p G_E^s + A_M^p G_M^s + A_A^p G_A^s$$
where the coefficients are
\begin{eqnarray*}
A_0^p & = & -K^p\left\{
\begin{array}{l}
~~~\epsilon G_E^p \left[(1-4\sin^2\theta_W)(1+R_V^p)G_E^p - (1+R_V^n)G_E^n\right] \\
+\tau G_M^p \left[(1-4\sin^2\theta_W)(1+R_V^p)G_M^p - (1+R_V^n)G_M^n\right] \\
-\epsilon' G_M^p(1-4\sin^2\theta_W)
                \left[(1+R_A^{T=1})G^{CC}_A + 
                      \sqrt{3}R_A^{T=0}G_A^8\right] \end{array} \right\} \\
A_E^p & = & K^p\left\{\epsilon G_E^p(1+R_V^0)\right\} \\
A_M^p & = & K^p\left\{\tau G_M^p(1+R_V^0)\right\} \\
A_A^p & = & K^p\left\{\epsilon'G_M^p(1-4\sin^2\theta_W)(1+R_A^0)\right\} \\
K^p & = & \frac{G_F Q^2}{4\pi\sqrt{2}\alpha} \frac{1}{\epsilon (G_E^p)^2 + \tau(G_M^p)^2}.
\end{eqnarray*}
This expression is used with the PVES data on hydrogen, listed in Tables~\ref{PV_data_forward} and \ref{PV_data_backward}.

It is well to note that the axial term in this asymmetry
is suppressed by the weak electron charge $(1-4\sin^2\theta_W \approx 0.075)$, and
at forward angles it is suppressed additionally by the kinematic
factor $\epsilon'$.  This might seem a disadvantage, since this
strongly suppresses the sensitivity to the strange axial form factor
in $G_A^e$; however, it simultaneously suppresses the uncertainty in
the radiative corrections in $G_A^e$ which are significant in
magnitude and have an unknown $Q^2$-dependence.  Therefore, the
parity-violating asymmetry data serve to provide a necessary constraint
among the strange vector form factors, with only a little sensitivity
to the strange axial form factor.

\subsection{Parity-violating Asymmetries in Quasi-Elastic $\vec{e}N$ Scattering in Deuterium}

In this case the asymmetry that is observed is for quasi-elastic electron-nucleon
scattering within the deuterium nucleus, with the electron detected at backward
angles. In all the experiments that used deuterium targets (SAMPLE, G0, PVA4) only
the final state electron is detected, so we do not know which nucleon it interacted
with. We say ``quasi-elastic'' because the initial state nucleon is not at rest and will
very likely have a momentum transfer with the other nucleon after the interaction with
the electron.
To leading order, we may ignore the interactions between the proton and neutron in
the deuterium nucleus; this is called the ''static approximation.'' Then the parity-violating
asymmetry is a weighted combination of the asymmetries on the bare proton
and neutron:
\begin{equation}
A^d = \frac{\sigma_p A^p+\sigma_n A^n}{\sigma_p + \sigma_n}
\end{equation}
where $\sigma_p$ ($\sigma_n$) is the cross section for $ep$ ($en$) elastic scattering.  Then, as in the case of a proton target, 
the parity-violating asymmetry may be written as a linear combination
of the strange electric form factor ($G_E^s$), the strange magnetic
form factor ($G_M^s$), and the strange axial form factor ($G_A^s$), as
follows:
$$A^d_{PV} = A_0^d + A_E^d G_E^s + A_M^d G_M^s + A_A^d G_A^s$$
where the coefficients are
\begin{eqnarray*}
A_0^d & = & -K^d\left\{
\begin{array}{l}
~~~\epsilon (1-4\sin^2\theta_W)\left[(1+R_V^p)\left(\epsilon (G_E^p)^2+\tau(G_M^p)^2\right) + (1+R_V^n)\left(\epsilon(G_E^n)^2+\tau(G_M^n)^2\right)\right]\\
-(2+R_V^p+R_V^n)\left[\epsilon G_E^p G_E^n+\tau G_M^p G_M^n\right]\\
-\epsilon'(1-4\sin^2\theta_W)
     \left[(G_M^p-G_M^n)(1+R_A^{T=1})(-G^{CC}_A) + (G_M^p+G_M^n) \sqrt{3}R_A^{T=0}G_A^8\right] \end{array} \right\} \\
A_E^d & = & K^d\left\{\epsilon (G_E^p+G_E^n)(1+R_V^0)\right\} \\
A_M^d & = & K^d\left\{\tau (G_M^p+G_M^n)(1+R_V^0)\right\} \\
A_A^d & = & K^d\left\{\epsilon'(G_M^p+G_M^n)(1-4\sin^2\theta_W)(1+R_A^0)\right\} \\
K^d & = & \frac{G_F Q^2}{4\pi\sqrt{2}\alpha} \frac{1}{\epsilon (G_E^p)^2 + \tau(G_M^p)^2
+\epsilon (G_E^n)^2 + \tau(G_M^n)^2}.
\end{eqnarray*}
This expression is used for the inclusion of the PVA4 backward-angle deuterium data~\cite{BalaguerRios:2016ftd} in our fit, listed in Table~\ref{PV_data_deuterium}.

For a more accurate interpretation of the measured parity-violating asymmetries on deuterium, a nuclear model calculation is required.  The SAMPLE~\cite{Beise:2004py} and G0-Backward~\cite{Androic:2009zu} experiments used calculations performed by R. Schiavilla and collaborators (as described in Refs.~\cite{PhysRevC.70.044007, PhysRevC.65.035502}) that were tailored to the kinematics and detector acceptance of those experiments.  The parity-violating asymmetry is then written in this form:
$$A_{PV}^d = b_0+b_1G_E^s + b_2G_M^s + b_3(1+R_A^{T=1})(-G_A^{CC}) + b_4(1+R_A^0)G_A^s$$
with the $b$ coefficients coming from the calculations mentioned; these are listed in Table~\ref{table:backward_coefficients}.  This expression is used with the SAMPLE and G0 deuterium data listed in Table~\ref{PV_data_deuterium}.

\begin{table}[ht]
\caption{Asymmetry coefficients used in the interpretation of parity-violating backward-scattering $\vec{e}d$ data.  Values are from SAMPLE~\cite{Beise:2004py} and G0~\cite{Androic:2009zu,G0_Backward}}.
\label{table:backward_coefficients}
\begin{tabular}{c|c|c|c|c|c|c}
    Experiment & $Q^2$ (GeV$^2$) & $b_0$ & $b_1$ & $b_2$ & $b_3$ & $b_4$ \\
    \hline
    SAMPLE & 0.091 & -7.06 & 1.52 & 0.72 & 1.66 & 0.325 \\
    SAMPLE & 0.038 & -2.14 & 1.13 & 0.27 & 0.76 & 0.149 \\
    G0 & 0.221 & -15.671 & 7.075 & 1.994 & 2.921 & 0.571 \\
    G0 & 0.628 & -53.295 & 12.124 & 12.492 & 9.504 & 1.891
\end{tabular}
\end{table}

\subsection{Parity-violating Asymmetries in Quasi-Elastic $\vec{e}N$ Scattering in Helium-4}

This is also a case of quasi-elastic scattering from nucleons within a nuclear target.
Since helium-4 is isoscalar, then the magnetic and axial contributions to the parity-violating
asymmetry cancel in the static approximation.  Ref.~\cite{Aniol:2005zf} discusses contributions to this asymmetry in great detail.
\begin{equation}
\label{eqn_PVeHe}
A^{^4\rm{He}}_{PV} 
= \frac{G_F Q^2}{4\pi\alpha\sqrt{2}}\left(4\sin^2\theta_W + \frac{G_E^s}{\frac{1}{2}(G_E^{\gamma p}+G_E^{\gamma n})}\right)
\end{equation}
This expression is used in the interpretation of the HAPPEx data using helium-4, listed in Table~\ref{PV_data_4He}.

\subsection{Neutral-Current $\nu p$ 
and $\bar{\nu} p$ Elastic Scattering Cross Sections}

The cross section for
$\nu p$ and $\bar{\nu} p$ elastic scattering is given by~\cite{Ahrens:1986xe}
\begin{equation}
\label{eqn_ncxs}
\frac{d\sigma^{NC}}{dQ^2} = \frac{G_F^2}{2\pi} \frac{Q^2}{E_\nu^2} (A\pm BW + CW^2)
\end{equation}
where the $+$ ($-$) sign is for $\nu$ ($\bar{\nu}$) scattering, and
\begin{eqnarray*}
W    &=& 4(E_\nu /M_p - \tau) \\
\tau &=& Q^2/4M_p^2 \\
A    &=& \frac{1}{4}\left[(G_A^Z)^2(1+\tau)-\left((F_1^Z)^2-\tau(F_2^Z)^2\right)(1-\tau)
          +4\tau{F_1^Z}{F_2^Z}\right] \\
B    &=& -\frac{1}{4}G_A^Z(F_1^Z + F_2^Z) \\
C    &=& \frac{1}{64\tau}\left[(G_A^Z)^2 + (F_1^Z)^2 + \tau(F_2^Z)^2\right].
\end{eqnarray*}
This expression is used in the interpretation of the BNL E734 data, listed in Table~\ref{E734_table}.

\section{Models for the Strangeness Form Factors}
 The existing data on $G_E^s$ and $G_M^s$ (see Fig.~\ref{fig:old_points}) are rather featureless and do not contain information on the $Q^2$-dependence of those form factors, and so we chose a very simple zeroth-order model for them:
 $$ G_E^s = \rho_s\tau ~~~~~~ G_M^s = \mu_s$$
 where $\rho_s \equiv (dG_E^s/d\tau)|_{\tau=0}$ is the strangeness radius, and $\mu_s$ is the strangeness magnetic moment.
 
 By contrast, the data on $G_A^s$ shows a definite $Q^2$-dependence, and for this form factor we have chosen to use two different 3-parameter models.
\begin{itemize}
\item The Modified-Dipole Model:  The expression used for the strangeness axial form factor is:
$$G_A^s = \frac{\Delta s + S_A Q^2}{(1+Q^2/\Lambda_A^2)^2}$$
where $\Delta s$ is the strange quark contribution to the proton spin, and $S_A$ and $\Lambda_A$ are parameters describing the $Q^2$-dependence of $G_A^s$.
This shape is referred to as a ``modified-dipole'' because of its similarity to the usual dipole shapes used to model other form factors.
\item The $z$-Expansion Model:  The modified-dipole model comes with a bias with respect to the $Q^2$-dependence of $G_A^s$.  The ``$z$-expansion'' technique \cite{Hill:2010yb,Lee:2015jqa} allows for a bias-free model because it is simply a power series, and the fit seeks to determine the coefficients of the series.  The power series is of the form
$$G_A^s(Q^2)=\sum_{k=0}^\infty a_k\left[z(Q^2)\right]^k$$
where $Q^2$ has been mapped onto the variable $z$ as follows:
$$z(Q^2,t_{\rm cut},t_0)=\frac{\sqrt{t_{\rm cut}+Q^2}-\sqrt{t_{\rm cut}-t_0}}{\sqrt{t_{\rm cut}+Q^2}+\sqrt{t_{\rm cut}-t_0}}.$$
Note that $|z|<1$.  The parameter $t_{\rm cut}$ is determined by the threshold of the relevant current, which in the case of the isoscalar axial current is $t_{\rm cut}=(4 m_\pi)^2$.  (As an example, the lightest decay mode of the f$_1$(1285) meson, with I$^{\rm G}$(J$^{\rm PC}$) = $0^+(1^{++})$, is four pions.)
The parameter $t_0$ is arbitrary, and can be adjusted 
to make the convergence of the series more rapid, but we have chosen simply to use $t_0=0$; this has the consequence that $a_0=\Delta s$.  Of course it is 
necessary to cut off the sum over $k$, and we have limited it to $k_{\rm max}=6$.  This would imply seven 
parameters for the description of $G_A^s$.  However, due to the fact that the form factor should behave 
like $1/Q^4$ at large values of $Q^2$, we have the following four conditions:
$$\left.\frac{d^n}{dz^n}G_A^s\right|_{z=1}=0~~~~~n=0,1,2,3.$$
This allows us to reduce the number of independent parameters from seven down to three:  $a_0$, $a_1$, and $a_2$.
\end{itemize}
Our approach to the $Q^2$-dependence of $G_A^s$ differs from previous workers, for example Refs.~\cite{Ahrens:1986xe,Garvey:1992cg,PhysRevC.88.024612,Butkevich:2011fu}, in that we do not assume $G_A^s$ has the same $Q^2$-dependence as $G_A^{CC}$.  We take the accepted value of $M_A = 1.014 \pm 0.014$ GeV to describe the $Q^2$-dependence of $G_A^{CC}$, and let the $Q^2$-dependence of $G_A^s$ be a free parameter.

\section{Nuclear models for carbon used in comparisons with MiniBooNE data}

The main interaction mechanism for (anti)neutrinos with energy around 1 GeV, at the core of the energy distribution for many neutrino experiments, is QuasiElastic (QE) scattering, where the incident neutrino or antineutrino directly interacts with a quasifree nucleon of the target, which is then ejected from the nucleus by a Direct KnockOut (DKO) mechanism.
The DKO mechanism is related to the Impulse Approximation (IA), which is based on the assumption that the incident particle interacts with the ejectile nucleon only through a one-body current and the recoiling residual nucleus acts as a spectator.

Most of the models available for QE $\nu(\bar\nu)$-nucleus scattering based on this mechanism were originally developed for QE electron-nucleus scattering and tested against the large amount of accurate electron scattering data that have been collected in different laboratories worldwide. The two situations present many similar aspects and the extension of the models developed for electron scattering to neutrino scattering is straightforward. In QE electron scattering models are available for the exclusive (e,e$^\prime$p) reaction, where the emitted proton is detected in coincidence with the scattered electron and the final nuclear state is completely determined, and for the inclusive (e,e$^\prime$) scattering, where only the scattered electron is detected, the final nuclear state is not determined, and the experimental cross section includes all available final nuclear states. In neutrino scattering coincidence measurements represent an extremely hard task, only either the scattered lepton
%, in Charged Current QE (CCQE) scattering,
or the ejected nucleon
%, in CCQE and Neutral Current QE (NCQE) scattering,
is detected, and so far it is mostly models for the inclusive (e,e$^\prime$) process that have been extended to neutrino scattering.

Models for the inclusive process are appropriate for CCQE scattering where, as in the (e,e$^\prime$) reaction, only the final lepton is detected, but may be less appropriate for the NCQE case, where only the emitted nucleon can be detected, the cross section is integrated over the energy and angle of the final lepton and the process is inclusive in the lepton sector but semi-inclusive in the hadronic sector. Also in this case the final nuclear state is not determined and the experimental cross section includes all available final nuclear states, but the number of final states can be lower than in the inclusive process where only the final lepton is detected. A specific calculation for NCQE scattering, capable of taking this fact into account, is so far unavailable.

In spite of many similar aspects, electron and neutrino scattering present some differences in the nuclear current and in the kinematic situation. In electron scattering experiments the incident electron energy is known and the energy and momentum transfer, $\omega$ and ${\bf q}$, are clearly determined. In contrast, in neutrino experiments the neutrino flux is uncertain, the beam energy is not known and  $\omega$ and ${\bf q}$ are not fixed. The beam energy reconstruction, and hence flux unfolding, is possible only in model-dependent ways. Therefore measurements produce flux-integrated cross sections which contain events for a wide range of kinematic situations, corresponding not only to the QE region, but to other kinematic regions, and contributions beyond the IA can be included in the experimental differential cross sections.
As a consequence, models developed for the QE (e,e$^\prime$) reaction could be unable to describe  data unless all other processes contributing to the experimental cross sections are taken into account. Models including, within different frameworks and approximations, contributions beyond the IA, such as, for instance, two-particle-two-hole (2p2h) excitations and two-body Meson-Exchange Currents (MEC), that can give a significant contribution to the calculated cross sections, have been developed and used for CC scattering. So far these models have not been extended to NC scattering,  with the exception of the calculation of Refs.~\cite{Rocco:2018mwt,Martini:2011wp} which, however, refer to the ideal situation of detecting the outgoing neutrino and cannot be compared with experimental data.

Among the available models~\cite{Giusti:2019cup}, all based on the IA, we have used for the present analysis three relatively simple models, that anyhow include the main aspects of the problem
%namely the Relativistic Fermi Gas (RFG), the SuperScaling approximation (SuSA), %and the Spectral Function (SF) models,
and that allow us to perform fast numerical calculations. Our choice is due to the fact that the global fit presented in this work requires a large amount of calculations and the use of more sophisticated models would be too computationally demanding  without leading to significantly different results in the present investigation.

For instance, we do not use the relativistic Green's function (RGF) model~\cite{Meucci:2003uy}, which would be able to give a better description of the experimental cross sections than other models based on the IA, which generally underpredict CCQE and NCQE experimental cross sections. In the RGF model, which is also based on the IA, the Final-State Interactions (FSI) between the emitted nucleon and the other nucleons of the target are taken into account by a complex energy-dependent relativistic optical potential where the imaginary part can recover contributions of nonelastic channels, such as, for instance, some multi-nucleon processes, rescattering, non nucleonic contributions, that are not included in other models based on the IA. The RGF is quite successful in the description of the experimental cross sections for the inclusive QE (e,e$^\prime$)~\cite{Meucci:2003uy,Meucci:2009nm} and CCQE~\cite{Meucci:2003cv,Meucci:2011pi,Meucci:2011vd,Meucci:2015bea} reactions and gives a reasonable description also of NCQE cross sections~\cite{Gonzalez-Jimenez:2013xpa,Meucci:2014pka,Ivanov:2015wpa,Giusti:2019cup}. However, it is a model for the inclusive scattering and its use can be less appropriate in the semi-inclusive NC scattering, where it recovers important contributions not included in other models based on the IA, but may include also channels that are present in the inclusive but not in a semi-inclusive process.
RGF calculations would be too time-consuming for the present analysis and would  not significantly change the ratios of cross sections used in this work with the aim of determining the strange quark contribution to the nucleon form factors. The differences given by our models on the calculated cross sections, which are due to the different treatments of FSI and other nuclear effects, are strongly reduced or almost cancelled in the ratios, where FSI and other nuclear effects can give a similar contribution to the numerator and to the denominator~\cite{Gonzalez-Jimenez:2013xpa,Ivanov:2015wpa,Giusti:2019cup}.

In the following we briefly describe the three nuclear models used in the present analysis: the Relativistic Fermi Gas (RFG), the Super Scaling Approximation model (SuSA), and the Spectral Function model (SF). All of these models are, exactly or approximately, relativistic, as required by the typical kinematics of the relevant experiments.

\subsection{The Relativistic Fermi Gas and the SuperScaling Approximation Models}

The simplest approach to a fully relativistic nuclear system is represented by the Relativistic Fermi Gas  model, in which the single-nucleon wave functions are free plane waves multiplied by Dirac spinors and the only correlations are the statistical ones induced by the Pauli principle. Each nucleus is characterized by a Fermi momentum $k_F$, usually fitted to the width of the QE Peak (QEP) in electron scattering data.

The procedure for calculating the lepton-nucleus QE cross section involves an integration over all unconstrained kinematic variables, namely those of the undetected outgoing lepton in the case of  NC neutrino scattering. 
In the case of neutrino scattering, due to the broad energy distribution of the beam, an extra integration over the experimental neutrino flux should be performed.
In the RFG model  the differential cross section with respect to the  momentum $(p_N)$ and solid scattering angle $(\Omega_N)$ of the outgoing nucleon can be written  - for a given energy of the incoming lepton - as~\cite{Barbaro:1996vd,Amaro:2006pr}:
\begin{equation}
\frac{d\sigma}{d\Omega_N dp_N} = \overline\sigma \, \frac{1}{k_F} f_{\rm RFG}(\psi)\,,
\label{eq:d2s}
\end{equation}
where $\overline\sigma$ is an effective single-nucleon cross section and $f(\psi)$ embodies the nuclear dynamics. The function
\begin{equation}
f_{\rm RFG}(\psi) = \frac{3}{4} \left(1-\psi^2\right) \theta\left(1-\psi^2\right)
\label{eq:fRFG}
\end{equation}
depends only upon one kinematic variable, $\psi$, instead of two as it would be expected, and  is independent of the specific nucleus - that is, independent of $k_F$. This occurrence is known as {\it super-scaling} and $f$ and $\psi$ are denoted as super-scaling function and scaling variable, respectively. Physically $\psi$ represents the (dimensionless) minimum kinetic energy required to a nucleon in the RFG ground state to participate in the reaction at  given kinematics (see \cite{PhysRevC.38.1801,Barbaro1998137}).
In general, the super-scaling function can be expressed as an integral of the spectral function $S$
\begin{equation}
 f(\psi) = \frac{1}{k_F} \iint_{\cal D}dp\, d{\cal E}  \frac{p}{E}  S(p,{\cal E})\,,
\label{eq:F}
\end{equation}
where $p$ is the momentum of the initial nucleon, $E$ the corresponding on-shell energy,  $\cal E$ the excitation energy of the residual nucleus and ${\cal D}$ the region in the $({\cal E},p)$-plane allowed by the kinematics.
In the RFG model the spectral function is simply given by
\begin{equation}
S_{\rm RFG}(p,{\cal E}) = \frac{3k_F}{4T_F}\,\theta(k_F-p)\,\delta\left({\cal E}-\sqrt{k_F^2+M^2}+\sqrt{p^2+M^2}\right) \,,
\end{equation}
where $T_F=\sqrt{k_F^2+M^2}-M$ is the Fermi kinetic energy.

The RFG model has the advantage of being exactly relativistic and therefore represents a suitable starting point for more sophisticated models, but it is well-known that it gives a poor description of electron scattering data. These, unlike neutrino data, are very abundant and precise and can be used as a benchmark in neutrino scattering studies. It was first suggested in Ref.~\cite{Amaro:2004bs} that the scaling behaviour of inclusive $(e,e')$ data can also be used as an input to get reliable predictions for CC neutrino-nucleus cross sections and the same approach was extended to NC reactions in Ref.~\cite{Amaro:2006pr}. This idea is at the basis of the SuSA model, which essentially amounts to replacing the RFG superscaling function in Equation~\eqref{eq:fRFG} by a phenomenological one, $f_{\rm SuSA}(\psi)$, extracted by the analysis of electron scattering data as the ratio between the double differential cross section and an appropriate single-nucleon function~\cite{Donnelly:1998xg,Donnelly:1999sw}.
The analysis of the longitudinal QE data shows that this function is indeed very weakly dependent on the momentum transfer $\bf q$ providing the latter is high enough (namely larger than about 400 MeV/c) to allow for the impulse approximation; this property is usually referred to as scaling of first kind. Moreover, the super-scaling function is almost independent of the specific nucleus for mass numbers $A$ ranging from 4 (helium) up to 198 (gold); this is known as scaling of second kind. Super-scaling is the simultaneous occurrence of the two kinds of scaling and is well respected by electron scattering data in the QEP region. Scaling violations occur in the transverse channel due to non-impulsive contributions, for example the excitation of 2p2h states, which are not included in the SuSA approach.

The phenomenological superscaling function $f_{\rm SuSA}$ incorporates effectively nucleon-nucleon (NN) correlations and FSI
%final state interactions 
and gives, by construction, a good agreement with $(e,e')$ data in a wide range of kinematics and mass numbers.
The parametrization used in this work is
\begin{equation}
f_{\rm  SuSA}(\psi) = \frac{\alpha}{\left[1+\beta^2\left(\psi+\gamma\right)^2\right] \left(1+e^{-\delta\psi}\right)} \,,
\label{eq:fSuSA}
\end{equation}
where the parameters are fitted to the electron scattering QE  world data analyzed in Ref.~\cite{Jourdan:1996np,Donnelly:1999sw} for all the experimentally available kinematics and nuclear targets.
Here we use the values $\alpha=1.5576$, $\beta=1.7720$, $\gamma=0.3014$ and $\delta=2.4291$, corresponding to the fit performed in Ref.~\cite{Amaro:2004bs}.
Two more parameters, the Fermi momentum  $k_F$ (228 MeV/c for carbon) and an energy shift $E_s$ (20 MeV), are fitted to the experimental width and position of the QEP~\cite{Maieron:2001it}.

The SuSA super-scaling function  is purely phenomenological. However, studies on its microscopic origin have shown that the shape and size of the $f_{\rm SuSA}$ can be reproduced with good accuracy by the relativistic mean field (RMF) model~\cite{Caballero:2005sj}. In particular, it was shown that the high-energy asymmetric tail displayed by $f_{\rm SuSA}$ can be mainly ascribed to FSI and it cannot be reproduced if the latter are neglected,  as in the Plane Wave Impulse Approximation (PWIA).
The RMF model was also exploited to construct a new version of the superscaling model (SuSAv2)~\cite{Gonzalez-Jimenez:2014eqa,Megias:2016lke}, where different scaling functions are used in each channel (longitudinal, transverse and axial, isoscalar and isovector), as predicted by the model in the quasielastic region. Although the differences between SuSA and SuSAv2 are not negligible, in this paper we stick to the original SuSA model, which employs the same scaling function in all channels. Further refinements of the model could be explored, but their impact on the present analysis is not expected to be significant.

\vskip 0.5cm

\subsection{The Spectral Function Model}

In a more fundamental approach, the superscaling function, constructed in the SuSA model by fitting the quasielastic $(e,e')$ data, can be evaluated microscopically using a realistic spectral function.

The area of analyses of the scaling function, the spectral function and their connection (see, \emph{e.g.}~\cite{PhysRevC.81.055502, PhysRevC.83.045504}) provides insight into the validity of the mean-field approximation (MFA) and the role of the NN correlations, as well as into the effects of FSI. Though in the MFA it is possible, in principle, to obtain the contributions of different shells to the spectral function $S(p,{\cal E})$ and to the momentum distribution $n(p)$ for each single-particle state, due to the residual interactions, the hole states are not eigenstates of the residual nucleus but mixtures of several single-particle states  leading to the spreading of the shell structure. As a consequence, a successful description of the results of the relevant experiments requires studies of the spectral function which make use of methods beyond the MFA.

In Ref.~\cite{PhysRevC.83.045504} a realistic spectral function $S(p,{\cal E})$ has been constructed in agreement with the phenomenological scaling function $f(\psi)$ obtained from ($e,e'$) data. For this purpose  effects beyond the MFA have been considered. The procedure takes into account the effects of a finite energy spread and of NN correlations, considering single-particle (s.p.)\ momentum distributions $n_i(p)$, that are components of $S(p,{\cal E})$ beyond the MFA, such as those related to the use of natural orbitals (NO's)~\cite{PhysRev.97.1474} for the single-particle wave functions, and occupation numbers within methods in which short-range NN correlations are included. For the latter the Jastrow correlation method~\cite{PhysRevC.48.74} has been considered. FSI  are taken into account in the spectral function 
%S(p,{\cal E})$
model~\cite{PhysRevC.83.045504} by a complex optical potential that leads to an asymmetric scaling function in accordance with the experimental analysis, thus showing the essential role of the FSI in the description of electron scattering data.

We adopt the following procedure for the SF model:
\begin{enumerate}

\item[(i)] The spectral function $S(p,{\cal E})$ is constructed in the form~\cite{PhysRevC.83.045504, PhysRevC.89.014607, Ivanov:2015wpa}:
 \begin{equation}\label{HF+lorent}
    S(p,{\cal E})=\sum_{i}2(2j_i+1) N_i n_i(p) L_{\Gamma_i}({\cal E} - {\cal E}_i),
\end{equation}
where the Lorentzian function is used:
\begin{equation}\label{lorent}
L_{\Gamma_i}({\cal E}-{\cal E}_i) = \dfrac{1}{\pi}\dfrac{\Gamma_i/2}{({\cal E}-{\cal E}_i)^2+(\Gamma_i/2)^2}\,
\end{equation}
${\Gamma_i}$ being the width of a given state. $\Gamma_{1p} = 6$~MeV and $\Gamma_{1s} = 20$~MeV are fixed to the experimental widths of the $1p$ and $1s$ states in $^{12}$C nucleus~\cite{Dutta:1999}.

\item [(ii)] In Equation~(\ref{HF+lorent})  the s.p.\ momentum distributions $n_i(p)$ correspond to natural orbitals s.p.\ wave functions $\varphi_i (\mathbf{r})$. The latter are defined in~\cite{PhysRev.97.1474} as the complete orthonormal set of s.p.\ wave functions that diagonalize the one-body density matrix $\rho(\mathbf{r},\mathbf{r'})$:
    \begin{equation}
    \rho (\mathbf{r},\mathbf{r}^{\prime} )=\sum_{i} N_{i} \varphi_{i}^{*}(\mathbf{r}) \varphi_{i}(\mathbf{r}^{\prime}) ,\label{defNO}
    \end{equation}
    where the eigenvalues $N_{i}$ ($0\leq N_{i}\leq 1$, $ \sum_{i} N_{i}=A$) are the natural occupation numbers. We use $\rho(\mathbf{r},\mathbf{r'})$ obtained within the lowest-order approximation of the Jastrow correlation methods~\cite{PhysRevC.48.74}.

\item[(iii)] For  given momentum transfer  $q$ and energy of the initial electron $\varepsilon$, we calculate the electron-nucleus ($^{12}$C) cross section by using the PWIA expression for the inclusive electron-nucleus scattering cross section
\begin{equation}\label{cr.s.}
\frac{d\sigma_t}{d\omega d |\mathbf{q}|}={2\pi\alpha^2}\frac{ |\mathbf{q}|}{\varepsilon^2} \int d{E}\:d^3p\:\frac{S_t(\mathbf{p}, {E})}{E_{\mathbf{p}}E_{\mathbf{ {p'}}}}\delta\big(\omega+M-{E}-E_{\mathbf{p'}}\big)L_{\mu\nu}^\text{em}H^{\mu\nu}_{\text{em, }t}\,.
\end{equation}
In Equation~(\ref{cr.s.}) the index $t$ 
denotes the nucleon isospin, $L_{\mu\nu}^\text{em}$ and $H^{\mu\nu}_{\text{em, }t}$ are the leptonic and hadronic tensors, respectively, and $S_t(\mathbf{p}, {E)}$ is the proton (neutron) spectral function.  We note that in the model a separate spectral function is calculated for protons and neutrons. The terms $E_{\mathbf{p}}$, $E_{\mathbf{{p'}}}$, and ${E}$ represent the energy of the nucleon inside the nucleus, the ejected nucleon energy, and the removal energy, respectively (see ~\cite{PhysRevC.77.044311} for details).
  
\item[(iv)] Following the approach of  Refs.~\cite{PhysRevC.22.1680,PhysRevC.77.044311}, we account for the FSI of the struck nucleon with the spectator system  by means of a time-independent optical potential (OP): $U=V-\imath W$. In this case the energy-conserving $\delta$-function in Equation~(\ref{cr.s.}) is replaced by
    \begin{equation}\label{deltaf}
    \delta(\omega+M-E-E_{\mathbf{p'}})
    \rightarrow \dfrac{W/\pi}{W^2+[\omega+M-E-E_{\mathbf{p'}}-V]^2} ,
    \end{equation}
    with $V$ and $W$ obtained from the Dirac OP~\cite{PhysRevC.73.024608}.

  \item[(v)] The corresponding superscaling function is calculated
as
\begin{equation} %F(q,\omega) =
f_{\rm SF}(\psi) = k_F
\dfrac{\left[d\sigma/d\varepsilon'd\Omega'\right]_{(e,e')}}{\overline{\sigma}^{eN}
(q,\omega;p=|y|,{\cal E}=0)}\,, \label{eq:fSF}
\end{equation}
where the electron single-nucleon cross section $\overline{\sigma}^{eN}$ is taken at $p = |y|$, the scaling variable $y$ being the smallest possible value of $p$ in electron-nucleus scattering for the smallest possible value of the excitation energy (${\cal E} = 0$). %By multiplying $F(q,\omega)$ by $k_F$ the superscaling function $f(\psi)$ is obtained, where the scaling variable  $\psi$ has been introduced (see \cite{PhysRevC.38.1801,Barbaro1998137}).

\item[(vi)] Finally, the nuclear responses are calculated by multiplying $f_{\rm SF}(\psi)$ by the appropriate single-nucleon functions given in~\cite{Amaro:2006pr}.

\end{enumerate}

\begin{figure}[ht]
\includegraphics[width=7.5 cm]{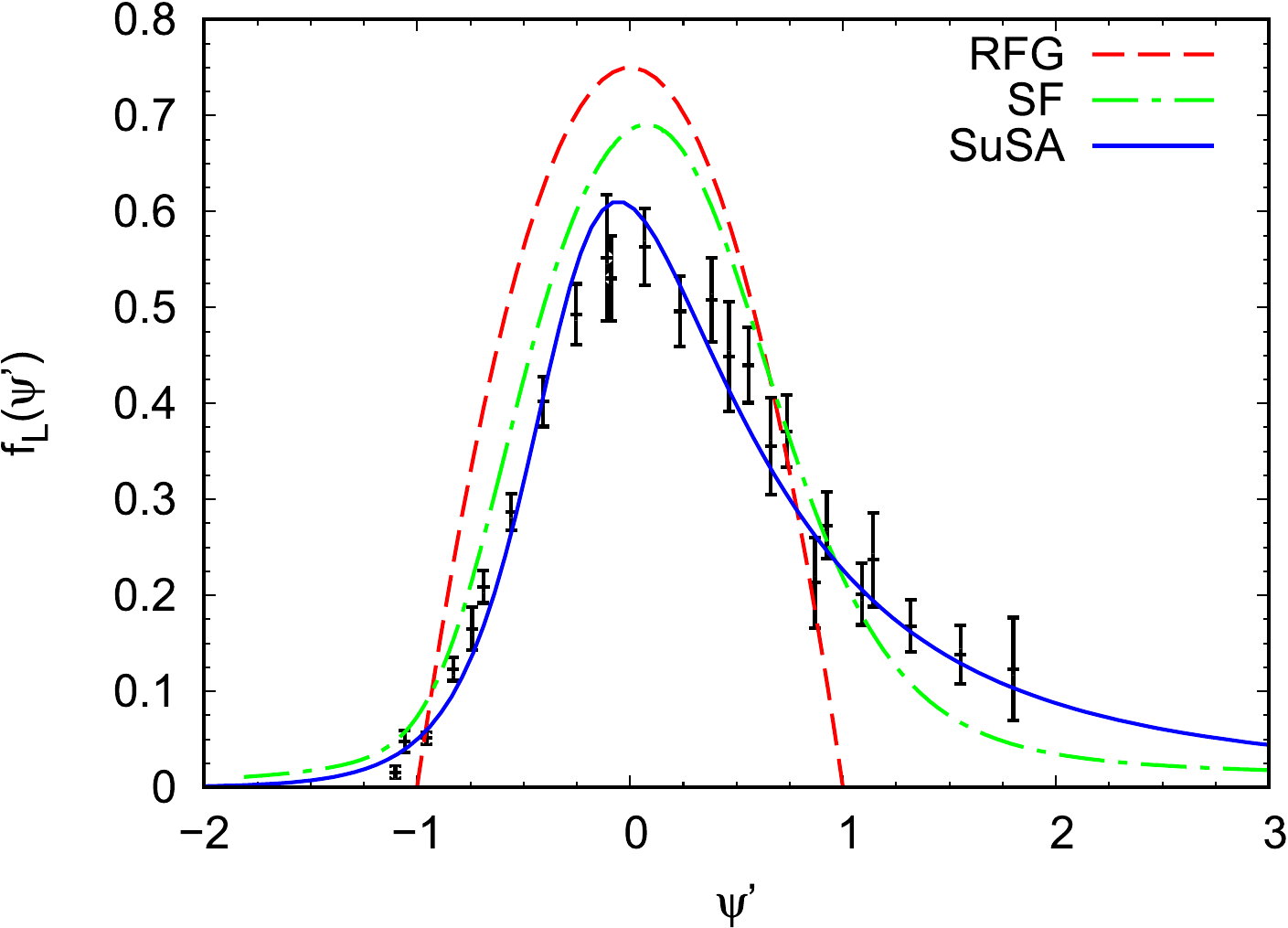}
\caption{The RFG, SuSA and SF scaling functions compared with the world averaged longitudinal inclusive electron scattering data~\cite{Jourdan:1996np}. \label{fig:scaling}}
\end{figure}

In Fig.~\ref{fig:scaling} the RFG, SuSA and SF scaling functions, Equations~\eqref{eq:fRFG}, \eqref{eq:fSuSA} and \eqref{eq:fSF}, are compared with the world averaged longitudinal (e,e') data~\footnote{Here the scaling variable is defined as $\psi'=\psi(|\vec q|,\omega-E_s; k_F)$ to incorporate the energy shift $E_s$.}. This comparison clearly shows that the RFG provides a rather poor description of electron scattering data, while  SuSA and SF  are more appropriate models for neutrino experimental analyses. They include, the former phenomenologically and the latter microscopically, the effects of NN correlations and FSI absent in the Fermi gas model.  It can also be noted that the agreement of the SF scaling function with the data is poorer than that of the SuSA model. However, $f_{\rm SF}$ is constructed starting from the total inclusive cross section (see Equation~\eqref{eq:fSF}), which is a linear combination of the longitudinal (L) and transverse (T) responses. The fact that $f_{\rm SF}$ is slightly higher than the longitudinal data reflects an enhancement of the transverse response, supported by the analysis of the separated L and T data~\cite{Donnelly:1999sw}.

\section{Results of our Fits}

The fitting consisted of a $\chi^2$-minimization procedure, implemented with the MINUIT tools available in the ROOT~\cite{ROOT} analysis system.  A value of $\chi^2$ was calculated for each data point, or for a collection of data points when correlations were known.  For example, for the BNL E734 results, a covariance matrix $C_{\rm E734}$ can be determined from Ref.~\cite{Ahrens:1986xe} and we used it in the calculation of the $\chi^2$,
$$\chi^2_{\rm E734} = \sum_{i=1}^{14}\sum_{j=1}^{14} (y_i - m_i)[C^{-1}_{\rm E734}]_{ij}(y_j - m_j)$$
where $y_i$ and $m_i$ are respectively the data and model value at the $i$th data point;
the 14 data points from the E734 measurements are listed in Table~\ref{E734_table}.  Covariance matrices also exist for the G0 data, one for forward-scattering and one for backward-scattering.  For all other PVES data, an uncorrelated calculation of the $\chi^2$ occurs, for example the HAPPEx helium-4 data,
$$\chi^2_{\rm 4He}=\sum_{i=1}^2 \frac{(y_i-m_i)^2}{(\Delta y_i)^2}$$
where $\Delta y_i$ is the uncertainty in the $i$th data point; the 2 HAPPEx helium-4 data points are listed in Table~\ref{PV_data_4He}.

%%%%%%%%%%%%%%%%%%%%%%%%%%%%%%%%%%%%
The MiniBooNE collaboration provided data releases for their measurements of neutrino~\cite{PhysRevD.82.092005, MB_neu_data} and anti-neutrino~\cite{PhysRevD.91.012004, MB_antineu_data} neutral current scattering, including covariance matrices. As mentioned earlier, MiniBooNE performed two different analyses with the neutrino-induced data and a single analysis with the antineutrino-induced data.
\begin{itemize}
    \item 
    The inclusive data from neutrino-induced NC scattering, which includes NC interactions with both protons and neutrons in the carbon nucleus, are reported as a yield as a function of reconstructed kinetic energy, $T_N$, along with a breakdown of the backgrounds.  In our fit, we add our prediction for the signal to the reported backgrounds to try to reproduce the yield.  The MiniBooNE collaboration provides instructions on how to smear the cross section with the MiniBooNE detector resolution and efficiency effects to get the reconstructed energy spectrum. In what follows we convert our theoretical true energy distribution to the reconstructed energy distribution. For that purpose, we follow the procedure described in Appendix B of Ref.~\cite{Perevalov_phdthesis}. Using the covariance matrices for the NCE event sample, one can calculate the $\chi^2$ in order to compare the theory prediction with the MiniBooNE data:
\begin{gather}\label{chi2CS}
\chi^2_{\nu\text{NCE/MiniBooNE/}}=\sum\limits_{i=1}^{n}\sum\limits_{j=1}^{n}
(\nu_i^\text{data}-\nu_i^{\text{model}})M_{ij}^{-1}(\nu_j^\text{data}-\nu_j^{\text{model}});\\
\nu_i=\text{events}_i;~\text{model = RFG, SuSA, or SF,}\notag
\end{gather}
where $M_{ij}$ is the covariance matrix for the NCE sample.

\item The exclusive data from MiniBooNE for neutrino-induced NC interactions, where the Cherenkov light has been used to isolate events with a single proton in the final state, are reported as a ratio of yields from protons to that on all nucleons, $\dfrac{\nu^\text{data,NCE(p)}}{\nu^\text{data,NCE(p+n)}}$, as function of the reconstructed kinetic energy.  For this data set, one has to calculate $\nu^{\text{model,NCE(p)}}$ and $\nu^{\text{model,NCE(n+p)}}$ for the NCE(p) and NCE(p+n) samples. Then the $\chi^2$ between data and theory prediction is
\begin{gather}\label{chi2ratio}
\chi^2_{\text{ratio/MiniBooNE/}}=\sum\limits_{i=1}^{n}\sum\limits_{j=1}^{n}
\left(\dfrac{\nu_i^\text{data,NCE(p)}}{\nu_i^\text{data,NCE(p+n)}}-\dfrac{\nu_i^\text{model,NCE(p)}}{\nu_i^\text{model,NCE(p+n)}}\right)
M_{ij}^{-1}
\left(\dfrac{\nu_j^\text{data,NCE(p)}}{\nu_j^\text{data,NCE(p+n)}}-\dfrac{\nu_j^\text{model,NCE(p)}}{\nu_j^\text{model,NCE(p+n)}}\right),
\end{gather}
with $M_{ij}$ being the covariance matrix of the ratio in this case.

\item
In the case of the antineutrino MiniBooNE NCE scattering~\cite{PhysRevD.91.012004, MB_antineu_data}, the data are presented as cross sections as a function of a {\em measured} $Q^2=2M\sum T$. The scintillation light in the event is taken to be a sum over all final state nucleons, and this is used to estimate the $Q^2$.  The model calculation uses the same approximation.  Then the $\chi^2$ calculation follows as
\begin{gather}\label{chi2aCS}
\chi^2_{\bar{\nu}\text{NCE/MiniBooNE/}}=\sum\limits_{i=1}^{n}\sum\limits_{j=1}^{n}
\left(\left(\frac{d\sigma}{dQ^2}\right)_i^\text{data}-\left(\frac{d\sigma}{dQ^2}\right)^{\text{model}}_i\right)M_{ij}^{-1}
\left(\left(\frac{d\sigma}{dQ^2}\right)_j^\text{data}-\left(\frac{d\sigma}{dQ^2}\right)^{\text{model}}_j\right),
\end{gather}
where $M_{ij}$ is the covariance matrix for the anti-neutrino NCE sample. The $\bar{\nu}$ cross section model prediction
\begin{eqnarray}\label{CS_MiniBooNE}
\frac{d\sigma}{dQ^{2}}=\frac{1}{7}C_{\bar \nu p, H}\frac{d\sigma_{\bar \nu p \rightarrow \bar \nu p, H}}{dQ^{2}}
+ \frac{3}{7}C_{\bar \nu p, C}\frac{d\sigma_{\bar \nu p \rightarrow \bar \nu p, C}}{dQ^{2}}
+\frac{3}{7}C_{\bar \nu n, C}\frac{d\sigma_{\bar \nu n \rightarrow \bar \nu n, C}}{dQ^{2}}
\end{eqnarray}
is a sum of three different processes: the antineutrino scattering off free protons in the hydrogen atom, the bound protons in the carbon atom, and the bound neutrons in the carbon atom. Each of the individual processes have different efficiencies in the MiniBooNE detector. The efficiency correction functions $C_{\bar \nu p, H}$, $C_{\bar \nu p, C}$, and $C_{\bar \nu n, C}$ for the three processes are given in Refs.~\cite{PhysRevD.91.012004, MB_antineu_data}.
\end{itemize}

%%%%%%%%%%%%%%%%%%%%%%%%%%%%%%%%%%%%

The MiniBooNE data extend into the region $Q^2>1.1$~GeV$^2$, beyond the range of the PVES data.  We found that including these large $Q^2$ NCES data, with no PVES data to balance them, distorted the fit results.  Also, the MiniBooNE data for anti-neutrino NC events included a point at $Q^2=0.066$ GeV$^2$ that is not included in the neutrino data; we removed this point from our fit because the nuclear models we use do not include correct modeling of Pauli-blocking effects that might be significant at this low $Q^2$.  So, the data we use from MiniBooNE all fall in the range $0.1<Q^2<1.1$ GeV$^2$.

There are altogether 49 data points from BNL E734, G0, SAMPLE, HAPPEx, and PVA4; these are listed in \Cref{E734_table,PV_data_forward,PV_data_backward,PV_data_deuterium,PV_data_4He}.  The inclusion of the NCES data from MiniBooNE brings the number of data points up to 128.

The results of our fits depend on the quantities listed in Table~\ref{table_params}, some of which have significant uncertainties.  This is a source of systematic error.  To measure the uncertainties in our results resulting from the uncertainties in Table~\ref{table_params}, we changed each of those quantities (for example $M_A$) by one standard deviation one at a time and repeated the fit.  The change in the best values of the fit parameters was noted.  This procedure was repeated for each quantity ($M_A$, $g_A$, and so on), and then those variations were added in quadrature, producing a total systematic error $\sigma^{\rm sys}_i$ for each fit parameter $i$.

The ROOT/MINUIT fitting routine produces a covariance matrix $C^{\rm fit}_{ij}$ containing the information about the fit errors and the correlations among the fit parameters.  Since the systematic uncertainties mentioned above were calculated using the same fitting procedure as the fit errors, we concluded that the systematic errors have the same correlations as the fit errors.  To include the systematic errors into the covariance matrix correctly, we first extracted the correlation matrix $\rho_{ij}$ from the fit covariance matrix:
$$\rho_{ij} = \frac{C^{\rm fit}_{ij}}{\sqrt{C^{\rm fit}_{ii}C^{\rm fit}_{jj}}}.$$
Adding the fit errors ($\sigma^{\rm fit}_i=\sqrt{C^{\rm fit}_{ii}}$) and systematic errors ($\sigma^{\rm sys}_i$) in quadrature, the total error for parameter $i$ is 
$$\sigma^{\rm total}_i=\sqrt{(\sigma^{\rm fit}_i)^2+(\sigma^{\rm sys}_i)^2}.$$  
Then the total covariance matrix is
$$C^{\rm total}_{ij} = \rho_{ij}\sigma^{\rm total}_i \sigma^{\rm total}_j.$$

\begin{table}[ht]
\caption{Summary of the results of the fits performed with three nuclear models (RFG, SuSA, and SF) and two strangeness axial form factor models (modified-dipole and $z$-expansion); also shown are the results when no MiniBooNE data are included. The central value and uncertainty is given for each fit parameter, and also the $\chi^2$ per number of degrees of freedom at the optimal fit point.  The first uncertainty is that arising from the fit itself, and the second uncertainty is a systematic due to the uncertainties in the quantities in Table~\ref{table_params} as described in the text.}
\label{table:results_summary}
\begin{tabular}{|c|c|c|c|c|c|c}
\hline
 \multicolumn{2}{|c|}{ } & RFG & SuSA & SF & w/o MiniBooNE Data\\
 \hline \hline
\multirow{6}*{Modified-Dipole} & 
              $\rho_s$ & $ -0.043 \pm 0.120 \pm 0.063 $ & $ -0.047 \pm 0.120 \pm 0.064 $ & $ -0.044 \pm 0.120 \pm 0.063 $ & $ -0.107 \pm 0.121 \pm 0.058 $\\
\cline{2-6} & $\mu_s$ & $ 0.045 \pm 0.036 \pm 0.032 $ & $ 0.047 \pm 0.036 \pm 0.032 $ & $ 0.045 \pm 0.036 \pm 0.032 $ & $ 0.065 \pm 0.036 \pm 0.030 $ \\
\cline{2-6} & $\Delta s$ & $ -0.203 \pm 0.115 \pm 0.030 $ & $ -0.386 \pm 0.155 \pm 0.055 $ & $ -0.224 \pm 0.121 \pm 0.033 $ & $ -0.267 \pm 0.393 \pm 0.156 $\\
\cline{2-6} & $\Lambda_A$ & $ 1.37 \pm 0.73 \pm 0.13 $ & $ 1.04 \pm 0.33 \pm 0.08 $ & $ 1.31 \pm 0.64 \pm 0.12 $ & $ 1.20 \pm 1.36 \pm 1.69 $ \\
\cline{2-6} & $S_A$ & $ 0.230 \pm 0.133 \pm 0.037 $ & $ 0.422 \pm 0.178 \pm 0.070 $ & $ 0.253 \pm 0.139 \pm 0.041 $ & $ 0.335 \pm 0.491 \pm 0.195$\\
\cline{2-6} & $\chi^2$/ndf & $ 133/123 $ & $ 144/123 $ & $ 134/123 $ & $ 55/44$ \\
\hline \hline
\multirow{6}*{$z$-Expansion} & 
              $\rho_s$ & $ -0.022 \pm 0.128 \pm 0.071 $ & $ -0.036 \pm 0.125 \pm 0.070 $ & $ -0.025 \pm 0.127 \pm 0.070 $ & $ -0.080 \pm 0.126 \pm 0.045 $\\
\cline{2-6} & $\mu_s$ & $ 0.038 \pm 0.038 \pm 0.034 $ & $ 0.044 \pm 0.037 \pm 0.034 $ & $ 0.040 \pm 0.038 \pm 0.034 $ & $ 0.055 \pm 0.038 \pm 0.024 $ \\
\cline{2-6} & $a_0$ & $ 0.403 \pm 0.222 \pm 0.183 $ & $ -0.087 \pm 0.199 \pm 0.150 $ & $ 0.323 \pm 0.220 \pm 0.191 $ & $ 1.07 \pm 0.33 \pm 1.39$ \\
\cline{2-6} & $a_1$ & $ -8.09 \pm 2.44 \pm 1.98 $ & $ -3.18 \pm 2.27 \pm 1.58 $ & $ -7.25 \pm 2.42 \pm 2.07 $ & $ -14.8 \pm 3.4 \pm 15.1 $ \\
\cline{2-6} & $a_2$ & $ 44.5 \pm 11.3 \pm 8.2 $ & $ 25.1 \pm 10.8 \pm 6.4 $ & $ 41.1 \pm 11.3 \pm 8.6 $ & $ 71.4 \pm 14.8 \pm 62.7 $ \\
\cline{2-6} & $\chi^2$/ndf & $ 130/123 $ & $ 143/123 $ & $ 131/123 $ & $ 53/44 $ \\
\hline
\end{tabular}
\end{table}

\begin{figure}[ht]
\includegraphics[scale=0.5]{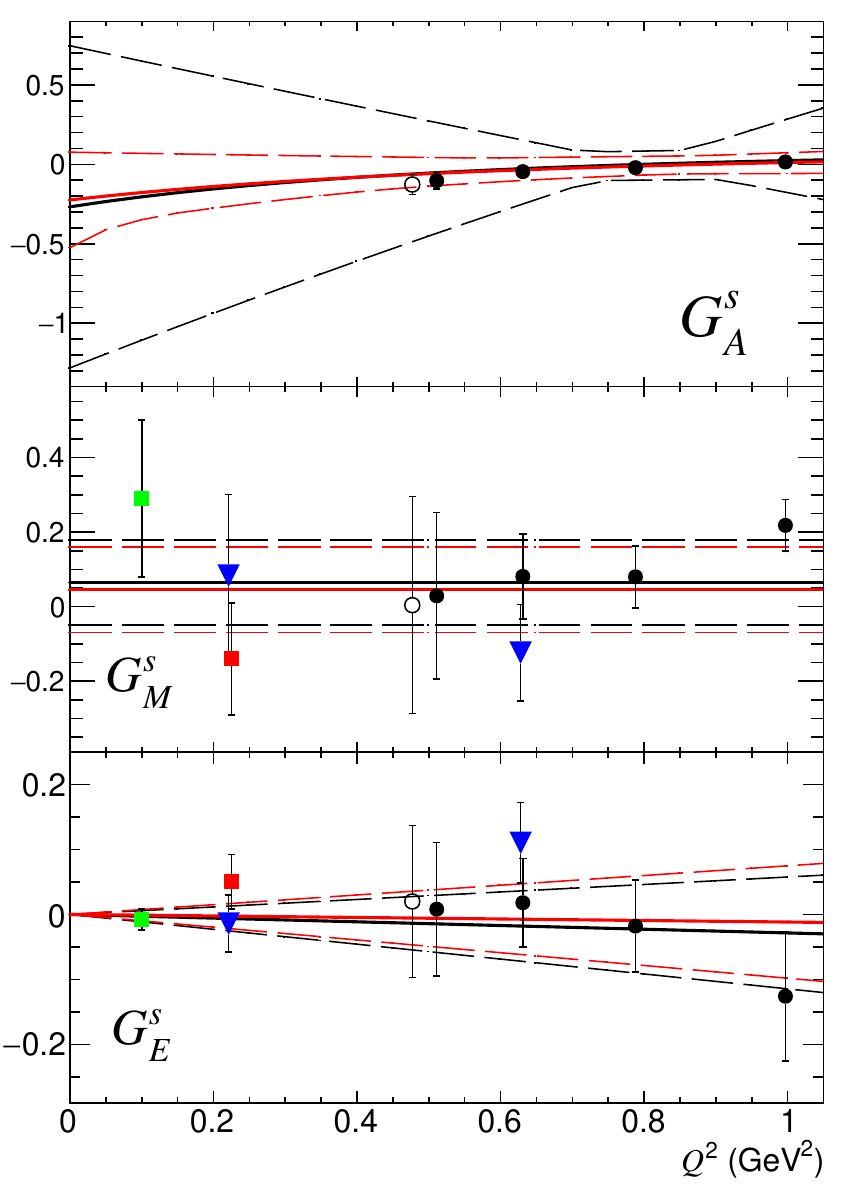}
\caption{An illustration of the effect of the introduction of the MiniBooNE neutral current data into our global fit.  The data points are the same as in Fig.~\ref{fig:old_points}.  The black solid line is the central value for the modified-dipole fit not using the MiniBooNE data. The red solid line includes the MiniBooNE data using the spectral function nuclear model.  The dashed lines represent the 70\% confidence limit for each fit. As mentioned in the text, the vector form factors fit is only slightly affected by the introduction of the MiniBooNE data, while the constraints on the axial form factor are greatly improved.}
\label{fig:reduction}
\end{figure}

Using the three nuclear models for carbon, and two models for the strangeness form factors, we performed 6 distinct fits.  In each fit, the five form factor parameters were varied to find the minimum $\chi^2$, and the behavior of the $\chi^2$ near the minimum was used to determine the uncertainties in the parameters.  The results are summarized in Table~\ref{table:results_summary}.
\begin{itemize}
    \item The results and the uncertainties for the parameters describing the strangeness vector form factors ($\rho_s$ and $\mu_s$) are not strongly affected by the inclusion of the MiniBooNE data.  The value of $\rho_s$ is slightly increased, and that of $\mu_s$ is slightly decreased, but both changes are within the fit and systematic uncertainties.
    \item The results and the uncertainties for the parameters describing the strangeness axial form factor are very strongly affected by the introduction of the MiniBooNE data.  The uncertainties, in particular, are reduced by 60-80\% in the case of the modified-dipole model, and by about 30\% in the case of the $z$-expansion model.
\end{itemize}
These two points are illustrated nicely by Fig.~\ref{fig:reduction}.  The total covariance matrix mentioned above has been used to calculate the 70\% confidence limit for each fit, and these limits are shown as the dashed lines in Figs.~\ref{fig:reduction} and \ref{fig:mod_vs_z}.

\begin{figure}[ht]
\includegraphics[scale=0.6]{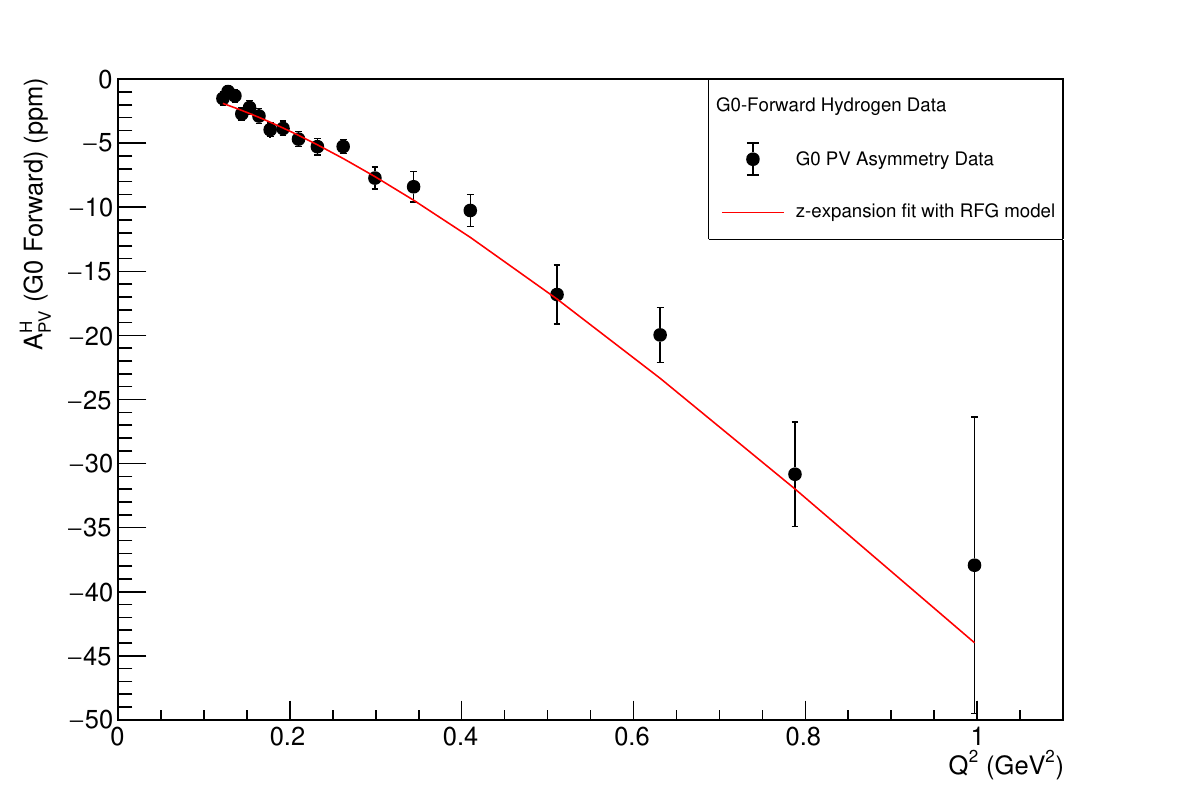}
\caption{Comparison between our fits and the PVES asymmetry data from the G0-Forward~\cite{Armstrong:2005hs} experiment.   The red line shows the results from a fit using the $z$-expansion model for $G_A^s$ and the RFG nuclear model.  The other five fits give nearly identical results for these data and so only one fit is shown here.}
\label{fig:G0}
\end{figure}

\begin{figure}[ht]
\includegraphics[scale=0.6]{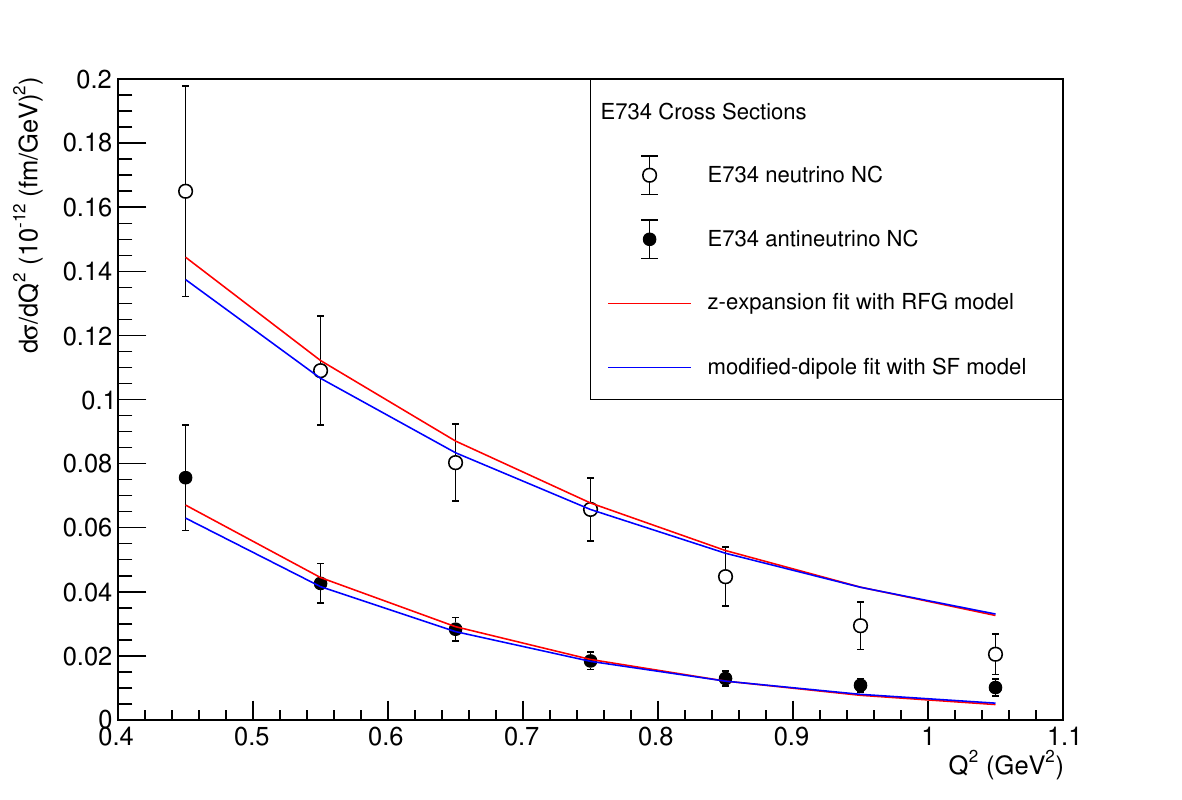}
\caption{Comparison between our fits and the NC neutrino-proton and antineutrino-proton data from the BNL E734~\cite{Ahrens:1986xe} experiment.  The red line shows the results of the fit using the $z$-expansion model for $G_A^s$ and the RFG nuclear model, while the blue line shows the result using the modified-dipole model for $G_A^s$ and the SF nuclear model. The other four fits show similar results for these data.}
\label{fig:E734}
\end{figure}

\begin{figure}[ht]
\includegraphics[scale=0.5]{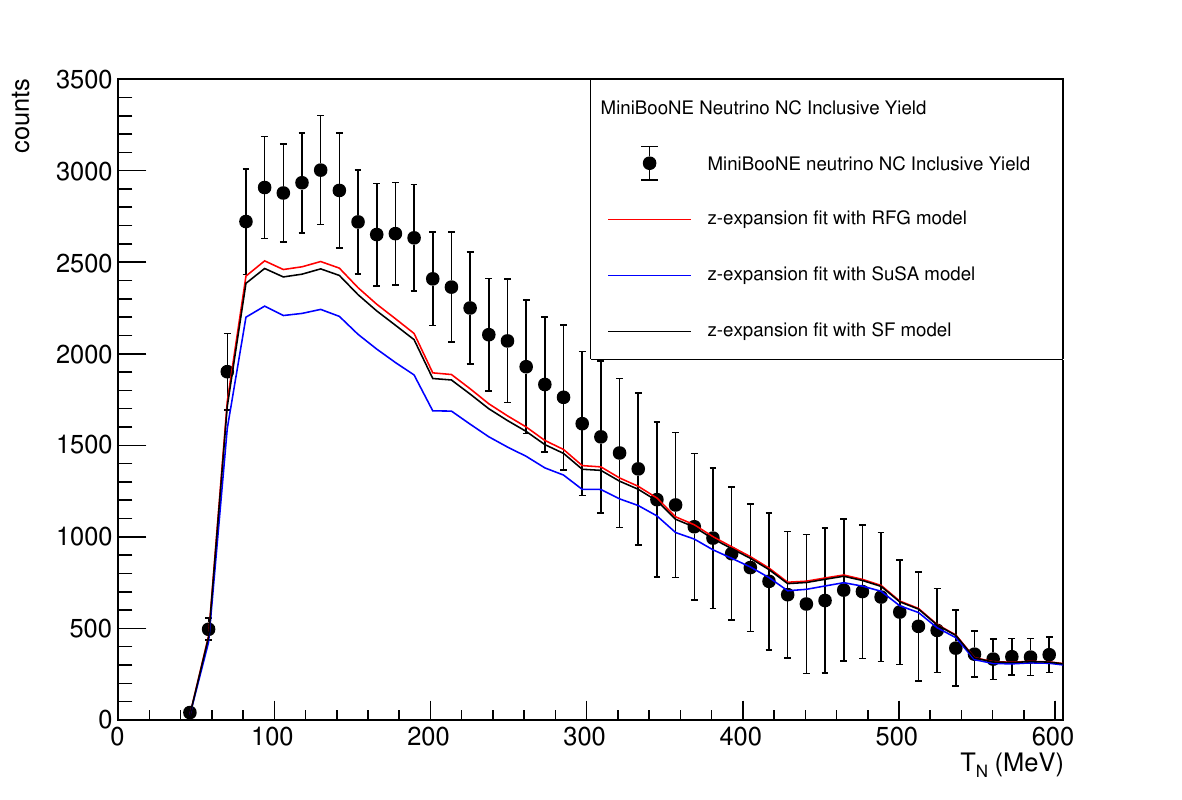} \\
\includegraphics[scale=0.5]{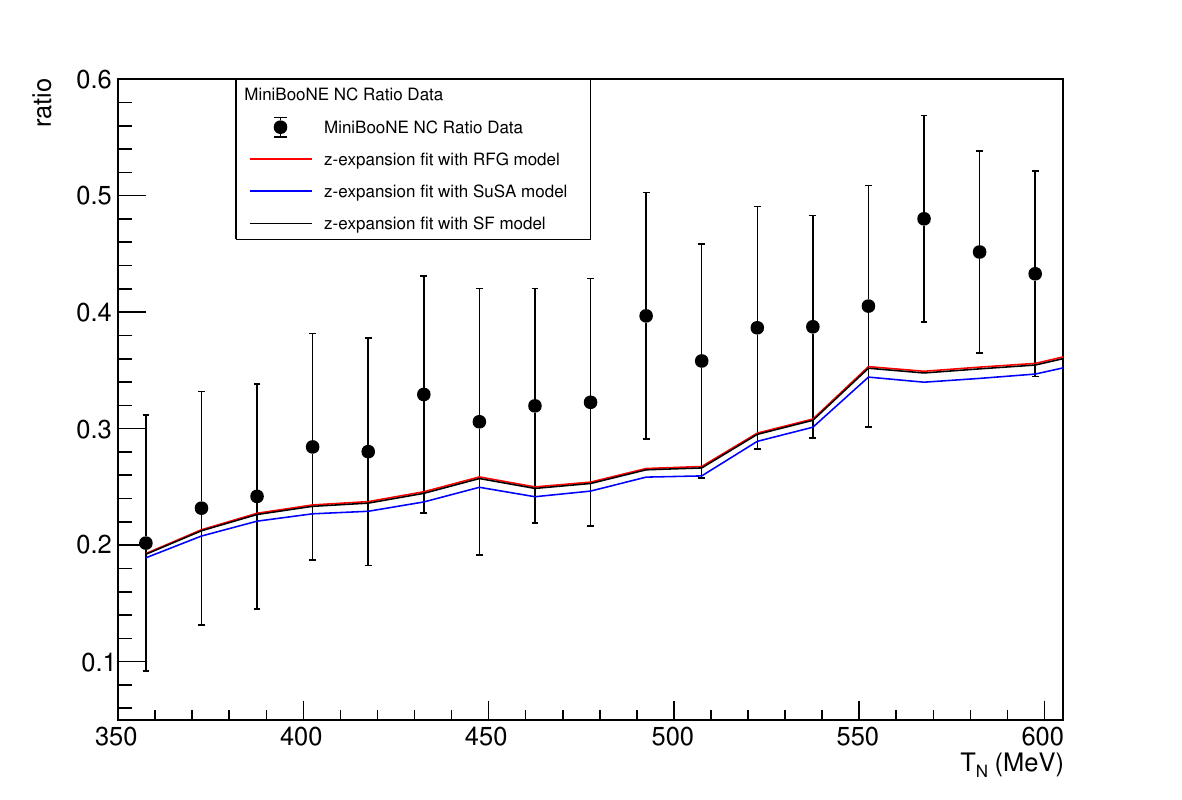} \\
\includegraphics[scale=0.5]{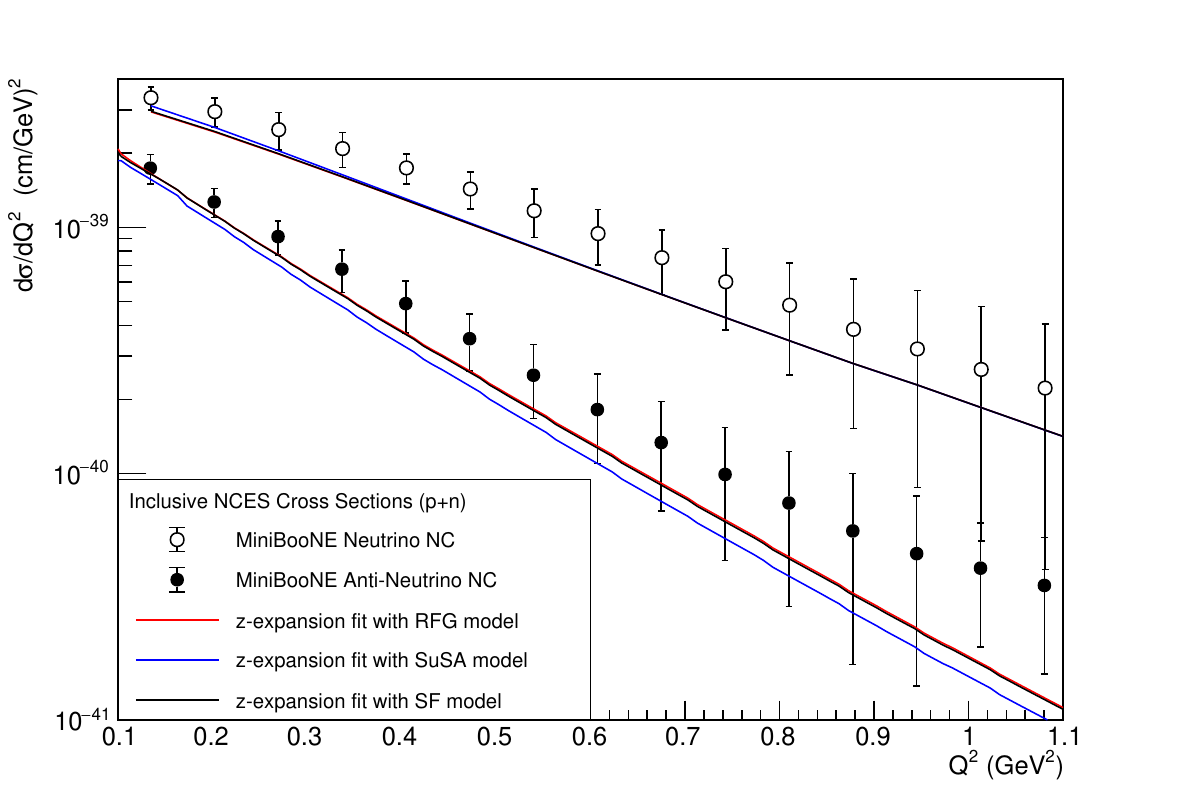}
\caption{Comparison between our fits and the NC scattering data from MiniBooNE~\cite{PhysRevD.82.092005,PhysRevD.91.012004}.  The upper panel shows the yield from the inclusive measurement of final state protons, the middle panel shows the $p/(p+n)$ yield ratio from the exclusive measurement, and the lower panel shows the cross sections from both the neutrino and antineutrino measurements.  All three fits shown use the $z$-expansion model for $G_A^s$.  The red line shows the results of the fit using the RFG nuclear model, the blue line shows the result using the SuSA nuclear model, and the black line is with the SF nuclear model. In all three cases the results for RFG (red) and SF (black) are very similar to each other and the lines almost overlap.  Fit results using instead the modified-dipole model for $G_A^s$ produce very similar results to those shown here.}
\label{fig:MB}
\end{figure}

\begin{figure}[ht]
\includegraphics[scale=0.45]{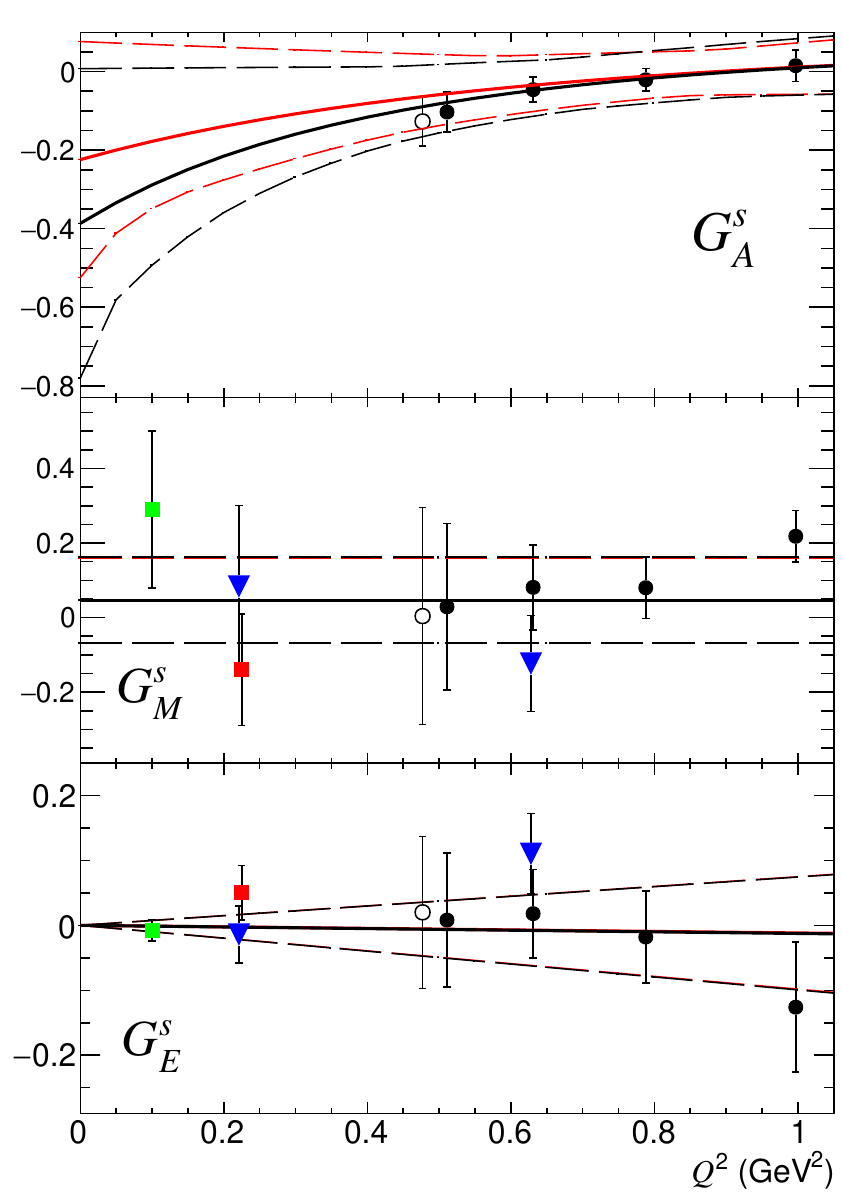}
~~~~~~~~
\includegraphics[scale=0.45]{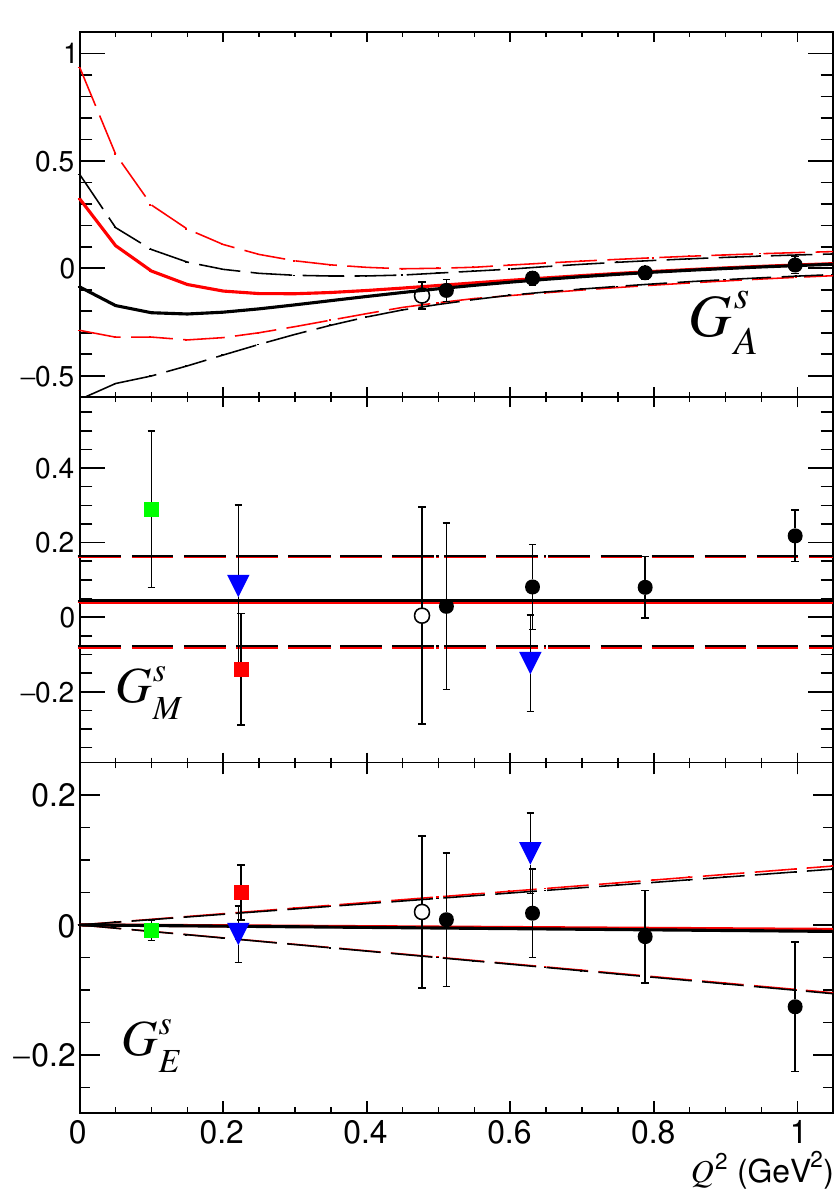}
\caption{Illustration of the vector and axial strangeness form factors from our global fits.  The fits in the left panel use the modified-dipole model for the strangeness axial form factor $G_A^s$, while in the right panel the $z$-expansion model is used.  The black solid line shows the fit with the SuSA nuclear model, and the red line is with the SF model.  The dashed lines represent the 70\% confidence limit for each fit. The results from the RFG and SF models are extremely similar, and so we have only shown the SF results in this figure.}
\label{fig:mod_vs_z}
\end{figure}

We stated earlier that the uncertainties in nucleon vector form factor models do not strongly affect the results presented here.  To illustrate this, we repeated one of the fits (SF nuclear model, and z-expansion model for $G_A^s$) using the Kelly vector form factors \cite{Kelly:2004hm} instead of the Arrington-Sick form factors, and we observed the following changes in the fit parameters:  $\Delta \rho_s$ = 0.012, $\Delta \mu_s$ = 0.004, $\Delta a_0$ = 0.006, $\Delta a_1$ = 0.10, and $\Delta a_2$ = 0.54.  In all five parameters, these changes are much less than the statistical and/or systematic errors.

\Cref{fig:G0,fig:E734,fig:MB} illustrate the quality of fitting to the wide variety of data we have used.
\Cref{fig:G0} shows just one of our fits, one using the $z$-expansion model for $G_A^s$ and the RFG nuclear model, compared to the PV asymmetry data from the G0 Forward~\cite{Armstrong:2005hs} experiment.  All six fits show very similar results for these data, as well as to other PVES data from HAPPEx, PVA4, and SAMPLE.  \Cref{fig:E734} compares two of our fits to the NC elastic data for neutrino and antineutrino scattering from BNL E734; all six fits show similar results for those data.

\Cref{fig:MB} displays all of the fits to the MiniBooNE~\cite{PhysRevD.82.092005,PhysRevD.91.012004} data using the $z$-expansion model and all three nuclear models; fits using the modified-dipole model are extremely similar and so are not shown here.  It is notable that the fits to the NC inclusive yield (top panel) and the exclusive $p/(p+n)$ ratio (middle panel) are not smooth curves, but instead have kinks.  These kinks arise from the backgrounds provided by the MiniBooNE collaboration, and can already be seen in Figures 4 and 12 of Ref.~\cite{PhysRevD.82.092005}.  These backgrounds are added to our signal calculation before comparison to the data; the kinks are an artefact of these backgrounds and not our model calculation.  

All the nuclear models employed in this study underestimate the NC neutrino cross section (upper panel in Fig.~\ref{fig:MB}) in the region of $T_N$ between 100 and 350 MeV. The disagreement between  theoretical predictions and cross section data can be explained by the absence of some nuclear effects in the RFG, SuSA, and SF models here adopted. For example, it has been shown in Ref.~\cite{Ivanov:2015wpa} that the inclusion of FSI in the SF model reduces the disagreement with the data, the same happens if the microscopic RMF model is used in place of the phenomenological SuSA approach, and in the RFG, where FSI give a reasonable agreement with the data. However, these improvements hardly affect the p/(p+n)  yield~\cite{Ivanov:2015wpa}, because the effects cancel in the ratio.

An important ingredient, which is missing in the present models, is the contribution of two-body currents, which can lead to the  excitation of 2p2h states. These contributions are not quasi-elastic, but they do contribute to the experimental signal represented in Fig.~\ref{fig:MB} and should be included in the calculation. In principle two-body currents could affect not only the cross sections but also the p/n ratios because of the isospin dependence of the current operator. However, while several calculations are now available for the 2p2h contribution to CC reactions, the corresponding calculations for NC scattering are very rare. In Ref.~\cite{Martini:2011wp} the 2p2h NC cross section was calculated in terms of the ``true'' $Q^2$ and, being based on an inclusive calculation where a sum over the final hadronic states was performed, the different isospin channels were not separated. Similarly, the 2p2h NC cross-section has been evaluated in  Refs.~\cite{Lovato:2014eva,Lovato:2017cux} for inclusive scattering $(\nu,\nu')$ as a function of the energy transfer $\omega$, an observable which is not experimentally accessible. Although both these calculations suggest that a better agreement with the experimental cross section is achieved by including the two-body currents in the model, none of them can provide predictions for the experimental ratios used in the extraction of the strange form factors. When a calculation for the 2p2h contribution to the $(\nu,\mu p)$ and $(\nu,\mu n)$ cross sections is available it will be possible to establish whether the two-body currents have an impact on the extraction of the strange form factors.

Finally, in Fig.~\ref{fig:mod_vs_z} are shown a sample of the six fits we have done for the strangeness vector and axial form factors, using both models for the form factors and two of the nuclear models.  The results from the RFG and SF nuclear models are very similar, so we have only shown the results from the SuSA and SF models in this Figure.  The results for the vector form factors $G_E^s$ and $G_M^s$ are nearly identical for both axial form factor models and all three nuclear models, but this is not a surprise since the NC neutrino scattering is not strongly dependent on them.  On the other hand, there is significant variation in the results for the strangeness axial form factor $G_A^s$.  The MiniBooNE data does greatly constrain the low-$Q^2$ behavior of $G_A^s$ (as shown in Fig.~\ref{fig:reduction}), but the lack of information on exclusive single-proton final states in the lowest $Q^2$ points means it cannot nail down a value of $\Delta s = G_A^s(Q^2=0)$.

\section{Conclusions and Next Steps}

We have performed a global fit of parity-violating electron-scattering data from the HAPPEx, SAMPLE, G0 and PVA4 experiments and of neutral-current elastic scattering data from the BNL E734 and Fermilab MiniBooNE experiments, a total of 128 data points in the momentum transfer range $0.1 < Q^2 < 1.1$ GeV$^2$, using two models for the strangeness form factors $G_E^s$, $G_M^s$, and $G_A^s$, and using three nuclear models to describe the interaction of neutrinos with the hydrocarbon target used in MiniBooNE. 
Our fits are in very good agreement with this collection of data, with $\chi^2$/ndf $\approx$ 1.1-1.2 for all fits.  

Depending on the model, we show a slightly negative value of the strangeness radius $\rho_s$ but also consistent with zero, and a slightly positive value for the strangeness magnetic moment $\mu_s$ also consistent with zero; we note this outcome is slightly at odds with other workers, for example Ref.~\cite{Gonzalez-Jimenez:2014bia}, who do not include neutrino NC scattering data into their fitting data set.  To quantify our conclusion that $\rho_s$ and $\mu_s$ are consistent with zero, we have taken $\rho_s=0$ and $\mu_s=0$ to be null hypotheses and then used our fit results for these quantities to calculate a corresponding p-value for each.  For the null hypothesis $\rho_s=0$ we find a p-value of 0.83; for the null hypothesis $\mu_s=0$ we find a p-value of 0.42.  These large p-values do not recommend a rejection of either of these null hypotheses.

The inclusion of the MiniBooNE neutral current data into the dataset has greatly improved the constraints on the strangeness axial form factor $G_A^s$, but still we cannot report a definite value for $\Delta s$ on the basis of these fits.
We can expect that a more refined model including two-body currents (which is currently not available but can hopefully become available in the future) would give a better description of the experimental NC cross section and might be helpful for an improved determination of the strange axial form factor, but presumably it should not change the main finding of our paper that the inclusion of the MiniBooNE neutral current data into the dataset greatly improves the constraints on $G_A^s$. 
Primarily, exclusive NCES data from proton interactions at low $Q^2$ are still needed for a complete determination of $G_A^s$, and we look forward to that data from MicroBooNE~\cite{Ren:2022qop} in the near future.

\begin{acknowledgements}
We are grateful to the following funding agencies:  US Department of Energy, Office of Science, Medium Energy Nuclear Physics Program, Grant DE-FG02-94ER40847; Bulgarian National Science Fund under Contract No. KP-06-N38/1; Istituto Nazionale di Fisica Nucleare under the National Project ``NUCSYS''; University of Turin local research funds BARM-RILO-22.
SFP is grateful for sabbatical support from both the Universities Research Association Visiting Scholars Program and the Los Alamos National Laboratory during 2022.
We are also grateful to R.~Dharmapalan for assistance in the interpretation of the MiniBooNE antineutrino NC data release.
\end{acknowledgements}

\clearpage

\appendix*
\section{Tables of Experimental Data}

\begin{table}[ht]
\caption{\label{E734_table} Differential cross section data from BNL
  E734~\protect\cite{Ahrens:1986xe}.  The uncertainties shown are
  total; they include statistical, $Q^2$-dependent systematic, and
  $Q^2$-independent systematic contributions, all added in quadrature.
  Also listed is the correlation 
  coefficient $\rho$ for the $\nu$ and $\bar{\nu}$
  data at each value of $Q^2$.}
\begin{tabular}{c|c|c|c}
$Q^2$ & $d\sigma/dQ^2(\nu p)$ & $d\sigma/dQ^2(\bar{\nu} p)$ & Correlation\\
GeV$^2$ &  $10^{-12}$~(fm/GeV)$^2$ & $10^{-12}$~(fm/GeV)$^2$ & Coefficient\\
\hline
0.45 & $0.165  \pm0.033$  & $0.0756 \pm0.0164$ & 0.13 \\
0.55 & $0.109  \pm0.017$  & $0.0426 \pm0.0062$ & 0.26 \\
0.65 & $0.0803 \pm0.0120$ & $0.0283 \pm0.0037$ & 0.29 \\
0.75 & $0.0657 \pm0.0098$ & $0.0184 \pm0.0027$ & 0.26 \\
0.85 & $0.0447 \pm0.0090$ & $0.0129 \pm0.0023$ & 0.16 \\
0.95 & $0.0294 \pm0.0073$ & $0.0108 \pm0.0022$ & 0.12 \\
1.05 & $0.0205 \pm0.0063$ & $0.0101 \pm0.0027$ & 0.07 \\
\end{tabular}
\end{table}

\begin{table}[ht]
\caption{Parity-violating asymmetries in forward-angle $\vec{e}p$ elastic
scattering from the PVA4, HAPPEx, and G0 experiments that have been used in this analysis.}
\label{PV_data_forward}
\begin{tabular}{c|c|c|c|c}
\hline
Experiment & $Q^2$   & $\theta_e$ & $A^H_{PV}$ & Reference\\
           & GeV$^2$ &            &  ppm       &          \\
\hline
PVA4   & 0.108 & $35.52^\circ$ & $ -1.36 \pm  0.32 $ & \cite{Maas:2004dh} \\
PVA4   & 0.230 & $35.45^\circ$ & $ -5.44 \pm  0.60 $ & \cite{Maas:2004ta} \\
HAPPEx & 0.099 & $  6.0^\circ$ & $ -1.14 \pm 0.25 $ & \cite{Aniol:2005zg} \\
HAPPEx & 0.109 & $  6.0^\circ$ & $ -1.58 \pm 0.13$ & \cite{Acha:2006my} \\
HAPPEx & 0.477 & $ 12.3^\circ$ & $-15.05 \pm  1.13 $ & \cite{Aniol:2004hp} \\
HAPPEx & 0.624 & $ 13.7^\circ$ & $-23.80 \pm 0.86 $ & \cite{Ahmed:2011vp} \\
    G0 & 0.122 & $ 6.68^\circ$ & $ -1.51 \pm  0.52 $ & \cite{Armstrong:2005hs} \\ 
    G0 & 0.128 & $ 6.84^\circ$ & $ -0.97 \pm  0.49 $ & \cite{Armstrong:2005hs} \\
    G0 & 0.136 & $ 7.06^\circ$ & $ -1.30 \pm  0.48 $ & \cite{Armstrong:2005hs} \\
    G0 & 0.144 & $ 7.27^\circ$ & $ -2.71 \pm  0.50 $ & \cite{Armstrong:2005hs} \\
    G0 & 0.153 & $ 7.50^\circ$ & $ -2.22 \pm  0.55 $ & \cite{Armstrong:2005hs} \\ 
    G0 & 0.164 & $ 7.77^\circ$ & $ -2.88 \pm  0.58 $ & \cite{Armstrong:2005hs} \\
    G0 & 0.177 & $ 8.09^\circ$ & $ -3.95 \pm  0.54 $ & \cite{Armstrong:2005hs} \\
    G0 & 0.192 & $ 8.43^\circ$ & $ -3.85 \pm  0.56 $ & \cite{Armstrong:2005hs} \\
    G0 & 0.210 & $ 8.83^\circ$ & $ -4.68 \pm  0.58 $ & \cite{Armstrong:2005hs} \\ 
    G0 & 0.232 & $ 9.31^\circ$ & $ -5.27 \pm  0.63 $ & \cite{Armstrong:2005hs} \\
    G0 & 0.262 & $ 9.92^\circ$ & $ -5.26 \pm  0.56 $ & \cite{Armstrong:2005hs} \\ 
    G0 & 0.299 & $10.6^\circ$  & $ -7.72 \pm  0.87 $ & \cite{Armstrong:2005hs} \\
    G0 & 0.344 & $11.4^\circ$  & $ -8.40 \pm  1.21 $ & \cite{Armstrong:2005hs} \\
    G0 & 0.410 & $12.5^\circ$  & $-10.25 \pm  1.24 $ & \cite{Armstrong:2005hs} \\
    G0 & 0.511 & $14.2^\circ$  & $-16.81 \pm  2.29 $ & \cite{Armstrong:2005hs} \\ 
    G0 & 0.631 & $15.9^\circ$  & $-19.96 \pm  2.14 $ & \cite{Armstrong:2005hs} \\
    G0 & 0.788 & $18.1^\circ$  & $-30.8  \pm  4.1  $ & \cite{Armstrong:2005hs} \\
    G0 & 0.997 & $20.9^\circ$  & $-37.9  \pm 11.5  $ & \cite{Armstrong:2005hs} \\
\hline
\end{tabular}
\end{table}

\begin{table}[ht]
\caption{\label{PV_data_backward} Parity-violating asymmetries in backward-angle $\vec{e}p$ elastic
scattering from the SAMPLE, PVA4, and G0 experiments that have been used in this analysis.}
\begin{tabular}{c|c|c|c|c}
\hline
Experiment & $Q^2$   & $\theta_e$ & $A^H_{PV}$ & Reference\\
           & GeV$^2$ &            &  ppm       &          \\
\hline
SAMPLE & 0.1   & $144.4^\circ$ & $ -5.61 \pm 1.11$ & \cite{Beise:2004py} \\
PVA4   & 0.224 & $144.8^\circ$ & $-17.23 \pm  1.21 $ & \cite{Baunack:2009gy} \\
    G0 & 0.221 & $106.5^\circ$ & $-11.25 \pm 0.99  $ & \cite{Androic:2009zu} \\
    G0 & 0.628 & $108.7^\circ$ & $-45.9  \pm 2.7   $ & \cite{Androic:2009zu} \\ 
\hline
\end{tabular}
\end{table}

\begin{table}[ht]
\caption{\label{PV_data_deuterium} Parity-violating asymmetries in $\vec{e}d$ quasi-elastic
scattering from the SAMPLE, PVA4, and G0 experiments that have been used in this analysis.}
\begin{tabular}{c|c|c|c|c}
\hline
Experiment & $Q^2$   & $\theta_e$ & $A^D_{PV}$ & Reference\\
           & GeV$^2$ &            &  ppm       &          \\
\hline
SAMPLE & 0.038 & $143.2^\circ$ & $  -3.51 \pm 0.81 $ & \cite{Beise:2004py} \\
SAMPLE & 0.091 & $144.0^\circ$ & $  -7.77 \pm 0.96 $ & \cite{Beise:2004py} \\
PVA4 & 0.224 & $144.6^\circ$ & $-20.11 \pm 1.35 $ & \cite{BalaguerRios:2016ftd} \\
G0     & 0.221 & $104.7^\circ$ & $ -16.93 \pm 0.93 $ & \cite{Androic:2009zu} \\
G0     & 0.628 & $107.8^\circ$ & $ -55.5  \pm 3.9  $ & \cite{Androic:2009zu} \\
\hline
\end{tabular}
\end{table}

\begin{table}[ht]
\caption{\label{PV_data_4He} Parity-violating asymmetries in $\vec{e}$-$^4$He quasi-elastic
scattering from the HAPPEx experiment that have been used in this analysis.}
\begin{tabular}{c|c|c|c|c}
\hline
Experiment & $Q^2$   & $\theta_e$ & $A^{^4\rm{He}}_{PV}$ & Reference\\
           & GeV$^2$ &            &  ppm       &          \\
\hline
HAPPEx & 0.077 & $5.8^\circ$ & $ 6.40 \pm 0.26$ & \cite{Acha:2006my} \\
HAPPEx & 0.091 & $5.7^\circ$ & $ 6.72 \pm 0.87 $ & \cite{Aniol:2005zf} \\
\hline
\end{tabular}
\end{table}

\clearpage

\bibliography{main}

\begin{thebibliography}{91}%
\makeatletter
\providecommand \@ifxundefined [1]{%
 \@ifx{#1\undefined}
}%
\providecommand \@ifnum [1]{%
 \ifnum #1\expandafter \@firstoftwo
 \else \expandafter \@secondoftwo
 \fi
}%
\providecommand \@ifx [1]{%
 \ifx #1\expandafter \@firstoftwo
 \else \expandafter \@secondoftwo
 \fi
}%
\providecommand \natexlab [1]{#1}%
\providecommand \enquote  [1]{``#1''}%
\providecommand \bibnamefont  [1]{#1}%
\providecommand \bibfnamefont [1]{#1}%
\providecommand \citenamefont [1]{#1}%
\providecommand \href@noop [0]{\@secondoftwo}%
\providecommand \href [0]{\begingroup \@sanitize@url \@href}%
\providecommand \@href[1]{\@@startlink{#1}\@@href}%
\providecommand \@@href[1]{\endgroup#1\@@endlink}%
\providecommand \@sanitize@url [0]{\catcode `\\12\catcode `\$12\catcode
  `\&12\catcode `\#12\catcode `\^12\catcode `\_12\catcode `\%12\relax}%
\providecommand \@@startlink[1]{}%
\providecommand \@@endlink[0]{}%
\providecommand \url  [0]{\begingroup\@sanitize@url \@url }%
\providecommand \@url [1]{\endgroup\@href {#1}{\urlprefix }}%
\providecommand \urlprefix  [0]{URL }%
\providecommand \Eprint [0]{\href }%
\providecommand \doibase [0]{http://dx.doi.org/}%
\providecommand \selectlanguage [0]{\@gobble}%
\providecommand \bibinfo  [0]{\@secondoftwo}%
\providecommand \bibfield  [0]{\@secondoftwo}%
\providecommand \translation [1]{[#1]}%
\providecommand \BibitemOpen [0]{}%
\providecommand \bibitemStop [0]{}%
\providecommand \bibitemNoStop [0]{.\EOS\space}%
\providecommand \EOS [0]{\spacefactor3000\relax}%
\providecommand \BibitemShut  [1]{\csname bibitem#1\endcsname}%
\let\auto@bib@innerbib\@empty
%</preamble>
\bibitem [{\citenamefont {Ashman}\ \emph {et~al.}(1988)\citenamefont {Ashman}
  \emph {et~al.}}]{AshmanEMC}%
  \BibitemOpen
  \bibfield  {author} {\bibinfo {author} {\bibfnamefont {J.}~\bibnamefont
  {Ashman}} \emph {et~al.},\ }\href {\doibase
  https://doi.org/10.1016/0370-2693(88)91523-7} {\bibfield  {journal} {\bibinfo
   {journal} {Physics Letters B}\ }\textbf {\bibinfo {volume} {206}},\ \bibinfo
  {pages} {364} (\bibinfo {year} {1988})}\BibitemShut {NoStop}%
\bibitem [{\citenamefont {Kaplan}\ and\ \citenamefont
  {Manohar}(1988)}]{KAPLAN1988527}%
  \BibitemOpen
  \bibfield  {author} {\bibinfo {author} {\bibfnamefont {D.~B.}\ \bibnamefont
  {Kaplan}}\ and\ \bibinfo {author} {\bibfnamefont {A.}~\bibnamefont
  {Manohar}},\ }\href {\doibase https://doi.org/10.1016/0550-3213(88)90090-9}
  {\bibfield  {journal} {\bibinfo  {journal} {Nuclear Physics B}\ }\textbf
  {\bibinfo {volume} {310}},\ \bibinfo {pages} {527} (\bibinfo {year}
  {1988})}\BibitemShut {NoStop}%
\bibitem [{\citenamefont {McKeown}(1989)}]{MCKEOWN1989140}%
  \BibitemOpen
  \bibfield  {author} {\bibinfo {author} {\bibfnamefont {R.}~\bibnamefont
  {McKeown}},\ }\href {\doibase https://doi.org/10.1016/0370-2693(89)90364-X}
  {\bibfield  {journal} {\bibinfo  {journal} {Physics Letters B}\ }\textbf
  {\bibinfo {volume} {219}},\ \bibinfo {pages} {140} (\bibinfo {year}
  {1989})}\BibitemShut {NoStop}%
\bibitem [{\citenamefont {Aidala}\ \emph {et~al.}(2013)\citenamefont {Aidala},
  \citenamefont {Bass}, \citenamefont {Hasch},\ and\ \citenamefont
  {Mallot}}]{RevModPhys.85.655}%
  \BibitemOpen
  \bibfield  {author} {\bibinfo {author} {\bibfnamefont {C.~A.}\ \bibnamefont
  {Aidala}}, \bibinfo {author} {\bibfnamefont {S.~D.}\ \bibnamefont {Bass}},
  \bibinfo {author} {\bibfnamefont {D.}~\bibnamefont {Hasch}}, \ and\ \bibinfo
  {author} {\bibfnamefont {G.~K.}\ \bibnamefont {Mallot}},\ }\href {\doibase
  10.1103/RevModPhys.85.655} {\bibfield  {journal} {\bibinfo  {journal} {Rev.
  Mod. Phys.}\ }\textbf {\bibinfo {volume} {85}},\ \bibinfo {pages} {655}
  (\bibinfo {year} {2013})}\BibitemShut {NoStop}%
\bibitem [{\citenamefont {Airapetian}\ \emph {et~al.}(2007)\citenamefont
  {Airapetian} \emph {et~al.}}]{Airapetian:2007mh}%
  \BibitemOpen
  \bibfield  {author} {\bibinfo {author} {\bibfnamefont {A.}~\bibnamefont
  {Airapetian}} \emph {et~al.} (\bibinfo {collaboration} {HERMES}),\ }\href
  {\doibase 10.1103/PhysRevD.75.012007} {\bibfield  {journal} {\bibinfo
  {journal} {Phys. Rev.}\ }\textbf {\bibinfo {volume} {D75}},\ \bibinfo {pages}
  {012007} (\bibinfo {year} {2007})},\ \Eprint
  {http://arxiv.org/abs/hep-ex/0609039} {arXiv:hep-ex/0609039} \BibitemShut
  {NoStop}%
%%CITATION = HEP-EX/0609039;%%
\bibitem [{\citenamefont {Airapetian}\ \emph {et~al.}(2005)\citenamefont
  {Airapetian} \emph {et~al.}}]{Airapetian:2004zf}%
  \BibitemOpen
  \bibfield  {author} {\bibinfo {author} {\bibfnamefont {A.}~\bibnamefont
  {Airapetian}} \emph {et~al.} (\bibinfo {collaboration} {HERMES}),\ }\href
  {\doibase 10.1103/PhysRevD.71.012003} {\bibfield  {journal} {\bibinfo
  {journal} {Phys. Rev.}\ }\textbf {\bibinfo {volume} {D71}},\ \bibinfo {pages}
  {012003} (\bibinfo {year} {2005})},\ \Eprint
  {http://arxiv.org/abs/hep-ex/0407032} {arXiv:hep-ex/0407032} \BibitemShut
  {NoStop}%
%%CITATION = HEP-EX/0407032;%%
\bibitem [{\citenamefont {Alekseev}\ \emph {et~al.}(2010)\citenamefont
  {Alekseev} \emph {et~al.}}]{Alekseev:2010ub}%
  \BibitemOpen
  \bibfield  {author} {\bibinfo {author} {\bibfnamefont {M.~G.}\ \bibnamefont
  {Alekseev}} \emph {et~al.} (\bibinfo {collaboration} {COMPASS}),\ }\href
  {\doibase 10.1016/j.physletb.2010.08.034} {\bibfield  {journal} {\bibinfo
  {journal} {Phys. Lett.}\ }\textbf {\bibinfo {volume} {B693}},\ \bibinfo
  {pages} {227} (\bibinfo {year} {2010})},\ \Eprint
  {http://arxiv.org/abs/1007.4061} {arXiv:1007.4061 [hep-ex]} \BibitemShut
  {NoStop}%
%%CITATION = 1007.4061;%%
\bibitem [{\citenamefont {de~Florian}\ \emph {et~al.}(2009)\citenamefont
  {de~Florian}, \citenamefont {Sassot}, \citenamefont {Stratmann},\ and\
  \citenamefont {Vogelsang}}]{deFlorian:2009vb}%
  \BibitemOpen
  \bibfield  {author} {\bibinfo {author} {\bibfnamefont {D.}~\bibnamefont
  {de~Florian}}, \bibinfo {author} {\bibfnamefont {R.}~\bibnamefont {Sassot}},
  \bibinfo {author} {\bibfnamefont {M.}~\bibnamefont {Stratmann}}, \ and\
  \bibinfo {author} {\bibfnamefont {W.}~\bibnamefont {Vogelsang}},\ }\href
  {\doibase 10.1103/PhysRevD.80.034030} {\bibfield  {journal} {\bibinfo
  {journal} {Phys. Rev.}\ }\textbf {\bibinfo {volume} {D80}},\ \bibinfo {pages}
  {034030} (\bibinfo {year} {2009})}\BibitemShut {NoStop}%
%%CITATION = 0904.3821;%%
\bibitem [{\citenamefont {Ahrens}\ \emph {et~al.}(1987)\citenamefont {Ahrens}
  \emph {et~al.}}]{Ahrens:1986xe}%
  \BibitemOpen
  \bibfield  {author} {\bibinfo {author} {\bibfnamefont {L.~A.}\ \bibnamefont
  {Ahrens}} \emph {et~al.},\ }\href {\doibase 10.1103/PhysRevD.35.785}
  {\bibfield  {journal} {\bibinfo  {journal} {Phys. Rev.}\ }\textbf {\bibinfo
  {volume} {D35}},\ \bibinfo {pages} {785} (\bibinfo {year}
  {1987})}\BibitemShut {NoStop}%
%%CITATION = PHRVA,D35,785;%%
\bibitem [{\citenamefont {Garvey}(1995)}]{GARVEY1995245}%
  \BibitemOpen
  \bibfield  {author} {\bibinfo {author} {\bibfnamefont {G.}~\bibnamefont
  {Garvey}},\ }\href {\doibase https://doi.org/10.1016/0146-6410(95)00021-A}
  {\bibfield  {journal} {\bibinfo  {journal} {Progress in Particle and Nuclear
  Physics}\ }\textbf {\bibinfo {volume} {34}},\ \bibinfo {pages} {245}
  (\bibinfo {year} {1995})}\BibitemShut {NoStop}%
\bibitem [{\citenamefont {Aguilar-Arevalo}\ \emph {et~al.}(2010)\citenamefont
  {Aguilar-Arevalo} \emph {et~al.}}]{PhysRevD.82.092005}%
  \BibitemOpen
  \bibfield  {author} {\bibinfo {author} {\bibfnamefont {A.~A.}\ \bibnamefont
  {Aguilar-Arevalo}} \emph {et~al.} (\bibinfo {collaboration} {MiniBooNE
  Collaboration}),\ }\href {\doibase 10.1103/PhysRevD.82.092005} {\bibfield
  {journal} {\bibinfo  {journal} {Phys. Rev. D}\ }\textbf {\bibinfo {volume}
  {82}},\ \bibinfo {pages} {092005} (\bibinfo {year} {2010})}\BibitemShut
  {NoStop}%
\bibitem [{\citenamefont {Aguilar-Arevalo}\ \emph {et~al.}(2015)\citenamefont
  {Aguilar-Arevalo} \emph {et~al.}}]{PhysRevD.91.012004}%
  \BibitemOpen
  \bibfield  {author} {\bibinfo {author} {\bibfnamefont {A.~A.}\ \bibnamefont
  {Aguilar-Arevalo}} \emph {et~al.} (\bibinfo {collaboration} {MiniBooNE
  Collaboration}),\ }\href {\doibase 10.1103/PhysRevD.91.012004} {\bibfield
  {journal} {\bibinfo  {journal} {Phys. Rev. D}\ }\textbf {\bibinfo {volume}
  {91}},\ \bibinfo {pages} {012004} (\bibinfo {year} {2015})}\BibitemShut
  {NoStop}%
\bibitem [{\citenamefont {Beise}\ \emph {et~al.}(2005)\citenamefont {Beise},
  \citenamefont {Pitt},\ and\ \citenamefont {Spayde}}]{Beise:2004py}%
  \BibitemOpen
  \bibfield  {author} {\bibinfo {author} {\bibfnamefont {E.~J.}\ \bibnamefont
  {Beise}}, \bibinfo {author} {\bibfnamefont {M.~L.}\ \bibnamefont {Pitt}}, \
  and\ \bibinfo {author} {\bibfnamefont {D.~T.}\ \bibnamefont {Spayde}},\
  }\href {\doibase 10.1016/j.ppnp.2004.07.002} {\bibfield  {journal} {\bibinfo
  {journal} {Prog. Part. Nucl. Phys.}\ }\textbf {\bibinfo {volume} {54}},\
  \bibinfo {pages} {289} (\bibinfo {year} {2005})},\ \Eprint
  {http://arxiv.org/abs/nucl-ex/0412054} {arXiv:nucl-ex/0412054} \BibitemShut
  {NoStop}%
%%CITATION = NUCL-EX/0412054;%%
\bibitem [{\citenamefont {Armstrong}\ \emph {et~al.}(2005)\citenamefont
  {Armstrong} \emph {et~al.}}]{Armstrong:2005hs}%
  \BibitemOpen
  \bibfield  {author} {\bibinfo {author} {\bibfnamefont {D.~S.}\ \bibnamefont
  {Armstrong}} \emph {et~al.} (\bibinfo {collaboration} {G0}),\ }\href
  {\doibase 10.1103/PhysRevLett.95.092001} {\bibfield  {journal} {\bibinfo
  {journal} {Phys. Rev. Lett.}\ }\textbf {\bibinfo {volume} {95}},\ \bibinfo
  {pages} {092001} (\bibinfo {year} {2005})},\ \Eprint
  {http://arxiv.org/abs/nucl-ex/0506021} {arXiv:nucl-ex/0506021} \BibitemShut
  {NoStop}%
%%CITATION = NUCL-EX/0506021;%%
\bibitem [{\citenamefont {Androi\'{c}}\ \emph {et~al.}(2010)\citenamefont
  {Androi\'{c}} \emph {et~al.}}]{Androic:2009zu}%
  \BibitemOpen
  \bibfield  {author} {\bibinfo {author} {\bibfnamefont {D.}~\bibnamefont
  {Androi\'{c}}} \emph {et~al.} (\bibinfo {collaboration} {G0}),\ }\href
  {\doibase 10.1103/PhysRevLett.104.012001} {\bibfield  {journal} {\bibinfo
  {journal} {Phys. Rev. Lett.}\ }\textbf {\bibinfo {volume} {104}},\ \bibinfo
  {pages} {012001} (\bibinfo {year} {2010})},\ \Eprint
  {http://arxiv.org/abs/0909.5107} {arXiv:0909.5107 [nucl-ex]} \BibitemShut
  {NoStop}%
%%CITATION = 0909.5107;%%
\bibitem [{\citenamefont {Aniol}\ \emph {et~al.}(2004)\citenamefont {Aniol}
  \emph {et~al.}}]{Aniol:2004hp}%
  \BibitemOpen
  \bibfield  {author} {\bibinfo {author} {\bibfnamefont {K.~A.}\ \bibnamefont
  {Aniol}} \emph {et~al.} (\bibinfo {collaboration} {HAPPEx}),\ }\href
  {\doibase 10.1103/PhysRevC.69.065501} {\bibfield  {journal} {\bibinfo
  {journal} {Phys. Rev.}\ }\textbf {\bibinfo {volume} {C69}},\ \bibinfo {pages}
  {065501} (\bibinfo {year} {2004})},\ \Eprint
  {http://arxiv.org/abs/nucl-ex/0402004} {arXiv:nucl-ex/0402004} \BibitemShut
  {NoStop}%
%%CITATION = NUCL-EX/0402004;%%
\bibitem [{\citenamefont {Aniol}\ \emph
  {et~al.}(2006{\natexlab{a}})\citenamefont {Aniol} \emph
  {et~al.}}]{Aniol:2005zf}%
  \BibitemOpen
  \bibfield  {author} {\bibinfo {author} {\bibfnamefont {K.~A.}\ \bibnamefont
  {Aniol}} \emph {et~al.} (\bibinfo {collaboration} {HAPPEx}),\ }\href
  {\doibase 10.1103/PhysRevLett.96.022003} {\bibfield  {journal} {\bibinfo
  {journal} {Phys. Rev. Lett.}\ }\textbf {\bibinfo {volume} {96}},\ \bibinfo
  {pages} {022003} (\bibinfo {year} {2006}{\natexlab{a}})},\ \Eprint
  {http://arxiv.org/abs/nucl-ex/0506010} {arXiv:nucl-ex/0506010} \BibitemShut
  {NoStop}%
%%CITATION = NUCL-EX/0506010;%%
\bibitem [{\citenamefont {Acha}\ \emph {et~al.}(2007)\citenamefont {Acha} \emph
  {et~al.}}]{Acha:2006my}%
  \BibitemOpen
  \bibfield  {author} {\bibinfo {author} {\bibfnamefont {A.}~\bibnamefont
  {Acha}} \emph {et~al.} (\bibinfo {collaboration} {HAPPEx}),\ }\href {\doibase
  10.1103/PhysRevLett.98.032301} {\bibfield  {journal} {\bibinfo  {journal}
  {Phys. Rev. Lett.}\ }\textbf {\bibinfo {volume} {98}},\ \bibinfo {pages}
  {032301} (\bibinfo {year} {2007})},\ \Eprint
  {http://arxiv.org/abs/nucl-ex/0609002} {arXiv:nucl-ex/0609002} \BibitemShut
  {NoStop}%
%%CITATION = NUCL-EX/0609002;%%
\bibitem [{\citenamefont {Aniol}\ \emph
  {et~al.}(2006{\natexlab{b}})\citenamefont {Aniol} \emph
  {et~al.}}]{Aniol:2005zg}%
  \BibitemOpen
  \bibfield  {author} {\bibinfo {author} {\bibfnamefont {K.~A.}\ \bibnamefont
  {Aniol}} \emph {et~al.} (\bibinfo {collaboration} {HAPPEx}),\ }\href
  {\doibase 10.1016/j.physletb.2006.03.011} {\bibfield  {journal} {\bibinfo
  {journal} {Phys. Lett.}\ }\textbf {\bibinfo {volume} {B635}},\ \bibinfo
  {pages} {275} (\bibinfo {year} {2006}{\natexlab{b}})},\ \Eprint
  {http://arxiv.org/abs/nucl-ex/0506011} {arXiv:nucl-ex/0506011} \BibitemShut
  {NoStop}%
%%CITATION = NUCL-EX/0506011;%%
\bibitem [{\citenamefont {Ahmed}\ \emph {et~al.}(2012)\citenamefont {Ahmed}
  \emph {et~al.}}]{Ahmed:2011vp}%
  \BibitemOpen
  \bibfield  {author} {\bibinfo {author} {\bibfnamefont {Z.}~\bibnamefont
  {Ahmed}} \emph {et~al.} (\bibinfo {collaboration} {HAPPEX collaboration}),\
  }\href {\doibase 10.1103/PhysRevLett.108.102001} {\bibfield  {journal}
  {\bibinfo  {journal} {Phys.Rev.Lett.}\ }\textbf {\bibinfo {volume} {108}},\
  \bibinfo {pages} {102001} (\bibinfo {year} {2012})},\ \Eprint
  {http://arxiv.org/abs/1107.0913} {arXiv:1107.0913 [nucl-ex]} \BibitemShut
  {NoStop}%
%%CITATION = ARXIV:1107.0913;%%
\bibitem [{\citenamefont {Maas}\ \emph {et~al.}(2005)\citenamefont {Maas} \emph
  {et~al.}}]{Maas:2004dh}%
  \BibitemOpen
  \bibfield  {author} {\bibinfo {author} {\bibfnamefont {F.~E.}\ \bibnamefont
  {Maas}} \emph {et~al.} (\bibinfo {collaboration} {A4}),\ }\href {\doibase
  10.1103/PhysRevLett.94.152001} {\bibfield  {journal} {\bibinfo  {journal}
  {Phys. Rev. Lett.}\ }\textbf {\bibinfo {volume} {94}},\ \bibinfo {pages}
  {152001} (\bibinfo {year} {2005})},\ \Eprint
  {http://arxiv.org/abs/nucl-ex/0412030} {arXiv:nucl-ex/0412030} \BibitemShut
  {NoStop}%
%%CITATION = NUCL-EX/0412030;%%
\bibitem [{\citenamefont {Maas}\ \emph {et~al.}(2004)\citenamefont {Maas} \emph
  {et~al.}}]{Maas:2004ta}%
  \BibitemOpen
  \bibfield  {author} {\bibinfo {author} {\bibfnamefont {F.~E.}\ \bibnamefont
  {Maas}} \emph {et~al.} (\bibinfo {collaboration} {A4}),\ }\href {\doibase
  10.1103/PhysRevLett.93.022002} {\bibfield  {journal} {\bibinfo  {journal}
  {Phys. Rev. Lett.}\ }\textbf {\bibinfo {volume} {93}},\ \bibinfo {pages}
  {022002} (\bibinfo {year} {2004})},\ \Eprint
  {http://arxiv.org/abs/nucl-ex/0401019} {arXiv:nucl-ex/0401019} \BibitemShut
  {NoStop}%
%%CITATION = NUCL-EX/0401019;%%
\bibitem [{\citenamefont {Baunack}\ \emph {et~al.}(2009)\citenamefont {Baunack}
  \emph {et~al.}}]{Baunack:2009gy}%
  \BibitemOpen
  \bibfield  {author} {\bibinfo {author} {\bibfnamefont {S.}~\bibnamefont
  {Baunack}} \emph {et~al.} (\bibinfo {collaboration} {A4}),\ }\href {\doibase
  10.1103/PhysRevLett.102.151803} {\bibfield  {journal} {\bibinfo  {journal}
  {Phys. Rev. Lett.}\ }\textbf {\bibinfo {volume} {102}},\ \bibinfo {pages}
  {151803} (\bibinfo {year} {2009})},\ \Eprint {http://arxiv.org/abs/0903.2733}
  {arXiv:0903.2733 [nucl-ex]} \BibitemShut {NoStop}%
%%CITATION = 0903.2733;%%
\bibitem [{\citenamefont {Balaguer~R\'\i{}os}\ \emph
  {et~al.}(2016)\citenamefont {Balaguer~R\'\i{}os} \emph
  {et~al.}}]{BalaguerRios:2016ftd}%
  \BibitemOpen
  \bibfield  {author} {\bibinfo {author} {\bibfnamefont {D.}~\bibnamefont
  {Balaguer~R\'\i{}os}} \emph {et~al.},\ }\href {\doibase
  10.1103/PhysRevD.94.051101} {\bibfield  {journal} {\bibinfo  {journal} {Phys.
  Rev. D}\ }\textbf {\bibinfo {volume} {94}},\ \bibinfo {pages} {051101}
  (\bibinfo {year} {2016})}\BibitemShut {NoStop}%
\bibitem [{\citenamefont {Pate}(2004)}]{Pate:2003rk}%
  \BibitemOpen
  \bibfield  {author} {\bibinfo {author} {\bibfnamefont {S.~F.}\ \bibnamefont
  {Pate}},\ }\href {\doibase 10.1103/PhysRevLett.92.082002} {\bibfield
  {journal} {\bibinfo  {journal} {Phys. Rev. Lett.}\ }\textbf {\bibinfo
  {volume} {92}},\ \bibinfo {pages} {082002} (\bibinfo {year} {2004})},\
  \Eprint {http://arxiv.org/abs/hep-ex/0310052} {arXiv:hep-ex/0310052}
  \BibitemShut {NoStop}%
\bibitem [{\citenamefont {Pate}\ \emph {et~al.}(2008)\citenamefont {Pate},
  \citenamefont {McKee},\ and\ \citenamefont {Papavassiliou}}]{Pate:2008va}%
  \BibitemOpen
  \bibfield  {author} {\bibinfo {author} {\bibfnamefont {S.~F.}\ \bibnamefont
  {Pate}}, \bibinfo {author} {\bibfnamefont {D.~W.}\ \bibnamefont {McKee}}, \
  and\ \bibinfo {author} {\bibfnamefont {V.}~\bibnamefont {Papavassiliou}},\
  }\href {\doibase 10.1103/PhysRevC.78.015207} {\bibfield  {journal} {\bibinfo
  {journal} {Phys. Rev.}\ }\textbf {\bibinfo {volume} {C78}},\ \bibinfo {pages}
  {015207} (\bibinfo {year} {2008})},\ \Eprint {http://arxiv.org/abs/0805.2889}
  {arXiv:0805.2889 [hep-ex]} \BibitemShut {NoStop}%
%%CITATION = 0805.2889;%%
\bibitem [{\citenamefont {Armstrong}\ and\ \citenamefont
  {McKeown}(2012)}]{Armstrong:2012bi}%
  \BibitemOpen
  \bibfield  {author} {\bibinfo {author} {\bibfnamefont {D.}~\bibnamefont
  {Armstrong}}\ and\ \bibinfo {author} {\bibfnamefont {R.}~\bibnamefont
  {McKeown}},\ }\href {\doibase 10.1146/annurev-nucl-102010-130419} {\bibfield
  {journal} {\bibinfo  {journal} {Ann.Rev.Nucl.Part.Sci.}\ }\textbf {\bibinfo
  {volume} {62}},\ \bibinfo {pages} {337} (\bibinfo {year} {2012})},\ \Eprint
  {http://arxiv.org/abs/1207.5238} {arXiv:1207.5238 [nucl-ex]} \BibitemShut
  {NoStop}%
%%CITATION = ARXIV:1207.5238;%%
\bibitem [{\citenamefont {Maas}\ and\ \citenamefont
  {Paschke}(2017)}]{MAAS2017209}%
  \BibitemOpen
  \bibfield  {author} {\bibinfo {author} {\bibfnamefont {F.}~\bibnamefont
  {Maas}}\ and\ \bibinfo {author} {\bibfnamefont {K.}~\bibnamefont {Paschke}},\
  }\href {\doibase https://doi.org/10.1016/j.ppnp.2016.11.001} {\bibfield
  {journal} {\bibinfo  {journal} {Progress in Particle and Nuclear Physics}\
  }\textbf {\bibinfo {volume} {95}},\ \bibinfo {pages} {209} (\bibinfo {year}
  {2017})}\BibitemShut {NoStop}%
\bibitem [{\citenamefont {Liu}\ \emph {et~al.}(2007)\citenamefont {Liu},
  \citenamefont {McKeown},\ and\ \citenamefont {Ramsey-Musolf}}]{Liu:2007yi}%
  \BibitemOpen
  \bibfield  {author} {\bibinfo {author} {\bibfnamefont {J.}~\bibnamefont
  {Liu}}, \bibinfo {author} {\bibfnamefont {R.~D.}\ \bibnamefont {McKeown}}, \
  and\ \bibinfo {author} {\bibfnamefont {M.~J.}\ \bibnamefont
  {Ramsey-Musolf}},\ }\href {\doibase 10.1103/PhysRevC.76.025202} {\bibfield
  {journal} {\bibinfo  {journal} {Phys. Rev.}\ }\textbf {\bibinfo {volume}
  {C76}},\ \bibinfo {pages} {025202} (\bibinfo {year} {2007})},\ \Eprint
  {http://arxiv.org/abs/0706.0226} {arXiv:0706.0226 [nucl-ex]} \BibitemShut
  {NoStop}%
%%CITATION = 0706.0226;%%
\bibitem [{\citenamefont {Pate}\ and\ \citenamefont
  {Trujillo}(2014)}]{Pate:2013wra}%
  \BibitemOpen
  \bibfield  {author} {\bibinfo {author} {\bibfnamefont {S.}~\bibnamefont
  {Pate}}\ and\ \bibinfo {author} {\bibfnamefont {D.}~\bibnamefont
  {Trujillo}},\ }\href {\doibase 10.1051/epjconf/20146606018} {\bibfield
  {journal} {\bibinfo  {journal} {EPJ Web Conf.}\ }\textbf {\bibinfo {volume}
  {66}},\ \bibinfo {pages} {06018} (\bibinfo {year} {2014})},\ \Eprint
  {http://arxiv.org/abs/1308.5694} {arXiv:1308.5694 [hep-ph]} \BibitemShut
  {NoStop}%
\bibitem [{\citenamefont {Cabibbo}\ \emph {et~al.}(2003)\citenamefont
  {Cabibbo}, \citenamefont {Swallow},\ and\ \citenamefont
  {Winston}}]{Cabibbo:2003cu}%
  \BibitemOpen
  \bibfield  {author} {\bibinfo {author} {\bibfnamefont {N.}~\bibnamefont
  {Cabibbo}}, \bibinfo {author} {\bibfnamefont {E.~C.}\ \bibnamefont
  {Swallow}}, \ and\ \bibinfo {author} {\bibfnamefont {R.}~\bibnamefont
  {Winston}},\ }\href {\doibase 10.1146/annurev.nucl.53.013103.155258}
  {\bibfield  {journal} {\bibinfo  {journal} {Ann. Rev. Nucl. Part. Sci.}\
  }\textbf {\bibinfo {volume} {53}},\ \bibinfo {pages} {39} (\bibinfo {year}
  {2003})},\ \Eprint {http://arxiv.org/abs/hep-ph/0307298}
  {arXiv:hep-ph/0307298} \BibitemShut {NoStop}%
\bibitem [{\citenamefont {Bodek}\ \emph {et~al.}(2008)\citenamefont {Bodek},
  \citenamefont {Avvakumov}, \citenamefont {Bradford},\ and\ \citenamefont
  {Budd}}]{Bodek:2007ym}%
  \BibitemOpen
  \bibfield  {author} {\bibinfo {author} {\bibfnamefont {A.}~\bibnamefont
  {Bodek}}, \bibinfo {author} {\bibfnamefont {S.}~\bibnamefont {Avvakumov}},
  \bibinfo {author} {\bibfnamefont {R.}~\bibnamefont {Bradford}}, \ and\
  \bibinfo {author} {\bibfnamefont {H.~S.}\ \bibnamefont {Budd}},\ }\href
  {\doibase 10.1140/epjc/s10052-007-0491-4} {\bibfield  {journal} {\bibinfo
  {journal} {Eur. Phys. J. C}\ }\textbf {\bibinfo {volume} {53}},\ \bibinfo
  {pages} {349} (\bibinfo {year} {2008})},\ \Eprint
  {http://arxiv.org/abs/0708.1946} {arXiv:0708.1946 [hep-ex]} \BibitemShut
  {NoStop}%
\bibitem [{\citenamefont {Marciano}\ and\ \citenamefont
  {Sirlin}(1980)}]{Marciano:1980pb}%
  \BibitemOpen
  \bibfield  {author} {\bibinfo {author} {\bibfnamefont {W.~J.}\ \bibnamefont
  {Marciano}}\ and\ \bibinfo {author} {\bibfnamefont {A.}~\bibnamefont
  {Sirlin}},\ }\href {\doibase 10.1103/PhysRevD.22.2695} {\bibfield  {journal}
  {\bibinfo  {journal} {Phys. Rev. D}\ }\textbf {\bibinfo {volume} {22}},\
  \bibinfo {pages} {2695} (\bibinfo {year} {1980})},\ \bibinfo {note}
  {[Erratum: Phys.Rev.D 31, 213 (1985)]}\BibitemShut {NoStop}%
\bibitem [{\citenamefont {Bass}\ \emph {et~al.}(2002)\citenamefont {Bass},
  \citenamefont {Crewther}, \citenamefont {Steffens},\ and\ \citenamefont
  {Thomas}}]{Bass:2002mv}%
  \BibitemOpen
  \bibfield  {author} {\bibinfo {author} {\bibfnamefont {S.~D.}\ \bibnamefont
  {Bass}}, \bibinfo {author} {\bibfnamefont {R.~J.}\ \bibnamefont {Crewther}},
  \bibinfo {author} {\bibfnamefont {F.~M.}\ \bibnamefont {Steffens}}, \ and\
  \bibinfo {author} {\bibfnamefont {A.~W.}\ \bibnamefont {Thomas}},\ }\href
  {\doibase 10.1103/PhysRevD.66.031901} {\bibfield  {journal} {\bibinfo
  {journal} {Phys. Rev. D}\ }\textbf {\bibinfo {volume} {66}},\ \bibinfo
  {pages} {031901} (\bibinfo {year} {2002})},\ \Eprint
  {http://arxiv.org/abs/hep-ph/0207071} {arXiv:hep-ph/0207071} \BibitemShut
  {NoStop}%
\bibitem [{\citenamefont {Beringer}\ \emph {et~al.}(2012)\citenamefont
  {Beringer} \emph {et~al.}}]{PDG2012}%
  \BibitemOpen
  \bibfield  {author} {\bibinfo {author} {\bibfnamefont {J.}~\bibnamefont
  {Beringer}} \emph {et~al.} (\bibinfo {collaboration} {Particle Data Group}),\
  }\href {\doibase 10.1103/PhysRevD.86.010001} {\bibfield  {journal} {\bibinfo
  {journal} {Phys. Rev. D}\ }\textbf {\bibinfo {volume} {86}},\ \bibinfo
  {pages} {010001} (\bibinfo {year} {2012})}\BibitemShut {NoStop}%
\bibitem [{\citenamefont {Musolf}\ \emph {et~al.}(1994)\citenamefont {Musolf}
  \emph {et~al.}}]{Musolf:1994tb}%
  \BibitemOpen
  \bibfield  {author} {\bibinfo {author} {\bibfnamefont {M.~J.}\ \bibnamefont
  {Musolf}} \emph {et~al.},\ }\href@noop {} {\bibfield  {journal} {\bibinfo
  {journal} {Phys. Rept.}\ }\textbf {\bibinfo {volume} {239}},\ \bibinfo
  {pages} {1} (\bibinfo {year} {1994})}\BibitemShut {NoStop}%
%%CITATION = PRPLC,239,1;%%
\bibitem [{\citenamefont {Eidelman}\ \emph {et~al.}(2004)\citenamefont
  {Eidelman} \emph {et~al.}}]{PDG2004}%
  \BibitemOpen
  \bibfield  {author} {\bibinfo {author} {\bibfnamefont {S.}~\bibnamefont
  {Eidelman}} \emph {et~al.} (\bibinfo {collaboration} {Particle Data Group}),\
  }\href@noop {} {\bibfield  {journal} {\bibinfo  {journal} {Phys. Lett.}\
  }\textbf {\bibinfo {volume} {B592}},\ \bibinfo {pages} {1} (\bibinfo {year}
  {2004})}\BibitemShut {NoStop}%
%%CITATION = PHLTA,B592,1;%%
\bibitem [{\citenamefont {Arrington}\ and\ \citenamefont
  {Sick}(2006)}]{Arrington2006PreciseDO}%
  \BibitemOpen
  \bibfield  {author} {\bibinfo {author} {\bibfnamefont {J.~R.}\ \bibnamefont
  {Arrington}}\ and\ \bibinfo {author} {\bibfnamefont {I.}~\bibnamefont
  {Sick}},\ }\href {https://api.semanticscholar.org/CorpusID:53559899}
  {\bibfield  {journal} {\bibinfo  {journal} {Physical Review C}\ }\textbf
  {\bibinfo {volume} {76}},\ \bibinfo {pages} {035201} (\bibinfo {year}
  {2006})}\BibitemShut {NoStop}%
\bibitem [{\citenamefont {Gran}\ \emph {et~al.}(2006)\citenamefont {Gran} \emph
  {et~al.}}]{Gran:2006jn}%
  \BibitemOpen
  \bibfield  {author} {\bibinfo {author} {\bibfnamefont {R.}~\bibnamefont
  {Gran}} \emph {et~al.} (\bibinfo {collaboration} {K2K}),\ }\href {\doibase
  10.1103/PhysRevD.74.052002} {\bibfield  {journal} {\bibinfo  {journal} {Phys.
  Rev.}\ }\textbf {\bibinfo {volume} {D74}},\ \bibinfo {pages} {052002}
  (\bibinfo {year} {2006})},\ \Eprint {http://arxiv.org/abs/hep-ex/0603034}
  {arXiv:hep-ex/0603034} \BibitemShut {NoStop}%
%%CITATION = HEP-EX/0603034;%%
\bibitem [{\citenamefont {Aguilar-Arevalo}\ \emph {et~al.}(2008)\citenamefont
  {Aguilar-Arevalo} \emph {et~al.}}]{AguilarArevalo:2007ru}%
  \BibitemOpen
  \bibfield  {author} {\bibinfo {author} {\bibfnamefont {A.~A.}\ \bibnamefont
  {Aguilar-Arevalo}} \emph {et~al.} (\bibinfo {collaboration} {MiniBooNE}),\
  }\href {\doibase 10.1103/PhysRevLett.100.032301} {\bibfield  {journal}
  {\bibinfo  {journal} {Phys. Rev. Lett.}\ }\textbf {\bibinfo {volume} {100}},\
  \bibinfo {pages} {032301} (\bibinfo {year} {2008})},\ \Eprint
  {http://arxiv.org/abs/0706.0926} {arXiv:0706.0926 [hep-ex]} \BibitemShut
  {NoStop}%
%%CITATION = 0706.0926;%%
\bibitem [{\citenamefont {Kuzmin}\ \emph {et~al.}(2008)\citenamefont {Kuzmin},
  \citenamefont {Lyubushkin},\ and\ \citenamefont {Naumov}}]{Kuzmin:2007kr}%
  \BibitemOpen
  \bibfield  {author} {\bibinfo {author} {\bibfnamefont {K.~S.}\ \bibnamefont
  {Kuzmin}}, \bibinfo {author} {\bibfnamefont {V.~V.}\ \bibnamefont
  {Lyubushkin}}, \ and\ \bibinfo {author} {\bibfnamefont {V.~A.}\ \bibnamefont
  {Naumov}},\ }\href {\doibase 10.1140/epjc/s10052-008-0582-x} {\bibfield
  {journal} {\bibinfo  {journal} {Eur. Phys. J. C}\ }\textbf {\bibinfo {volume}
  {54}},\ \bibinfo {pages} {517} (\bibinfo {year} {2008})},\ \Eprint
  {http://arxiv.org/abs/0712.4384} {arXiv:0712.4384 [hep-ph]} \BibitemShut
  {NoStop}%
\bibitem [{\citenamefont {Schiavilla}\ \emph {et~al.}(2004)\citenamefont
  {Schiavilla}, \citenamefont {Carlson},\ and\ \citenamefont
  {Paris}}]{PhysRevC.70.044007}%
  \BibitemOpen
  \bibfield  {author} {\bibinfo {author} {\bibfnamefont {R.}~\bibnamefont
  {Schiavilla}}, \bibinfo {author} {\bibfnamefont {J.}~\bibnamefont {Carlson}},
  \ and\ \bibinfo {author} {\bibfnamefont {M.}~\bibnamefont {Paris}},\ }\href
  {\doibase 10.1103/PhysRevC.70.044007} {\bibfield  {journal} {\bibinfo
  {journal} {Phys. Rev. C}\ }\textbf {\bibinfo {volume} {70}},\ \bibinfo
  {pages} {044007} (\bibinfo {year} {2004})}\BibitemShut {NoStop}%
\bibitem [{\citenamefont {Carlson}\ \emph {et~al.}(2002)\citenamefont
  {Carlson}, \citenamefont {Schiavilla}, \citenamefont {Brown},\ and\
  \citenamefont {Gibson}}]{PhysRevC.65.035502}%
  \BibitemOpen
  \bibfield  {author} {\bibinfo {author} {\bibfnamefont {J.}~\bibnamefont
  {Carlson}}, \bibinfo {author} {\bibfnamefont {R.}~\bibnamefont {Schiavilla}},
  \bibinfo {author} {\bibfnamefont {V.~R.}\ \bibnamefont {Brown}}, \ and\
  \bibinfo {author} {\bibfnamefont {B.~F.}\ \bibnamefont {Gibson}},\ }\href
  {\doibase 10.1103/PhysRevC.65.035502} {\bibfield  {journal} {\bibinfo
  {journal} {Phys. Rev. C}\ }\textbf {\bibinfo {volume} {65}},\ \bibinfo
  {pages} {035502} (\bibinfo {year} {2002})}\BibitemShut {NoStop}%
\bibitem [{G0_(2010)}]{G0_Backward}%
  \BibitemOpen
  \href@noop {} {\enquote {\bibinfo {title} {G0 backward scattering results},}\
  }\bibinfo {howpublished}
  {\url{http://research.npl.illinois.edu/exp/G0/backward/}} (\bibinfo {year}
  {2010})\BibitemShut {NoStop}%
\bibitem [{\citenamefont {Hill}\ and\ \citenamefont {Paz}(2010)}]{Hill:2010yb}%
  \BibitemOpen
  \bibfield  {author} {\bibinfo {author} {\bibfnamefont {R.~J.}\ \bibnamefont
  {Hill}}\ and\ \bibinfo {author} {\bibfnamefont {G.}~\bibnamefont {Paz}},\
  }\href {\doibase 10.1103/PhysRevD.82.113005} {\bibfield  {journal} {\bibinfo
  {journal} {Phys. Rev. D}\ }\textbf {\bibinfo {volume} {82}},\ \bibinfo
  {pages} {113005} (\bibinfo {year} {2010})},\ \Eprint
  {http://arxiv.org/abs/1008.4619} {arXiv:1008.4619 [hep-ph]} \BibitemShut
  {NoStop}%
\bibitem [{\citenamefont {Lee}\ \emph {et~al.}(2015)\citenamefont {Lee},
  \citenamefont {Arrington},\ and\ \citenamefont {Hill}}]{Lee:2015jqa}%
  \BibitemOpen
  \bibfield  {author} {\bibinfo {author} {\bibfnamefont {G.}~\bibnamefont
  {Lee}}, \bibinfo {author} {\bibfnamefont {J.~R.}\ \bibnamefont {Arrington}},
  \ and\ \bibinfo {author} {\bibfnamefont {R.~J.}\ \bibnamefont {Hill}},\
  }\href {\doibase 10.1103/PhysRevD.92.013013} {\bibfield  {journal} {\bibinfo
  {journal} {Phys. Rev. D}\ }\textbf {\bibinfo {volume} {92}},\ \bibinfo
  {pages} {013013} (\bibinfo {year} {2015})},\ \Eprint
  {http://arxiv.org/abs/1505.01489} {arXiv:1505.01489 [hep-ph]} \BibitemShut
  {NoStop}%
\bibitem [{\citenamefont {Garvey}\ \emph {et~al.}(1993)\citenamefont {Garvey},
  \citenamefont {Louis},\ and\ \citenamefont {White}}]{Garvey:1992cg}%
  \BibitemOpen
  \bibfield  {author} {\bibinfo {author} {\bibfnamefont {G.~T.}\ \bibnamefont
  {Garvey}}, \bibinfo {author} {\bibfnamefont {W.~C.}\ \bibnamefont {Louis}}, \
  and\ \bibinfo {author} {\bibfnamefont {D.~H.}\ \bibnamefont {White}},\ }\href
  {\doibase 10.1103/PhysRevC.48.761} {\bibfield  {journal} {\bibinfo  {journal}
  {Phys. Rev. C}\ }\textbf {\bibinfo {volume} {48}},\ \bibinfo {pages} {761}
  (\bibinfo {year} {1993})}\BibitemShut {NoStop}%
\bibitem [{\citenamefont {Golan}\ \emph {et~al.}(2013)\citenamefont {Golan},
  \citenamefont {Graczyk}, \citenamefont {Juszczak},\ and\ \citenamefont
  {Sobczyk}}]{PhysRevC.88.024612}%
  \BibitemOpen
  \bibfield  {author} {\bibinfo {author} {\bibfnamefont {T.}~\bibnamefont
  {Golan}}, \bibinfo {author} {\bibfnamefont {K.~M.}\ \bibnamefont {Graczyk}},
  \bibinfo {author} {\bibfnamefont {C.}~\bibnamefont {Juszczak}}, \ and\
  \bibinfo {author} {\bibfnamefont {J.~T.}\ \bibnamefont {Sobczyk}},\ }\href
  {\doibase 10.1103/PhysRevC.88.024612} {\bibfield  {journal} {\bibinfo
  {journal} {Phys. Rev. C}\ }\textbf {\bibinfo {volume} {88}},\ \bibinfo
  {pages} {024612} (\bibinfo {year} {2013})}\BibitemShut {NoStop}%
\bibitem [{\citenamefont {Butkevich}\ and\ \citenamefont
  {Perevalov}(2011)}]{Butkevich:2011fu}%
  \BibitemOpen
  \bibfield  {author} {\bibinfo {author} {\bibfnamefont {A.~V.}\ \bibnamefont
  {Butkevich}}\ and\ \bibinfo {author} {\bibfnamefont {D.}~\bibnamefont
  {Perevalov}},\ }\href {\doibase 10.1103/PhysRevC.84.015501} {\bibfield
  {journal} {\bibinfo  {journal} {Phys. Rev. C}\ }\textbf {\bibinfo {volume}
  {84}},\ \bibinfo {pages} {015501} (\bibinfo {year} {2011})},\ \Eprint
  {http://arxiv.org/abs/1106.0976} {arXiv:1106.0976 [hep-ph]} \BibitemShut
  {NoStop}%
\bibitem [{\citenamefont {Rocco}\ \emph {et~al.}(2019)\citenamefont {Rocco},
  \citenamefont {Barbieri}, \citenamefont {Benhar}, \citenamefont {De~Pace},\
  and\ \citenamefont {Lovato}}]{Rocco:2018mwt}%
  \BibitemOpen
  \bibfield  {author} {\bibinfo {author} {\bibfnamefont {N.}~\bibnamefont
  {Rocco}}, \bibinfo {author} {\bibfnamefont {C.}~\bibnamefont {Barbieri}},
  \bibinfo {author} {\bibfnamefont {O.}~\bibnamefont {Benhar}}, \bibinfo
  {author} {\bibfnamefont {A.}~\bibnamefont {De~Pace}}, \ and\ \bibinfo
  {author} {\bibfnamefont {A.}~\bibnamefont {Lovato}},\ }\href {\doibase
  10.1103/PhysRevC.99.025502} {\bibfield  {journal} {\bibinfo  {journal} {Phys.
  Rev. C}\ }\textbf {\bibinfo {volume} {99}},\ \bibinfo {pages} {025502}
  (\bibinfo {year} {2019})},\ \Eprint {http://arxiv.org/abs/1810.07647}
  {arXiv:1810.07647 [nucl-th]} \BibitemShut {NoStop}%
\bibitem [{\citenamefont {Martini}\ \emph {et~al.}(2011)\citenamefont
  {Martini}, \citenamefont {Ericson},\ and\ \citenamefont
  {Chanfray}}]{Martini:2011wp}%
  \BibitemOpen
  \bibfield  {author} {\bibinfo {author} {\bibfnamefont {M.}~\bibnamefont
  {Martini}}, \bibinfo {author} {\bibfnamefont {M.}~\bibnamefont {Ericson}}, \
  and\ \bibinfo {author} {\bibfnamefont {G.}~\bibnamefont {Chanfray}},\ }\href
  {\doibase 10.1103/PhysRevC.84.055502} {\bibfield  {journal} {\bibinfo
  {journal} {Phys. Rev. C}\ }\textbf {\bibinfo {volume} {84}},\ \bibinfo
  {pages} {055502} (\bibinfo {year} {2011})},\ \Eprint
  {http://arxiv.org/abs/1110.0221} {arXiv:1110.0221 [nucl-th]} \BibitemShut
  {NoStop}%
\bibitem [{\citenamefont {Giusti}\ and\ \citenamefont
  {Ivanov}(2020)}]{Giusti:2019cup}%
  \BibitemOpen
  \bibfield  {author} {\bibinfo {author} {\bibfnamefont {C.}~\bibnamefont
  {Giusti}}\ and\ \bibinfo {author} {\bibfnamefont {M.~V.}\ \bibnamefont
  {Ivanov}},\ }\href {\doibase 10.1088/1361-6471/ab5251} {\bibfield  {journal}
  {\bibinfo  {journal} {J. Phys. G}\ }\textbf {\bibinfo {volume} {47}},\
  \bibinfo {pages} {024001} (\bibinfo {year} {2020})},\ \Eprint
  {http://arxiv.org/abs/1908.08603} {arXiv:1908.08603 [hep-ph]} \BibitemShut
  {NoStop}%
\bibitem [{\citenamefont {Meucci}\ \emph {et~al.}(2003)\citenamefont {Meucci},
  \citenamefont {Capuzzi}, \citenamefont {Giusti},\ and\ \citenamefont
  {Pacati}}]{Meucci:2003uy}%
  \BibitemOpen
  \bibfield  {author} {\bibinfo {author} {\bibfnamefont {A.}~\bibnamefont
  {Meucci}}, \bibinfo {author} {\bibfnamefont {F.}~\bibnamefont {Capuzzi}},
  \bibinfo {author} {\bibfnamefont {C.}~\bibnamefont {Giusti}}, \ and\ \bibinfo
  {author} {\bibfnamefont {F.~D.}\ \bibnamefont {Pacati}},\ }\href {\doibase
  10.1103/PhysRevC.67.054601} {\bibfield  {journal} {\bibinfo  {journal} {Phys.
  Rev. C}\ }\textbf {\bibinfo {volume} {67}},\ \bibinfo {pages} {054601}
  (\bibinfo {year} {2003})},\ \Eprint {http://arxiv.org/abs/nucl-th/0301084}
  {arXiv:nucl-th/0301084} \BibitemShut {NoStop}%
\bibitem [{\citenamefont {Meucci}\ \emph {et~al.}(2009)\citenamefont {Meucci},
  \citenamefont {Caballero}, \citenamefont {Giusti}, \citenamefont {Pacati},\
  and\ \citenamefont {Udias}}]{Meucci:2009nm}%
  \BibitemOpen
  \bibfield  {author} {\bibinfo {author} {\bibfnamefont {A.}~\bibnamefont
  {Meucci}}, \bibinfo {author} {\bibfnamefont {J.~A.}\ \bibnamefont
  {Caballero}}, \bibinfo {author} {\bibfnamefont {C.}~\bibnamefont {Giusti}},
  \bibinfo {author} {\bibfnamefont {F.~D.}\ \bibnamefont {Pacati}}, \ and\
  \bibinfo {author} {\bibfnamefont {J.~M.}\ \bibnamefont {Udias}},\ }\href
  {\doibase 10.1103/PhysRevC.80.024605} {\bibfield  {journal} {\bibinfo
  {journal} {Phys. Rev. C}\ }\textbf {\bibinfo {volume} {80}},\ \bibinfo
  {pages} {024605} (\bibinfo {year} {2009})},\ \Eprint
  {http://arxiv.org/abs/0906.2645} {arXiv:0906.2645 [nucl-th]} \BibitemShut
  {NoStop}%
\bibitem [{\citenamefont {Meucci}\ \emph {et~al.}(2004)\citenamefont {Meucci},
  \citenamefont {Giusti},\ and\ \citenamefont {Pacati}}]{Meucci:2003cv}%
  \BibitemOpen
  \bibfield  {author} {\bibinfo {author} {\bibfnamefont {A.}~\bibnamefont
  {Meucci}}, \bibinfo {author} {\bibfnamefont {C.}~\bibnamefont {Giusti}}, \
  and\ \bibinfo {author} {\bibfnamefont {F.~D.}\ \bibnamefont {Pacati}},\
  }\href {\doibase 10.1016/j.nuclphysa.2004.04.108} {\bibfield  {journal}
  {\bibinfo  {journal} {Nucl. Phys. A}\ }\textbf {\bibinfo {volume} {739}},\
  \bibinfo {pages} {277} (\bibinfo {year} {2004})},\ \Eprint
  {http://arxiv.org/abs/nucl-th/0311081} {arXiv:nucl-th/0311081} \BibitemShut
  {NoStop}%
\bibitem [{\citenamefont {Meucci}\ \emph
  {et~al.}(2011{\natexlab{a}})\citenamefont {Meucci}, \citenamefont
  {Caballero}, \citenamefont {Giusti},\ and\ \citenamefont
  {Udias}}]{Meucci:2011pi}%
  \BibitemOpen
  \bibfield  {author} {\bibinfo {author} {\bibfnamefont {A.}~\bibnamefont
  {Meucci}}, \bibinfo {author} {\bibfnamefont {J.~A.}\ \bibnamefont
  {Caballero}}, \bibinfo {author} {\bibfnamefont {C.}~\bibnamefont {Giusti}}, \
  and\ \bibinfo {author} {\bibfnamefont {J.~M.}\ \bibnamefont {Udias}},\ }\href
  {\doibase 10.1103/PhysRevC.83.064614} {\bibfield  {journal} {\bibinfo
  {journal} {Phys. Rev. C}\ }\textbf {\bibinfo {volume} {83}},\ \bibinfo
  {pages} {064614} (\bibinfo {year} {2011}{\natexlab{a}})},\ \Eprint
  {http://arxiv.org/abs/1103.0636} {arXiv:1103.0636 [nucl-th]} \BibitemShut
  {NoStop}%
\bibitem [{\citenamefont {Meucci}\ \emph
  {et~al.}(2011{\natexlab{b}})\citenamefont {Meucci}, \citenamefont {Barbaro},
  \citenamefont {Caballero}, \citenamefont {Giusti},\ and\ \citenamefont
  {Udias}}]{Meucci:2011vd}%
  \BibitemOpen
  \bibfield  {author} {\bibinfo {author} {\bibfnamefont {A.}~\bibnamefont
  {Meucci}}, \bibinfo {author} {\bibfnamefont {M.~B.}\ \bibnamefont {Barbaro}},
  \bibinfo {author} {\bibfnamefont {J.~A.}\ \bibnamefont {Caballero}}, \bibinfo
  {author} {\bibfnamefont {C.}~\bibnamefont {Giusti}}, \ and\ \bibinfo {author}
  {\bibfnamefont {J.~M.}\ \bibnamefont {Udias}},\ }\href {\doibase
  10.1103/PhysRevLett.107.172501} {\bibfield  {journal} {\bibinfo  {journal}
  {Phys. Rev. Lett.}\ }\textbf {\bibinfo {volume} {107}},\ \bibinfo {pages}
  {172501} (\bibinfo {year} {2011}{\natexlab{b}})},\ \Eprint
  {http://arxiv.org/abs/1107.5145} {arXiv:1107.5145 [nucl-th]} \BibitemShut
  {NoStop}%
\bibitem [{\citenamefont {Meucci}\ and\ \citenamefont
  {Giusti}(2015)}]{Meucci:2015bea}%
  \BibitemOpen
  \bibfield  {author} {\bibinfo {author} {\bibfnamefont {A.}~\bibnamefont
  {Meucci}}\ and\ \bibinfo {author} {\bibfnamefont {C.}~\bibnamefont
  {Giusti}},\ }\href {\doibase 10.1103/PhysRevD.91.093004} {\bibfield
  {journal} {\bibinfo  {journal} {Phys. Rev. D}\ }\textbf {\bibinfo {volume}
  {91}},\ \bibinfo {pages} {093004} (\bibinfo {year} {2015})},\ \Eprint
  {http://arxiv.org/abs/1501.03213} {arXiv:1501.03213 [nucl-th]} \BibitemShut
  {NoStop}%
\bibitem [{\citenamefont {Gonz\'alez-Jim\'enez}\ \emph
  {et~al.}(2013)\citenamefont {Gonz\'alez-Jim\'enez}, \citenamefont
  {Caballero}, \citenamefont {Meucci}, \citenamefont {Giusti}, \citenamefont
  {Barbaro}, \citenamefont {Ivanov},\ and\ \citenamefont
  {Ud\'\i{}as}}]{Gonzalez-Jimenez:2013xpa}%
  \BibitemOpen
  \bibfield  {author} {\bibinfo {author} {\bibfnamefont {R.}~\bibnamefont
  {Gonz\'alez-Jim\'enez}}, \bibinfo {author} {\bibfnamefont {J.~A.}\
  \bibnamefont {Caballero}}, \bibinfo {author} {\bibfnamefont {A.}~\bibnamefont
  {Meucci}}, \bibinfo {author} {\bibfnamefont {C.}~\bibnamefont {Giusti}},
  \bibinfo {author} {\bibfnamefont {M.~B.}\ \bibnamefont {Barbaro}}, \bibinfo
  {author} {\bibfnamefont {M.~V.}\ \bibnamefont {Ivanov}}, \ and\ \bibinfo
  {author} {\bibfnamefont {J.~M.}\ \bibnamefont {Ud\'\i{}as}},\ }\href
  {\doibase 10.1103/PhysRevC.88.025502} {\bibfield  {journal} {\bibinfo
  {journal} {Phys. Rev. C}\ }\textbf {\bibinfo {volume} {88}},\ \bibinfo
  {pages} {025502} (\bibinfo {year} {2013})},\ \Eprint
  {http://arxiv.org/abs/1307.4309} {arXiv:1307.4309 [nucl-th]} \BibitemShut
  {NoStop}%
\bibitem [{\citenamefont {Meucci}\ and\ \citenamefont
  {Giusti}(2014)}]{Meucci:2014pka}%
  \BibitemOpen
  \bibfield  {author} {\bibinfo {author} {\bibfnamefont {A.}~\bibnamefont
  {Meucci}}\ and\ \bibinfo {author} {\bibfnamefont {C.}~\bibnamefont
  {Giusti}},\ }\href {\doibase 10.1103/PhysRevD.89.057302} {\bibfield
  {journal} {\bibinfo  {journal} {Phys. Rev. D}\ }\textbf {\bibinfo {volume}
  {89}},\ \bibinfo {pages} {057302} (\bibinfo {year} {2014})},\ \Eprint
  {http://arxiv.org/abs/1401.3650} {arXiv:1401.3650 [nucl-th]} \BibitemShut
  {NoStop}%
\bibitem [{\citenamefont {Ivanov}\ \emph {et~al.}(2015)\citenamefont {Ivanov},
  \citenamefont {Antonov}, \citenamefont {Barbaro}, \citenamefont {Giusti},
  \citenamefont {Meucci}, \citenamefont {Caballero}, \citenamefont
  {Gonz\'alez-Jim\'enez}, \citenamefont {Moya~de Guerra},\ and\ \citenamefont
  {Ud\'\i{}as}}]{Ivanov:2015wpa}%
  \BibitemOpen
  \bibfield  {author} {\bibinfo {author} {\bibfnamefont {M.~V.}\ \bibnamefont
  {Ivanov}}, \bibinfo {author} {\bibfnamefont {A.~N.}\ \bibnamefont {Antonov}},
  \bibinfo {author} {\bibfnamefont {M.~B.}\ \bibnamefont {Barbaro}}, \bibinfo
  {author} {\bibfnamefont {C.}~\bibnamefont {Giusti}}, \bibinfo {author}
  {\bibfnamefont {A.}~\bibnamefont {Meucci}}, \bibinfo {author} {\bibfnamefont
  {J.~A.}\ \bibnamefont {Caballero}}, \bibinfo {author} {\bibfnamefont
  {R.}~\bibnamefont {Gonz\'alez-Jim\'enez}}, \bibinfo {author} {\bibfnamefont
  {E.}~\bibnamefont {Moya~de Guerra}}, \ and\ \bibinfo {author} {\bibfnamefont
  {J.~M.}\ \bibnamefont {Ud\'\i{}as}},\ }\href {\doibase
  10.1103/PhysRevC.91.034607} {\bibfield  {journal} {\bibinfo  {journal} {Phys.
  Rev. C}\ }\textbf {\bibinfo {volume} {91}},\ \bibinfo {pages} {034607}
  (\bibinfo {year} {2015})},\ \Eprint {http://arxiv.org/abs/1503.00053}
  {arXiv:1503.00053 [nucl-th]} \BibitemShut {NoStop}%
\bibitem [{\citenamefont {Barbaro}\ \emph {et~al.}(1996)\citenamefont
  {Barbaro}, \citenamefont {De~Pace}, \citenamefont {Donnelly}, \citenamefont
  {Molinari},\ and\ \citenamefont {Musolf}}]{Barbaro:1996vd}%
  \BibitemOpen
  \bibfield  {author} {\bibinfo {author} {\bibfnamefont {M.~B.}\ \bibnamefont
  {Barbaro}}, \bibinfo {author} {\bibfnamefont {A.}~\bibnamefont {De~Pace}},
  \bibinfo {author} {\bibfnamefont {T.~W.}\ \bibnamefont {Donnelly}}, \bibinfo
  {author} {\bibfnamefont {A.}~\bibnamefont {Molinari}}, \ and\ \bibinfo
  {author} {\bibfnamefont {M.~J.}\ \bibnamefont {Musolf}},\ }\href {\doibase
  10.1103/PhysRevC.54.1954} {\bibfield  {journal} {\bibinfo  {journal} {Phys.
  Rev. C}\ }\textbf {\bibinfo {volume} {54}},\ \bibinfo {pages} {1954}
  (\bibinfo {year} {1996})},\ \Eprint {http://arxiv.org/abs/nucl-th/9605020}
  {arXiv:nucl-th/9605020} \BibitemShut {NoStop}%
\bibitem [{\citenamefont {Amaro}\ \emph {et~al.}(2006)\citenamefont {Amaro},
  \citenamefont {Barbaro}, \citenamefont {Caballero},\ and\ \citenamefont
  {Donnelly}}]{Amaro:2006pr}%
  \BibitemOpen
  \bibfield  {author} {\bibinfo {author} {\bibfnamefont {J.~E.}\ \bibnamefont
  {Amaro}}, \bibinfo {author} {\bibfnamefont {M.~B.}\ \bibnamefont {Barbaro}},
  \bibinfo {author} {\bibfnamefont {J.~A.}\ \bibnamefont {Caballero}}, \ and\
  \bibinfo {author} {\bibfnamefont {T.~W.}\ \bibnamefont {Donnelly}},\ }\href
  {\doibase 10.1103/PhysRevC.73.035503} {\bibfield  {journal} {\bibinfo
  {journal} {Phys. Rev. C}\ }\textbf {\bibinfo {volume} {73}},\ \bibinfo
  {pages} {035503} (\bibinfo {year} {2006})},\ \Eprint
  {http://arxiv.org/abs/nucl-th/0602053} {arXiv:nucl-th/0602053} \BibitemShut
  {NoStop}%
\bibitem [{\citenamefont {Alberico}\ \emph {et~al.}(1988)\citenamefont
  {Alberico}, \citenamefont {Molinari}, \citenamefont {Donnelly}, \citenamefont
  {Kronenberg},\ and\ \citenamefont {Van~Orden}}]{PhysRevC.38.1801}%
  \BibitemOpen
  \bibfield  {author} {\bibinfo {author} {\bibfnamefont {W.~M.}\ \bibnamefont
  {Alberico}}, \bibinfo {author} {\bibfnamefont {A.}~\bibnamefont {Molinari}},
  \bibinfo {author} {\bibfnamefont {T.~W.}\ \bibnamefont {Donnelly}}, \bibinfo
  {author} {\bibfnamefont {E.~L.}\ \bibnamefont {Kronenberg}}, \ and\ \bibinfo
  {author} {\bibfnamefont {J.~W.}\ \bibnamefont {Van~Orden}},\ }\href {\doibase
  10.1103/PhysRevC.38.1801} {\bibfield  {journal} {\bibinfo  {journal} {Phys.
  Rev. C}\ }\textbf {\bibinfo {volume} {38}},\ \bibinfo {pages} {1801}
  (\bibinfo {year} {1988})}\BibitemShut {NoStop}%
\bibitem [{\citenamefont {Barbaro}\ \emph {et~al.}(1998)\citenamefont
  {Barbaro}, \citenamefont {Cenni}, \citenamefont {Pace}, \citenamefont
  {Donnelly},\ and\ \citenamefont {Molinari}}]{Barbaro1998137}%
  \BibitemOpen
  \bibfield  {author} {\bibinfo {author} {\bibfnamefont {M.}~\bibnamefont
  {Barbaro}}, \bibinfo {author} {\bibfnamefont {R.}~\bibnamefont {Cenni}},
  \bibinfo {author} {\bibfnamefont {A.~D.}\ \bibnamefont {Pace}}, \bibinfo
  {author} {\bibfnamefont {T.}~\bibnamefont {Donnelly}}, \ and\ \bibinfo
  {author} {\bibfnamefont {A.}~\bibnamefont {Molinari}},\ }\href {\doibase
  10.1016/S0375-9474(98)00443-6} {\bibfield  {journal} {\bibinfo  {journal}
  {Nuclear Physics A}\ }\textbf {\bibinfo {volume} {643}},\ \bibinfo {pages}
  {137 } (\bibinfo {year} {1998})}\BibitemShut {NoStop}%
\bibitem [{\citenamefont {Amaro}\ \emph {et~al.}(2005)\citenamefont {Amaro},
  \citenamefont {Barbaro}, \citenamefont {Caballero}, \citenamefont {Donnelly},
  \citenamefont {Molinari},\ and\ \citenamefont {Sick}}]{Amaro:2004bs}%
  \BibitemOpen
  \bibfield  {author} {\bibinfo {author} {\bibfnamefont {J.~E.}\ \bibnamefont
  {Amaro}}, \bibinfo {author} {\bibfnamefont {M.~B.}\ \bibnamefont {Barbaro}},
  \bibinfo {author} {\bibfnamefont {J.~A.}\ \bibnamefont {Caballero}}, \bibinfo
  {author} {\bibfnamefont {T.~W.}\ \bibnamefont {Donnelly}}, \bibinfo {author}
  {\bibfnamefont {A.}~\bibnamefont {Molinari}}, \ and\ \bibinfo {author}
  {\bibfnamefont {I.}~\bibnamefont {Sick}},\ }\href {\doibase
  10.1103/PhysRevC.71.015501} {\bibfield  {journal} {\bibinfo  {journal} {Phys.
  Rev. C}\ }\textbf {\bibinfo {volume} {71}},\ \bibinfo {pages} {015501}
  (\bibinfo {year} {2005})},\ \Eprint {http://arxiv.org/abs/nucl-th/0409078}
  {arXiv:nucl-th/0409078} \BibitemShut {NoStop}%
\bibitem [{\citenamefont {Donnelly}\ and\ \citenamefont
  {Sick}(1999{\natexlab{a}})}]{Donnelly:1998xg}%
  \BibitemOpen
  \bibfield  {author} {\bibinfo {author} {\bibfnamefont {T.~W.}\ \bibnamefont
  {Donnelly}}\ and\ \bibinfo {author} {\bibfnamefont {I.}~\bibnamefont
  {Sick}},\ }\href {\doibase 10.1103/PhysRevLett.82.3212} {\bibfield  {journal}
  {\bibinfo  {journal} {Phys. Rev. Lett.}\ }\textbf {\bibinfo {volume} {82}},\
  \bibinfo {pages} {3212} (\bibinfo {year} {1999}{\natexlab{a}})},\ \Eprint
  {http://arxiv.org/abs/nucl-th/9809063} {arXiv:nucl-th/9809063} \BibitemShut
  {NoStop}%
\bibitem [{\citenamefont {Donnelly}\ and\ \citenamefont
  {Sick}(1999{\natexlab{b}})}]{Donnelly:1999sw}%
  \BibitemOpen
  \bibfield  {author} {\bibinfo {author} {\bibfnamefont {T.~W.}\ \bibnamefont
  {Donnelly}}\ and\ \bibinfo {author} {\bibfnamefont {I.}~\bibnamefont
  {Sick}},\ }\href {\doibase 10.1103/PhysRevC.60.065502} {\bibfield  {journal}
  {\bibinfo  {journal} {Phys. Rev. C}\ }\textbf {\bibinfo {volume} {60}},\
  \bibinfo {pages} {065502} (\bibinfo {year} {1999}{\natexlab{b}})},\ \Eprint
  {http://arxiv.org/abs/nucl-th/9905060} {arXiv:nucl-th/9905060} \BibitemShut
  {NoStop}%
\bibitem [{\citenamefont {Jourdan}(1996)}]{Jourdan:1996np}%
  \BibitemOpen
  \bibfield  {author} {\bibinfo {author} {\bibfnamefont {J.}~\bibnamefont
  {Jourdan}},\ }\href {\doibase 10.1016/0375-9474(96)00143-1} {\bibfield
  {journal} {\bibinfo  {journal} {Nucl. Phys. A}\ }\textbf {\bibinfo {volume}
  {603}},\ \bibinfo {pages} {117} (\bibinfo {year} {1996})}\BibitemShut
  {NoStop}%
\bibitem [{\citenamefont {Maieron}\ \emph {et~al.}(2002)\citenamefont
  {Maieron}, \citenamefont {Donnelly},\ and\ \citenamefont
  {Sick}}]{Maieron:2001it}%
  \BibitemOpen
  \bibfield  {author} {\bibinfo {author} {\bibfnamefont {C.}~\bibnamefont
  {Maieron}}, \bibinfo {author} {\bibfnamefont {T.~W.}\ \bibnamefont
  {Donnelly}}, \ and\ \bibinfo {author} {\bibfnamefont {I.}~\bibnamefont
  {Sick}},\ }\href {\doibase 10.1103/PhysRevC.65.025502} {\bibfield  {journal}
  {\bibinfo  {journal} {Phys. Rev. C}\ }\textbf {\bibinfo {volume} {65}},\
  \bibinfo {pages} {025502} (\bibinfo {year} {2002})},\ \Eprint
  {http://arxiv.org/abs/nucl-th/0109032} {arXiv:nucl-th/0109032} \BibitemShut
  {NoStop}%
\bibitem [{\citenamefont {Caballero}\ \emph {et~al.}(2005)\citenamefont
  {Caballero}, \citenamefont {Amaro}, \citenamefont {Barbaro}, \citenamefont
  {Donnelly}, \citenamefont {Maieron},\ and\ \citenamefont
  {Udias}}]{Caballero:2005sj}%
  \BibitemOpen
  \bibfield  {author} {\bibinfo {author} {\bibfnamefont {J.~A.}\ \bibnamefont
  {Caballero}}, \bibinfo {author} {\bibfnamefont {J.~E.}\ \bibnamefont
  {Amaro}}, \bibinfo {author} {\bibfnamefont {M.~B.}\ \bibnamefont {Barbaro}},
  \bibinfo {author} {\bibfnamefont {T.~W.}\ \bibnamefont {Donnelly}}, \bibinfo
  {author} {\bibfnamefont {C.}~\bibnamefont {Maieron}}, \ and\ \bibinfo
  {author} {\bibfnamefont {J.~M.}\ \bibnamefont {Udias}},\ }\href {\doibase
  10.1103/PhysRevLett.95.252502} {\bibfield  {journal} {\bibinfo  {journal}
  {Phys. Rev. Lett.}\ }\textbf {\bibinfo {volume} {95}},\ \bibinfo {pages}
  {252502} (\bibinfo {year} {2005})},\ \Eprint
  {http://arxiv.org/abs/nucl-th/0504040} {arXiv:nucl-th/0504040} \BibitemShut
  {NoStop}%
\bibitem [{\citenamefont {Gonzal\'ez-Jim\'enez}\ \emph
  {et~al.}(2014)\citenamefont {Gonzal\'ez-Jim\'enez}, \citenamefont {Megias},
  \citenamefont {Barbaro}, \citenamefont {Caballero},\ and\ \citenamefont
  {Donnelly}}]{Gonzalez-Jimenez:2014eqa}%
  \BibitemOpen
  \bibfield  {author} {\bibinfo {author} {\bibfnamefont {R.}~\bibnamefont
  {Gonzal\'ez-Jim\'enez}}, \bibinfo {author} {\bibfnamefont {G.~D.}\
  \bibnamefont {Megias}}, \bibinfo {author} {\bibfnamefont {M.~B.}\
  \bibnamefont {Barbaro}}, \bibinfo {author} {\bibfnamefont {J.~A.}\
  \bibnamefont {Caballero}}, \ and\ \bibinfo {author} {\bibfnamefont {T.~W.}\
  \bibnamefont {Donnelly}},\ }\href {\doibase 10.1103/PhysRevC.90.035501}
  {\bibfield  {journal} {\bibinfo  {journal} {Phys. Rev. C}\ }\textbf {\bibinfo
  {volume} {90}},\ \bibinfo {pages} {035501} (\bibinfo {year} {2014})},\
  \Eprint {http://arxiv.org/abs/1407.8346} {arXiv:1407.8346 [nucl-th]}
  \BibitemShut {NoStop}%
\bibitem [{\citenamefont {Megias}\ \emph {et~al.}(2016)\citenamefont {Megias},
  \citenamefont {Amaro}, \citenamefont {Barbaro}, \citenamefont {Caballero},\
  and\ \citenamefont {Donnelly}}]{Megias:2016lke}%
  \BibitemOpen
  \bibfield  {author} {\bibinfo {author} {\bibfnamefont {G.~D.}\ \bibnamefont
  {Megias}}, \bibinfo {author} {\bibfnamefont {J.~E.}\ \bibnamefont {Amaro}},
  \bibinfo {author} {\bibfnamefont {M.~B.}\ \bibnamefont {Barbaro}}, \bibinfo
  {author} {\bibfnamefont {J.~A.}\ \bibnamefont {Caballero}}, \ and\ \bibinfo
  {author} {\bibfnamefont {T.~W.}\ \bibnamefont {Donnelly}},\ }\href {\doibase
  10.1103/PhysRevD.94.013012} {\bibfield  {journal} {\bibinfo  {journal} {Phys.
  Rev. D}\ }\textbf {\bibinfo {volume} {94}},\ \bibinfo {pages} {013012}
  (\bibinfo {year} {2016})},\ \Eprint {http://arxiv.org/abs/1603.08396}
  {arXiv:1603.08396 [nucl-th]} \BibitemShut {NoStop}%
\bibitem [{\citenamefont {Caballero}\ \emph {et~al.}(2010)\citenamefont
  {Caballero}, \citenamefont {Barbaro}, \citenamefont {Antonov}, \citenamefont
  {Ivanov},\ and\ \citenamefont {Donnelly}}]{PhysRevC.81.055502}%
  \BibitemOpen
  \bibfield  {author} {\bibinfo {author} {\bibfnamefont {J.~A.}\ \bibnamefont
  {Caballero}}, \bibinfo {author} {\bibfnamefont {M.~B.}\ \bibnamefont
  {Barbaro}}, \bibinfo {author} {\bibfnamefont {A.~N.}\ \bibnamefont
  {Antonov}}, \bibinfo {author} {\bibfnamefont {M.~V.}\ \bibnamefont {Ivanov}},
  \ and\ \bibinfo {author} {\bibfnamefont {T.~W.}\ \bibnamefont {Donnelly}},\
  }\href {\doibase 10.1103/PhysRevC.81.055502} {\bibfield  {journal} {\bibinfo
  {journal} {Phys. Rev. C}\ }\textbf {\bibinfo {volume} {81}},\ \bibinfo
  {pages} {055502} (\bibinfo {year} {2010})}\BibitemShut {NoStop}%
\bibitem [{\citenamefont {Antonov}\ \emph {et~al.}(2011)\citenamefont
  {Antonov}, \citenamefont {Ivanov}, \citenamefont {Caballero}, \citenamefont
  {Barbaro}, \citenamefont {Udias}, \citenamefont {Moya~de Guerra},\ and\
  \citenamefont {Donnelly}}]{PhysRevC.83.045504}%
  \BibitemOpen
  \bibfield  {author} {\bibinfo {author} {\bibfnamefont {A.~N.}\ \bibnamefont
  {Antonov}}, \bibinfo {author} {\bibfnamefont {M.~V.}\ \bibnamefont {Ivanov}},
  \bibinfo {author} {\bibfnamefont {J.~A.}\ \bibnamefont {Caballero}}, \bibinfo
  {author} {\bibfnamefont {M.~B.}\ \bibnamefont {Barbaro}}, \bibinfo {author}
  {\bibfnamefont {J.~M.}\ \bibnamefont {Udias}}, \bibinfo {author}
  {\bibfnamefont {E.}~\bibnamefont {Moya~de Guerra}}, \ and\ \bibinfo {author}
  {\bibfnamefont {T.~W.}\ \bibnamefont {Donnelly}},\ }\href {\doibase
  10.1103/PhysRevC.83.045504} {\bibfield  {journal} {\bibinfo  {journal} {Phys.
  Rev. C}\ }\textbf {\bibinfo {volume} {83}},\ \bibinfo {pages} {045504}
  (\bibinfo {year} {2011})}\BibitemShut {NoStop}%
\bibitem [{\citenamefont {L\"owdin}(1955)}]{PhysRev.97.1474}%
  \BibitemOpen
  \bibfield  {author} {\bibinfo {author} {\bibfnamefont {P.-O.}\ \bibnamefont
  {L\"owdin}},\ }\href {\doibase 10.1103/PhysRev.97.1474} {\bibfield  {journal}
  {\bibinfo  {journal} {Phys. Rev.}\ }\textbf {\bibinfo {volume} {97}},\
  \bibinfo {pages} {1474} (\bibinfo {year} {1955})}\BibitemShut {NoStop}%
\bibitem [{\citenamefont {Stoitsov}\ \emph {et~al.}(1993)\citenamefont
  {Stoitsov}, \citenamefont {Antonov},\ and\ \citenamefont
  {Dimitrova}}]{PhysRevC.48.74}%
  \BibitemOpen
  \bibfield  {author} {\bibinfo {author} {\bibfnamefont {M.~V.}\ \bibnamefont
  {Stoitsov}}, \bibinfo {author} {\bibfnamefont {A.~N.}\ \bibnamefont
  {Antonov}}, \ and\ \bibinfo {author} {\bibfnamefont {S.~S.}\ \bibnamefont
  {Dimitrova}},\ }\href {\doibase 10.1103/PhysRevC.48.74} {\bibfield  {journal}
  {\bibinfo  {journal} {Phys. Rev. C}\ }\textbf {\bibinfo {volume} {48}},\
  \bibinfo {pages} {74} (\bibinfo {year} {1993})}\BibitemShut {NoStop}%
\bibitem [{\citenamefont {Ivanov}\ \emph {et~al.}(2014)\citenamefont {Ivanov},
  \citenamefont {Antonov}, \citenamefont {Caballero}, \citenamefont {Megias},
  \citenamefont {Barbaro}, \citenamefont {de~Guerra},\ and\ \citenamefont
  {Ud\'{\i}as}}]{PhysRevC.89.014607}%
  \BibitemOpen
  \bibfield  {author} {\bibinfo {author} {\bibfnamefont {M.~V.}\ \bibnamefont
  {Ivanov}}, \bibinfo {author} {\bibfnamefont {A.~N.}\ \bibnamefont {Antonov}},
  \bibinfo {author} {\bibfnamefont {J.~A.}\ \bibnamefont {Caballero}}, \bibinfo
  {author} {\bibfnamefont {G.~D.}\ \bibnamefont {Megias}}, \bibinfo {author}
  {\bibfnamefont {M.~B.}\ \bibnamefont {Barbaro}}, \bibinfo {author}
  {\bibfnamefont {E.~M.}\ \bibnamefont {de~Guerra}}, \ and\ \bibinfo {author}
  {\bibfnamefont {J.~M.}\ \bibnamefont {Ud\'{\i}as}},\ }\href {\doibase
  10.1103/PhysRevC.89.014607} {\bibfield  {journal} {\bibinfo  {journal} {Phys.
  Rev. C}\ }\textbf {\bibinfo {volume} {89}},\ \bibinfo {pages} {014607}
  (\bibinfo {year} {2014})}\BibitemShut {NoStop}%
\bibitem [{\citenamefont {Dutta}(1999)}]{Dutta:1999}%
  \BibitemOpen
  \bibfield  {author} {\bibinfo {author} {\bibfnamefont {D.}~\bibnamefont
  {Dutta}},\ }\emph {\bibinfo {title} {The ($e,e'p$) Reaction Mechanism in the
  Quasi-Elastic Region}},\ \href@noop {} {Ph.D. thesis},\ \bibinfo  {school}
  {Northwestern University} (\bibinfo {year} {1999})\BibitemShut {NoStop}%
\bibitem [{\citenamefont {Ankowski}\ and\ \citenamefont
  {Sobczyk}(2008)}]{PhysRevC.77.044311}%
  \BibitemOpen
  \bibfield  {author} {\bibinfo {author} {\bibfnamefont {A.~M.}\ \bibnamefont
  {Ankowski}}\ and\ \bibinfo {author} {\bibfnamefont {J.~T.}\ \bibnamefont
  {Sobczyk}},\ }\href {\doibase 10.1103/PhysRevC.77.044311} {\bibfield
  {journal} {\bibinfo  {journal} {Phys. Rev. C}\ }\textbf {\bibinfo {volume}
  {77}},\ \bibinfo {pages} {044311} (\bibinfo {year} {2008})}\BibitemShut
  {NoStop}%
\bibitem [{\citenamefont {Horikawa}\ \emph {et~al.}(1980)\citenamefont
  {Horikawa}, \citenamefont {Lenz},\ and\ \citenamefont
  {Mukhopadhyay}}]{PhysRevC.22.1680}%
  \BibitemOpen
  \bibfield  {author} {\bibinfo {author} {\bibfnamefont {Y.}~\bibnamefont
  {Horikawa}}, \bibinfo {author} {\bibfnamefont {F.}~\bibnamefont {Lenz}}, \
  and\ \bibinfo {author} {\bibfnamefont {N.~C.}\ \bibnamefont {Mukhopadhyay}},\
  }\href {\doibase 10.1103/PhysRevC.22.1680} {\bibfield  {journal} {\bibinfo
  {journal} {Phys. Rev. C}\ }\textbf {\bibinfo {volume} {22}},\ \bibinfo
  {pages} {1680} (\bibinfo {year} {1980})}\BibitemShut {NoStop}%
\bibitem [{\citenamefont {Clark}\ \emph {et~al.}(2006)\citenamefont {Clark},
  \citenamefont {Cooper},\ and\ \citenamefont {Hama}}]{PhysRevC.73.024608}%
  \BibitemOpen
  \bibfield  {author} {\bibinfo {author} {\bibfnamefont {B.~C.}\ \bibnamefont
  {Clark}}, \bibinfo {author} {\bibfnamefont {E.~D.}\ \bibnamefont {Cooper}}, \
  and\ \bibinfo {author} {\bibfnamefont {S.}~\bibnamefont {Hama}},\ }\href
  {\doibase 10.1103/PhysRevC.73.024608} {\bibfield  {journal} {\bibinfo
  {journal} {Phys. Rev. C}\ }\textbf {\bibinfo {volume} {73}},\ \bibinfo
  {pages} {024608} (\bibinfo {year} {2006})}\BibitemShut {NoStop}%
\bibitem [{ROO(2023)}]{ROOT}%
  \BibitemOpen
  \href@noop {} {\enquote {\bibinfo {title} {{ROOT/MINUIT}},}\ }\bibinfo
  {howpublished} {\url{https://root.cern.ch/doc/master/classTMinuit.html}}
  (\bibinfo {year} {2023})\BibitemShut {NoStop}%
\bibitem [{MB_(2010)}]{MB_neu_data}%
  \BibitemOpen
  \href@noop {} {\enquote {\bibinfo {title} {{MiniBooNE NCE neutrino data
  release}},}\ }\bibinfo {howpublished}
  {\url{https://www-boone.fnal.gov/for\_physicists/data\_release/ncel/}}
  (\bibinfo {year} {2010})\BibitemShut {NoStop}%
\bibitem [{MB_(2015)}]{MB_antineu_data}%
  \BibitemOpen
  \href@noop {} {\enquote {\bibinfo {title} {{MiniBooNE NCE anti-neutrino data
  release}},}\ }\bibinfo {howpublished}
  {\url{https://www-boone.fnal.gov/for\_physicists/data\_release/ncel\_nubar/}}
  (\bibinfo {year} {2015})\BibitemShut {NoStop}%
\bibitem [{\citenamefont {Perevalov}(2009)}]{Perevalov_phdthesis}%
  \BibitemOpen
  \bibfield  {author} {\bibinfo {author} {\bibfnamefont {D.}~\bibnamefont
  {Perevalov}},\ }\emph {\bibinfo {title} {Neutrino-nucleus neutral current
  elastic interactions measurement in MiniBooNE}},\ \href
  {https://www-boone.fnal.gov/publications/Papers/denis_thesis.pdf} {Ph.D.
  thesis},\ \bibinfo  {school} {The University of Alabama}, \bibinfo {address}
  {Tuscaloosa, Alabama} (\bibinfo {year} {2009})\BibitemShut {NoStop}%
\bibitem [{\citenamefont {Kelly}(2004)}]{Kelly:2004hm}%
  \BibitemOpen
  \bibfield  {author} {\bibinfo {author} {\bibfnamefont {J.~J.}\ \bibnamefont
  {Kelly}},\ }\href {\doibase 10.1103/PhysRevC.70.068202} {\bibfield  {journal}
  {\bibinfo  {journal} {Phys. Rev. C}\ }\textbf {\bibinfo {volume} {70}},\
  \bibinfo {pages} {068202} (\bibinfo {year} {2004})}\BibitemShut {NoStop}%
\bibitem [{\citenamefont {Lovato}\ \emph {et~al.}(2014)\citenamefont {Lovato},
  \citenamefont {Gandolfi}, \citenamefont {Carlson}, \citenamefont {Pieper},\
  and\ \citenamefont {Schiavilla}}]{Lovato:2014eva}%
  \BibitemOpen
  \bibfield  {author} {\bibinfo {author} {\bibfnamefont {A.}~\bibnamefont
  {Lovato}}, \bibinfo {author} {\bibfnamefont {S.}~\bibnamefont {Gandolfi}},
  \bibinfo {author} {\bibfnamefont {J.}~\bibnamefont {Carlson}}, \bibinfo
  {author} {\bibfnamefont {S.~C.}\ \bibnamefont {Pieper}}, \ and\ \bibinfo
  {author} {\bibfnamefont {R.}~\bibnamefont {Schiavilla}},\ }\href {\doibase
  10.1103/PhysRevLett.112.182502} {\bibfield  {journal} {\bibinfo  {journal}
  {Phys. Rev. Lett.}\ }\textbf {\bibinfo {volume} {112}},\ \bibinfo {pages}
  {182502} (\bibinfo {year} {2014})},\ \Eprint {http://arxiv.org/abs/1401.2605}
  {arXiv:1401.2605 [nucl-th]} \BibitemShut {NoStop}%
\bibitem [{\citenamefont {Lovato}\ \emph {et~al.}(2018)\citenamefont {Lovato},
  \citenamefont {Gandolfi}, \citenamefont {Carlson}, \citenamefont {Lusk},
  \citenamefont {Pieper},\ and\ \citenamefont {Schiavilla}}]{Lovato:2017cux}%
  \BibitemOpen
  \bibfield  {author} {\bibinfo {author} {\bibfnamefont {A.}~\bibnamefont
  {Lovato}}, \bibinfo {author} {\bibfnamefont {S.}~\bibnamefont {Gandolfi}},
  \bibinfo {author} {\bibfnamefont {J.}~\bibnamefont {Carlson}}, \bibinfo
  {author} {\bibfnamefont {E.}~\bibnamefont {Lusk}}, \bibinfo {author}
  {\bibfnamefont {S.~C.}\ \bibnamefont {Pieper}}, \ and\ \bibinfo {author}
  {\bibfnamefont {R.}~\bibnamefont {Schiavilla}},\ }\href {\doibase
  10.1103/PhysRevC.97.022502} {\bibfield  {journal} {\bibinfo  {journal} {Phys.
  Rev. C}\ }\textbf {\bibinfo {volume} {97}},\ \bibinfo {pages} {022502}
  (\bibinfo {year} {2018})},\ \Eprint {http://arxiv.org/abs/1711.02047}
  {arXiv:1711.02047 [nucl-th]} \BibitemShut {NoStop}%
\bibitem [{\citenamefont {Gonz\'alez-Jim\'enez}\ \emph
  {et~al.}(2014)\citenamefont {Gonz\'alez-Jim\'enez}, \citenamefont
  {Caballero},\ and\ \citenamefont {Donnelly}}]{Gonzalez-Jimenez:2014bia}%
  \BibitemOpen
  \bibfield  {author} {\bibinfo {author} {\bibfnamefont {R.}~\bibnamefont
  {Gonz\'alez-Jim\'enez}}, \bibinfo {author} {\bibfnamefont {J.~A.}\
  \bibnamefont {Caballero}}, \ and\ \bibinfo {author} {\bibfnamefont {T.~W.}\
  \bibnamefont {Donnelly}},\ }\href {\doibase 10.1103/PhysRevD.90.033002}
  {\bibfield  {journal} {\bibinfo  {journal} {Phys. Rev. D}\ }\textbf {\bibinfo
  {volume} {90}},\ \bibinfo {pages} {033002} (\bibinfo {year} {2014})},\
  \Eprint {http://arxiv.org/abs/1403.5119} {arXiv:1403.5119 [nucl-th]}
  \BibitemShut {NoStop}%
\bibitem [{\citenamefont {Ren}(2022)}]{Ren:2022qop}%
  \BibitemOpen
  \bibfield  {author} {\bibinfo {author} {\bibfnamefont {L.}~\bibnamefont
  {Ren}},\ }\href {\doibase 10.7566/JPSCP.37.020309} {\bibfield  {journal}
  {\bibinfo  {journal} {JPS Conf. Proc.}\ }\textbf {\bibinfo {volume} {37}},\
  \bibinfo {pages} {020309} (\bibinfo {year} {2022})}\BibitemShut {NoStop}%
\end{thebibliography}%

\end{document}